\documentclass[aps,pre,showpacs,preprintnumbers,onecolumn,twocolumn,nofootinbib,superscriptaddress]{revtex4-1}

\expandafter\let\csname equation*\endcsname\relax
\expandafter\let\csname endequation*\endcsname\relax
\usepackage{amsmath}

\usepackage{graphicx}
\usepackage{amssymb}
\usepackage{dsfont}
\usepackage{epstopdf}
\usepackage{color}
\usepackage{cancel}

\usepackage{mathtools}
\usepackage{hyperref}

\definecolor{red}{rgb}{1,0,0}
\definecolor{blue}{rgb}{0,0,1}
\definecolor{black}{rgb}{0,0,0}

\newcommand{\p}{\partial}
\newcommand{\eq}[1]{\begin{align}#1\end{align}}
\newcommand{\eqs}[1]{\begin{align*}#1\end{align*}}

\newcommand{\half}{\mbox{$\frac{1}{2}$}}
\newcommand{\OO}{\mathcal{O}}
\newcommand{\GG}{\mathcal{G}}

\newcommand{\PP}{\mathbb{P}}
\newcommand{\ZZ}{\mathbb{Z}}

\newcommand{\FF}{\mathcal{F}}
\newcommand{\TT}{\mathcal{T}}

\newcommand{\NN}{\mathcal{N}}

\newcommand{\ed}{\tilde\epsilon_d}
\newcommand{\es}{\tilde\epsilon_s}

\newcommand{\Ld}{L^\dagger}
\newcommand{\Rd}{R^\dagger}

\makeatletter
\g@addto@macro\@floatboxreset\centering
\makeatother

\begin{document}
\title{Scaling limit of the Random Language Model}
\author{Eric De \!Giuli}
\affiliation{Department of Physics, Toronto Metropolitan University, M5B 2K3, Toronto, Canada}
\affiliation{Capital Fund Management, 75007 Paris, France}

\begin{abstract}

We develop a quantitative theory of the Random Language Model (RLM), an ensemble of stochastic context-free grammars, in a scaling limit where the number of hidden symbols $N \to \infty$ while the grammar temperature $\ed \to 0$ at fixed $x = \ed \log N$. In this limit, the model admits a controlled description based on a large-deviation principle over rule-usage patterns. A semi-annealed approximation maps the problem to a class of Random Energy Models with nontrivial combinatorics.

We show that the RLM exhibits a freezing transition at a critical value $x_c=1/8$, below which rule usage condenses on a subset of preferred rules and language statistics acquire a nontrivial dependence on corpus length. A second characteristic scale at $x=1/2$, associated with the breakdown of Gaussian fluctuations, separates transition and babbling regimes. 
Across regimes, we derive explicit scaling laws for the number of distinct rules, entropy, and energy, identifying distinct frozen and critical regimes controlled by the interplay of grammar size, corpus length, and temperature. 

The theory resolves previous ambiguities regarding the existence of a thermodynamic transition and explains the slow approach to the large-$N$ limit as a consequence of the dependence on $\log N$. It further provides a unified framework in which universal statistical properties of language emerge from typical realizations of generative grammars, with implications for both natural language statistics and the behavior of large language models. Heaps' law and Zipf's law are both derived, and the dependence of predictive entropy on context length is rationalized. Comparison with empirical measurements of Heaps' and Zipf's laws yields parameter estimates for real languages.

\end{abstract}

\maketitle


\tableofcontents

\section{Introduction}

Despite its perennial interest, the quantitative study of language has focused on a small number of observables, among them: (1) the distribution of word rarity in a text, characterized by Zipf's law and Heaps' law \cite{Zipf13,Heaps78}; and (2) the entropy rate of text \cite{Shannon64,Shannon51, Schurmann96,Takahira16}. By now these are empirically well characterized: in its simplest form, Zipf's law states that when sorted by decreasing frequency in a text, the $k^{th}$ word will have relative frequency $1/k^\zeta$ with $\zeta \approx 1$; Heaps' law states that in a text of length $\ell$ the number of unique words scales as $n_* \sim \ell^{b}$ with $b \approx 0.5$; and the entropy rate in English using contexts of roughly 100 characters is $\approx 1.3$ bits/character. The mild language-dependence of these quantities has also been characterized, and will be discussed later in detail \cite{Petersen12,Ferrer-i-Cancho01,Corominas-Murtra10,Corral15,Bentz17}. They provide a nontrivial but coarse, low-dimensional view of language structure.


The simple universal character of these results stands in contrast to classical linguistic studies, which emphasize the structural diversity of syntax across natural languages, as developed in generative grammar and related traditions \cite{Chomsky02,Chomsky14,Chomsky14a,De-Saussure16}. While these approaches provide highly refined formal descriptions of grammatical structure, they do not, in general, yield predictions for coarse-grained observables such as frequency distributions or entropy rates. Partial bridges exist, for example in dependency and probabilistic grammars \cite{Jurafsky00,Ferrer-i-Cancho22}, but a unified quantitative framework linking syntactic structure to statistical regularities remains underdeveloped.

A conceptual bridge between statistical and generative approaches is obtained by considering an ensemble of languages, where the definition of the ensemble encodes the structural constraints, while the behavior of typical languages can give rise to universal laws. Such an ensemble was proposed in \cite{DeGiuli19} and dubbed the Random Language Model (RLM), taking the context-free structure of natural and computer languages as the basis for the ensemble. A language is created from a stochastic context-free grammar, which is a generative model for sequences with hidden tree structure, called derivations. Each grammar generates an infinite set of possible sequences, and assigns to each sequence a probability $P$ of occurrence. 
 The dynamic range of probabilities is controlled by two temperature-like parameters, $\ed$ and $\es$, which control the variability of grammar rules and symbol emissions, respectively. At large $\ed,\es$ the output distribution approaches maximal entropy and exhibits little structure, while at small $\ed,\es$ it becomes highly constrained. 

Context-free grammar-based codes are universal \cite{Kieffer00}, in that sequences from any stationary and ergodic source can be compressed by a grammar-based code whose compression rate approaches the entropy rate of the source. This justifies the use of context-free grammars as a parametrization of sequence generators. In particular,
they span the entire range from trivial output with maximal entropy to arbitrarily constrained sequences of vanishing entropy. 

Ref. \cite{DeGiuli19} showed that typical languages show a strong dependence on $\ed$ such that the entropy rate of sequences is nearly maximal above a certain value $\ed>\ed{}_*$ while it drops precipitously below $\ed{}_*$. A spin-glass order parameter was found to be consistent with this interpretation, although a controlled theoretical description was not obtained, and a single critical temperature was not identified.

Ref. \cite{DeGiuli19a} showed that the RLM has a field-theoretic representation, which was used to build a quantitative description of the high-temperature phase in the replica-symmetric approximation \cite{Mezard87}. It  predicted that the low-temperature regime will begin when $\ed \log N \lesssim \rm const$. While this could explain the spin-glass order parameter of \cite{DeGiuli19}, it ultimately did not lead to a complete description of the low-temperature phase.

In Ref. \cite{Nakaishi22}, the existence of a true thermodynamic phase transition in the RLM was questioned. By analogy with Heisenberg spin glasses, the Binder cumulant was used to diagnose a phase transition; the results suggest that a true thermodynamic phase transition could exist only in the limit where the number of hidden symbols $N$ diverges. This appears to be in tension with the empirical results of \cite{DeGiuli19}, where measured properties at different $N$ from $N=10$ to $N=40$ (and up to $N=80$ in \cite{Lalegani24}) showed a quantitative dependence on $N$ that could be collapsed. These correspond to grammar sizes $N^3$ from $1000$ to $64000$. Such finite-size scaling suggests that these $N$ are already in a scaling regime not qualitatively different from $N \to \infty$. 

Here, we resolve this apparent paradox by a detailed theoretical and empirical study of the RLM, in a new approximation that is applicable to the low-temperature phase. This {\it semi-annealed} approximation considers an annealed average over grammar weights, but over finite grammars and derivations. It respects both the wide dynamic range of the low-temperature phase and the finiteness of $N$ (and trees) in any putative application of the model. The essential step is to consider a double scaling limit in which $N \to \infty$ while $\ed \to 0$ such that $x=\ed\log N$ is finite; this maintains the ability to span the full range of entropy while enabling analytical control in the large $N$ limit. We will show that the RLM inherits critical behavior from the Random Energy Model (REM) of Derrida \cite{Derrida80,Derrida81}, but enriched by the tree structure of derivations. There is a critical condensation transition at $x_c=1/8$, below which the entropy rate is $\OO(1)$ in the large $N$ limit, while it is $\OO(\log N)$ above. It will follow in the course of the theory that the large $N$ behavior is primarily controlled by $\log N$, explaining the slowness of approach to the putative thermodynamic limit observed in numerical studies. 

We show that Heaps-like sublinear vocabulary growth and Zipf's rank-frequency relationship are predicted by the theory, derive the nontrivial dependence of their associated exponents on grammar temperatures, corpus-size, and vocabulary size, and use these results to infer model parameters from published data in real languages \cite{Petersen12,Gerlach13}. We also show that the dependence of predictive entropy (or code length) on context length is captured, under mild modelling assumptions. Overall, we show that universal statistical properties of language emerge from a minimal ensemble of generative grammars governed by a well-defined condensation transition.

A brief report of this theory has been presented in \cite{Giorlandino26}.

\subsection{Roadmap}

The outline of this paper is as follows. First, we establish by a simple example the connection to the REM, which motivates the definition of the scaling limit. Namely, RLM partition functions involve sums whose convergence via the Law of Large Numbers (LLN) can break down at low enough temperature. These sums take the form of partition sums of Random Energy Models, establishing the basic connection between the two models. 
This discussion motivates the semi-annealed approximation that follows. We rewrite the RLM partition function in terms of excitations called `patterns,' which are analogous to occupation numbers in statistical mechanics. We present the combinatorics of counting derivations, both for small and large corpora, and including both the tree structure and the surface layer. We then derive the most likely pattern and compare this prediction directly with numerics, finding good agreement. The theory shows that the number of unique hidden rules $p$ used in a corpus is a key observable. We present numerical results of $p$ versus $x$, $N$, and corpus length $\ell$, finding various scaling laws, which are characterized. Then we analytically find the leading-order behavior of $p$ in all regimes, showing that all numerical results are accounted for within the approximation. 

 We then use these results to derive the thermodynamics of the model, in particular the energy and entropy. Theory explains the change of scaling of entropy, and quantitatively explains the energy, which shows a clear change in scaling at a critical $x$ that tends to $x_c=1/8$ in the scaling limit. Finally, we apply the theory both to natural language and Large Language Models (LLMs). We show how the theory reproduces Heaps' law and Zipf's law, in agreement with previous work, and the dependence of predictive entropy on context length observed in LLMs. 

\subsection{Model definition}

\begin{figure}[t!]
\includegraphics[width=.9\columnwidth]{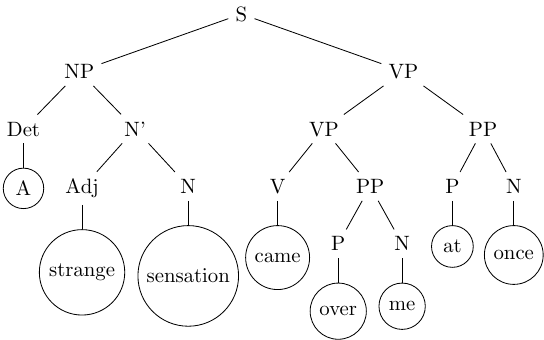}
\caption{ Illustrative derivation tree for a simple English sentence. Terminal symbols are encircled. 
}\label{fig1}
\end{figure}

The Random Language Model is defined as follows \cite{DeGiuli19}. A stochastic context-free grammar \cite{Chomsky02,Booth69,Wetherell80} is a generative model for sequences with a hidden tree structure; a grammar in Chomsky normal form defines a weight over binary trees, called derivations, in which the leaves of the tree represent the observed sequence; Fig. \ref{fig1} shows an example derivation of English phrase structure. The grammar $\GG$ is defined by two objects $\GG = \{M_{abc}, O_{aB} \}$ over sets of hidden (``nonterminal") symbols indexed by $a=1,\ldots,N$ and observable (``terminal") symbols indexed by $B=1,\ldots,T$ such that the rule $a \to bc$ has the weight $M_{abc}$ and the rule $a \to B$ has the weight $O_{aB}$. Conventionally, each derivation begins with a distinguished symbol, called the start symbol, which is the root. The model of \cite{DeGiuli19} defines weights for tree topologies based on a branching process with emission probability $p_E$, so that a binary tree $\TT$ with $\ell$ leaves has a probability $\PP(\TT | \GG)  = W_{tree}(\TT)/Z_{tree}$ with $W_{tree}(\TT) = p_E^{\ell} (1-p_E)^{\ell-1}$. The `emission probability' $p_E$ controls the size of trees \footnote{$Z_{tree} = 2p_E/(1+|2p_E-1|) = 1$ for $p_E>1/2$.}. The tree is defined by a set of interior branches $\Omega_\TT$ and surface branches $\p \Omega_\TT$. The emission probability $p_E$ needs to be tuned close to $1/2^+$ to obtain large trees; if we write $p_E = 1/2 + \varepsilon$ then the distribution of tree lengths is given by $\PP(\ell) \propto e^{-\ell/\xi}/\ell^{3/2}$ with $\xi = 1/(4\varepsilon^2)$. 

\begin{figure*}[t!]
\includegraphics[width=.9\textwidth]{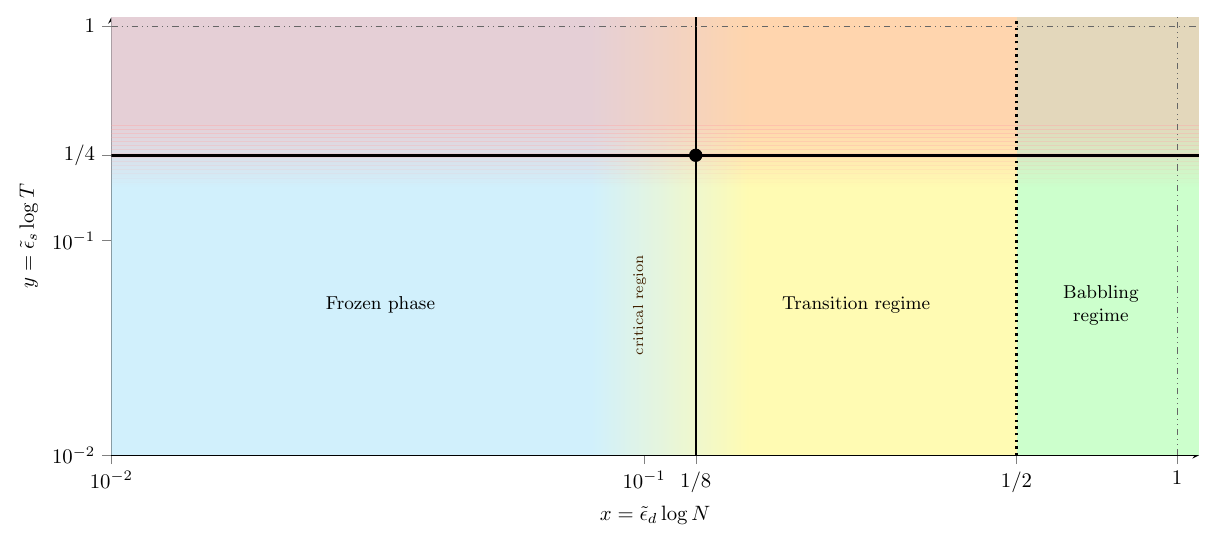}
\caption{ Phase diagram of the RLM as a function of $x = \ed\log N$ and $y = \es \log T$, in the scaling limit $N,T\to \infty$, at fixed $x,y$. In the scaling limit $N \to \infty$, there are three main regimes along each axis: frozen, transition, and babbling, separated by $x$ values $x_c=1/8$ and $x=1/2$ in $x$, and $y_c=1/4$ and $y=1$ in $y$. The gradients are schematic.
}\label{figphase}
\end{figure*}

Given the topology of a tree, $\GG$ defines a weight for each derivation on it. Writing the hidden symbols of a particular derivation as $\{\sigma_i\}$ and the observable symbols as $\{ o_t \}$ the weight is
\eq{ \label{w1}
W( \{\sigma_i, o_t \} | \TT, \GG) = \prod_{\alpha \in \Omega_\TT} M_{\sigma_{\alpha_1} \sigma_{\alpha_2} \sigma_{\alpha_3} } \prod_{\alpha \in \p\Omega_\TT} O_{\sigma_{\alpha_1} o_{\alpha_2} },
}
where each $\alpha=(\alpha_1,\alpha_2,\alpha_3)$ is a factor in the order $\sigma_{\alpha_1} \to \sigma_{\alpha_2} \sigma_{\alpha_3}$. We can write the weight in a Boltzmann form
\eq{ \label{w2}
W( \{\sigma_i, o_t \} | \TT, \GG) = e^{-E},
}
with
\eqs{
E = -  \sum_{\alpha \in \Omega_\TT} \log M_{\sigma_{\alpha_1} \sigma_{\alpha_2} \sigma_{\alpha_3} } -\sum_{\alpha \in \p\Omega_\TT} \log O_{\sigma_{\alpha_1} o_{\alpha_2} }
}
We can write this as 
\eqs{
E & = -\sum_{a,b,c} \pi_{abc}(\sigma; \TT) \log M_{abc} - \sum_{a,B} \rho_{aB}(\sigma,o; \TT ) \log O_{aB}
}
where the usage frequencies are
\eqs{
\pi_{abc}(\sigma; \TT) & = \sum_{\alpha \in \Omega_{\TT}} \delta_{\sigma_{\alpha_1},a} \delta_{\sigma_{\alpha_2},b} \delta_{\sigma_{\alpha_3},c} \\
\rho_{aB}(\sigma,o; \TT) & = \sum_{\alpha \in \p \Omega_{\TT}} \delta_{\sigma_{\alpha_1},a} \delta_{o_{\alpha_2},B}
}
For a given derivation, these quantities count how many times each grammar rule is used. The grammar can be written as $\log M_{abc} = \log \overline{M} + \log M_{abc}/\overline{M} \equiv \log \overline{M} + X_{abc}$ and similarly $\log O_{aB} = \log \overline{O} + Y_{aB}$, where $\overline{M}=1/N^2$ and $\overline{O}=1/T$. Using the sum rules $\sum_{a,b,c} \pi_{abc} = |\Omega| = \ell - 1$ and $\sum_{a,B} \rho_{aB} = |\p \Omega| = \ell$ we can extract a mean contribution, independent of the configuration $\sigma, o$. Thus
\eqs{
E & = E_0 -\sum_{a,b,c} \pi_{abc}(\sigma; \TT) X_{abc} - \sum_{a,B} \rho_{aB}(\sigma,o; \TT ) Y_{aB}
}
with $E_0 = -(\ell-1) \log \overline{M} - \ell \log \overline{O}$. The partition function of derivations on a fixed tree and with a fixed grammar is 
\eq{
Z(\TT,\GG) = \sum_{\{ \sigma_i, o_t \} } e^{-E(\sigma,o)} 
}
and the probability of a derivation, given the tree and the grammar, in this weighted CFG, is 
\eqs{
\PP(\sigma,o | \TT,\GG) = \frac{e^{-E(\sigma,o)}}{Z(\TT,\GG)} .
}
The above definitions are for a single grammar $\GG$. The Random Language Model is an ensemble of grammars in which the $X_{abc}$ and $Y_{aB}$ are themselves random. Two variations of the model were presented in \cite{DeGiuli19}: in model W, a weighted CFG, the $X_{abc}$ and $Y_{aB}$ are drawn from a Gaussian distribution, iid for each rule, with variance $1/(2\ed)$ and $1/(2\es)$ for $X$ and $Y$, respectively. In model P, a probabilistic CFG, the weights are drawn in the same way but then normalized such that $\sum_{b,c}M_{abc}=1$ and $\sum_B O_{aB}=1$ for all $a$: this amounts to $X_{abc} \to X_{abc} - \log [\overline{M}\sum_{b,c}e^{X_{abc}}]$ and $Y_{aB} \to Y_{aB} - \log [\overline{O}\sum_{B}e^{Y_{aB}}]$. This normalization introduces correlations among weights within each row.

Model P admits direct generative sampling and satisfies $Z(\TT,\GG)=1$ for any fixed tree and fixed root symbol, as follows by recursively summing configurations from the leaves to the root. Model W provides a simpler starting point for the theoretical treatment because its log weights are independent Gaussian variables.

We develop the semi-annealed theory for model W. The closure does not resolve the joint fluctuations of head visitation and the
head-dependent row sums $\tilde M_a=\sum_{b,c}M_{abc}$ and $\tilde O_a=\sum_B O_{aB}$. As a result, the predictions should equally apply to model P. We test the resulting scaling predictions primarily against simulations of model P, which exhibit smaller grammar-to-grammar fluctuations. Comparisons with model W, shown in Appendix \ref{app:WCFG} and referred to in the main text, show the same scaling behavior in the regimes examined, with larger fluctuations and differences in amplitude.



\section{Connection to the Random Energy Model}

To connect the RLM to Derrida's Random Energy Model, consider initially a simple tree with a single branch, in model W. The root, which takes the start symbol 1, branches to two hidden symbols, with indices $b$ and $c$. These emit visible symbols $B$ and $C$, respectively. The partition function is
\eqs{
Z_{\Delta} & = \sum_{b,c,B,C} M_{1bc}O_{bB}O_{cC} \\
& = \overline{M} \overline{O}^2 \sum_{b,c} e^{X_{1bc}} \left( \sum_B e^{Y_{bB}} \right)\left( \sum_C e^{Y_{cC}} \right) .
}
The sums appearing here take the form of Random Energy Model sums. Let
\eq{
Z_{REM}(\NN,\lambda) = \sum_{j=1}^{\NN} e^{-\beta E_j} 
}
where $\langle E_j \rangle = 0, \langle (\beta E_j)^2 \rangle = \lambda^2/2$. Derrida's result \cite{Derrida80,Derrida81} is that the LLN breaks down when 
\eq{
\frac{\beta_c}{\beta} = \frac{2\sqrt{\log 2}}{\beta J} = \frac{2\sqrt{\log \NN}}{\lambda} 
}
is smaller than unity. The corresponding parameters for the above sums are
\eqs{
\NN^{(1)} & = N^2, \lambda^{(1)} = \sqrt{1/\ed} \\
\NN^{(2)} & = T, \lambda^{(2)} = \sqrt{1/\es} 
}
The corresponding temperature ratios are 
\eq{
(\beta_c/\beta)_d & \equiv \sqrt{8 \ed \log N} \\
(\beta_c/\beta)_s & \equiv \sqrt{4 \es \log T} 
}
When both of these are larger than unity, the LLN applies, and $Z$ is trivial. However as soon as one of these ratios is smaller than unity, $Z$ depends nontrivially on the rule weights, especially in larger trees. 

Derrida's result holds in the limit $\NN \to \infty$ at finite $\beta_c/\beta$, so that $\lambda \sim \sqrt{\log \NN}$. Then, there is a competition between increasing $\NN$, which favours averaging, and increasing $\lambda$, which increases energy variance. In the RLM, this opens the possibility to take $N \to \infty$ while also taking $\ed \to 0$, such that $\ed \log N$ remains finite. In this {\it scaling limit}, we can exploit simplifications due to large $N$ while also retaining the ability to control entropy and other language properties, by tuning 
\eq{
x = \ed \log N. 
}
Similar considerations imply that we should fix 
\eq{
y = \es \log T
}
to obtain nontrivial scaling in the surface properties.
Then the thermodynamic limit of the REM corresponds to the scaling limit of the RLM. 


These breakdowns are all of the LLN. As in the REM there will be corresponding breakdowns in the CLT at larger temperatures \cite{Ben-Arous05}. For a sum over $N^K$ variables, the LLN breaks down at $x_{\rm LLN}=1/4K$ while the CLT breaks down at $x_{\rm CLT}=1/K$. Thus each CLT transition occurs at a value of $x$ that is larger by a factor 4.

Although illustrative and approximate, since it neglects shared symbols between branches, this simple argument extends naturally to any finite tree, and correctly identifies the hidden and surface freezing transitions of the RLM. It yields the phase diagram of the RLM sketched in Fig.\ref{figphase}, as a function of $x$ and $y$.  A more formal approach uses the field theory of the RLM \cite{DeGiuli19a}, itself related to belief propagation equations \cite{Mezard09}. This is presented in Appendix \ref{FieldTheory}. In the subsequent development of the semi-annealed theory, the field theory will be used for supplemental calculations, but not main results.

We consider the scaling limit both for finite trees and in the limit of a large corpus; in the numerics of \cite{DeGiuli19}, trees were chosen by a branching process with $\langle \ell \rangle \approx 15$. This of course does not preclude considering the limit of an infinite corpus. In large trees, or when considering a forest of many trees, the above argument is too crude: we must attend to rule repetition, since then the various REM sums do not involve independent weights. This will entail a coupling between the combinatorics of derivations and their energetic contributions. The semi-annealed theory treats these self-consistently.


\section{Scaling in the low-temperature regime}


Consider the partition sum of the RLM for a fixed forest $\FF$ of trees with lengths $\{ \ell_i, i = 1 \ldots m \}$ and grammar $\GG$, which is written as a sum over trees $\TT_i$:
\eq{
\ZZ(\FF; \GG) = \prod_i Z(\TT_i,\GG) .
}
with 
\eqs{
Z(\TT,\GG) = \underbrace{\overline{M}^{\ell-1} \overline{O}^\ell }_{Z_0(\ell)} \sum_{\{ \sigma_i, o_t \} } e^{-E'}
}
and 
\eqs{
E' = -\sum_{a,b,c} \pi_{abc}(\sigma; \TT) X_{abc} -\sum_{a,B} \rho_{aB}(\sigma,o; \TT ) Y_{aB} .
}

\subsection{Excitations}

Fix a forest $\FF$ and define the rule spectrum $\{p_k, q_k, k \geq 1\}$ that for each $k$ indicates how many rules are used $k$ times, with $p_k$ counting the number of hidden rules and $q_k$ the number of terminal rules. For example, the derivation in Fig.\ref{fig1} has 7 branches, of which 2 use the same rule, so that $p_1 = 5, p_2 = 1,$ and $p_k=0$ for $k \geq 3$, and $q_1=8,q_k= 0$ for $k \geq 2$. The set of all derivations consistent with a given spectrum is called a pattern, $P(\{p_k,q_k\}; \FF)$. 

%
%

The partition sum $Z$ can be organized as a sum over patterns, which are the relevant excitations of the model:
\eq{
\ZZ(\FF; \GG) = \ZZ_0 \sum_{\{ P  \}} \underbrace{\sum_{ (\sigma,o) \in P } e^{- E'(\sigma,o; \GG)}}_{\mathcal{P}(P; \GG)} ,
}
where $\ZZ_0 = \prod_i Z(\ell_i)$ depends only on the number of trees $m$ and the total length $\ell = \sum_i \ell_i$. 
For any derivation, applying a global permutation to the symbols results in another derivation with the same pattern, but different rules. Since the model is defined by iid rule weights, each pattern thus explores the space of grammars. 
 In other words, even though $\ZZ$ is computed for a fixed grammar, sums over symbol values amount to sums over possible grammar weights. This is a form of self-averaging, which we use to justify the semi-annealed approximation that follows. It is however essential to respect the rule spectrum of the pattern, since this controls the Boltzmann weight.

The organization of the partition function as a sum over patterns suggests a natural large-deviation structure. Each pattern $P$ contributes a weight that combines an energetic term, controlled by the rule multiplicities, and a combinatorial entropy that counts the number of derivations realizing that spectrum. Crucially, both contributions diverge in the scaling limit (in particular both are powers of $N$, as confirmed below). As a result, the partition function can be written
\eqs{
\ZZ(\FF; \GG)/\ZZ_0 = \sum_{\{ P  \}} e^{S(P)} e^{-\hat E(P)} 
}
where $S(P)$ encodes the entropy of patterns and $\hat E(P)$ is an effective energy after averaging over rule weights. In the scaling limit this will be dominated by a saddle point $P_*$ that minimizes the effective free energy $\hat E-S$. This provides a basis for a {\it semi-annealed approximation}: we perform an annealed average over rule weights within each pattern, while retaining the constraint that the pattern itself is selected by a saddle-point condition at fixed $N$. In this sense, the approximation is not a fully annealed treatment of the original quenched problem, but rather a coarse-grained saddle-point evaluation over patterns, which captures the dominant contributions in the scaling limit.

This perspective also clarifies the role of the rule spectrum $\{ p_k, q_k \}$ as an order-parameter-like object: different phases of the model correspond to qualitatively different saddle-point spectra, reflecting whether weight is distributed over many rules or concentrated onto a few. As we show below, this structure leads to a condensation transition analogous to that of the Random Energy Model.

There are two important ingredients to the semi-annealed approximation. First, since derivations can be constructed sequentially from the root without backtracking, for a given head $a$ there are $N^2$ possible rules $a \to bc$ to consider. This defines an elementary REM partition sum, one for each branch.


Second, although rule distributions are heavy-tailed at small $\ed$, at any finite $N$ we only sample a finite number of values from the distribution. Using extreme-value statistics, a sum over $N^2$ variables sampled from $P(M)$ will typically not see a value larger than $M^*$ where
\eq{
\frac{1}{N^2} & = \int_{M^*}^\infty dM P(M) \notag \\
& = \int_{M^*}^\infty \frac{dM}{M\sqrt{\pi/\ed}} e^{-\ed \log^2 (M/\overline{M}) } \notag \\
& = \int_{X_*}^\infty  \frac{dX}{\sqrt{\pi/\ed}} e^{-\ed X^2 } \notag \\
& = \half \text{erfc}(\sqrt{\ed} X_*)
}
with $X=\log (M/\overline{M})$ and $X_* = \log (M^*/\overline{M})$. Define an average restricted to the interval $M<M^*$, or equivalently $X<X_*$:
\eqs{
\langle f(X) \rangle_{X<X_*} = \frac{\int_{-\infty}^{X_*} dX P(X) f(X)}{\int_{-\infty}^{X_*} dX P(X)}
}
Averages restricted to $X<X_*$ will be crucial in what follows.

We evaluate the contribution of a pattern in two steps. 
For any given assignment of rules, we count how many patterns realize that assignment; this yields one contribution to $S(P)$. Then we count assignments of rule identities and evaluate the most likely energy of the pattern. This yields $\hat E(P)$, and a second entropic contribution. 

\subsection{Pattern entropy}

Exact counting of derivations is difficult in large trees, but we can obtain a simple estimate by counting degrees of freedom. We consider a forest of $m$ trees, of total length $\ell$. There are $n= \ell-m$ branches. We have $n = \sum_k k p_k$ and $\ell = \sum_k k q_k$. The total number of unique hidden rules used in the pattern is $p = \sum_j p_j$, while $q= \sum_j q_j$ is the total number of unique terminal rules. Eventually we will need to decide on the relative size of $\ell$ and $N$.

Suppose we have fixed the identities of all the rules. In general, we can create different derivations by placing them in different configurations.
This can be estimated due to two simplifying factors of the model: first, the tree structure allows configurations to be built sequentially, starting from the root, without backtracking. Second, since all rules are iid, if for example we have a rule $a \to bc$, then the likelihood of having a rule whose head is $b$ is independent of the value $(a)$ of the head. Thus, suppose we have a pile of $\ell_c$ rules, of which $p$ of them are unique. We assume that $\ell_c/p$ is large, so that a typical rule appears frequently. Then, starting from the root, we will have typically $p/N$ rules with the start symbol as head. Continuing to its children, we will have $p/N$ choices for the next branch, and so on. This yields a simple estimate $(p/N)^{\ell_c}$ for the number of unique derivations. 
It implies that the pile will yield a unique derivation if $p \approx N$. 

This estimate is refined by an exact calculation based on Eqs.\ref{SP} (see Appendix \ref{AppA}). It confirms the $(p/N)^{\ell_c}$ result and applies even when $p<N$, when the above argument is unreliable; an analysis of corrections to the leading order reveals that the estimate works best when trees are large. 

However, this estimate does not account for the depletion of rules as we build derivations. We account for this by dividing up the full $n$ rules (counted with multiplicity) into piles of $\ell_c$ rules, each of which has $N$ unique rules. Each pile will then typically have a unique derivation, and plays the role of a correlation volume.  The global average repetition rate of a rule is $n/p$; however, this may differ from the local repetition rate of a rule within a pile, since rules can be shared among several piles. Let $m_c=n/\ell_c$ be the number of piles. Below, we will fix $m_c$ self-consistently by requiring that a typical pile contains $N$ distinct branching rules.

For $p>N$, the leading counting of derivations is therefore
approximated by counting the number of ways to partition the
initial $n$ rules, with given multiplicity spectrum, into
$m_c$ correlation volumes.

Label the volumes 1 through $m_c$ and consider each rule type $r$, i.e. triple $r= (a,b,c)$ corresponding to $a \to bc$. Suppose this rule occurs $\pi$ times in the complete pile of $n$ rules. Then there are $\binom{\pi + m_c - 1}{m_c -1}$ ways to put these $\pi$ rules into the correlation volumes. Repeating this for all rules and recalling the definitions of $\{ p_k \}$ we find \footnote{This procedure does not strictly enforce that each pile has exactly $\ell_c$ rules, counted with multiplicity. However in the large $m_c$ limit this is the most likely outcome. The corrections are computable; see Appendix \ref{AppB} on Contingency Tables.}
\eq{
\prod_{k \geq 1} \binom{k + m_c - 1}{m_c -1}^{p_k} .
}
As a result, counting derivations gives a factor
\eq{ \label{OmD1}
\Omega_D = \begin{cases} (p/N)^{n} &\qquad p/N<1 \\ \prod_{k \geq 1} \binom{k + m_c - 1}{m_c -1}^{p_k} &\qquad p/N>1 \end{cases}
}
We can now fix $m_c$. Notice that if a rule appears $k$ times in total, the probability that it does not appear in a particular pile is 
\eqs{
\frac{\binom{k + m_c - 2}{m_c - 2} }{  \binom{k + m_c - 1}{m_c -1} }= \frac{(k+m_c-2)!(m_c-1)!}{(k+m_c-1)!(m_c-2)!} = \frac{m_c-1}{k + m_c-1} ,
}
so that the probability for the rule to appear in this particular pile is 
\eq{
u_k = \frac{k}{k+m_c-1}. 
}
The statement that the particular pile has $N$ distinct branching rules then becomes
\eq{ \label{mcoverlapmain}
N=\sum_{k\geq1}p_k u_k . 
}
This implicitly determines $m_c$. If the summand is dominated by the regime where $k \ll m_c$ then it becomes $m_c \approx n/N$, but here we leave it general. 

Note that when there is a single correlation volume, $m_c=1$, then Eq.\ref{mcoverlapmain} collapses to $p=N$; so $p/N=1$ indeed marks the boundary between the two regimes.

Now we choose and place the terminal rules. Consider the hidden symbols as given, for any particular realization of the pattern. We first need to know how many unique hidden symbols there are that become leaves; for the branching process considered in the model, this can be estimated, as discussed in Appendix \ref{app:surface_coverage}. The result is that if $p \gg N$, then typically all hidden symbols are used. For simplicity we will assume this. If $p \sim N$ then typically a finite fraction of hidden symbols appear, which will reduce the amount of novelty in the surface layer but does not change overall scalings. 


Assuming that all hidden symbols are represented, the argument is very similar: first we choose the rules, $\binom{NT}{q} \sim (NT)^q/q!$ if $q \ll NT$; then we distribute these according to the rarity distribution, giving a multinomial coefficient $\binom{q}{q_1 q_2 \cdots }$. Then we need to count sequences that are compatible. For each hidden symbol that becomes a leaf, there are typically $q/N$ terminal rules that are compatible. Thus, as before we estimate $(q/N)^\ell$ sequences if rule depletion is ignored. Taking into account rule depletion, we split the $\ell$ rules (counted with multiplicity) into $m_c'$ piles of $\ell_c'$ rules each. If each pile has $N$ unique rules, then there will typically be one compatible sequence. Because local and global repetition rates may differ, we deduce the number $m_c'$ of piles by self-consistency, requiring
\eq{ \label{mcpoverlapmain}
N = \sum_k q_k \frac{k}{k+m_c'-1}
}

The result from counting sequences is a factor
\eq{ \label{OmpD1}
\Omega_D' = \begin{cases} (q/N)^{\ell} &\qquad q/N<1 \\ \prod_{k \geq 1} \binom{k + m_c' - 1}{m_c' -1}^{q_k} &\qquad q/N>1 \end{cases}
}
%
%
%

\subsection{Pattern energy}

A rule of weight $X$ used $k$ times contributes $e^{k X}$ to the partition function. A naive semi-annealed approximation to $\ZZ$, incorporating extreme-value statistics, would obtain $\langle e^{k X} \rangle_{X<X_*}$ after weight averaging. However, in a large corpus and at low temperature,  the assignment of rule identities is intertwined with evaluating pattern energy and a many-body treatment is necessary. A key constraint is that any given rule takes only one value of multiplicity in the derivation. 

To evaluate $\hat E(P)$, fix the $N^3$ hidden-rule weights $X_r$, $r=1,\ldots,N^3$, and introduce an auxiliary activity $t_k$ for the class of rules used $k$ times. Then
\eq{ \label{Gdepgen}
\prod_{r=1}^{N^3}\left(1+\sum_{k\geq1}t_k e^{kX_r}\right)
}
contains all possible assignments of rule identities to multiplicity classes. The coefficient of $\prod_k t_k^{p_k}$ contains the desired assignment; call this coefficient $\mathcal G_d(\{p_k\})$. By Cauchy's formula, 
\eq{ \label{Gd0}
\mathcal G_d(\{p_k\}) = \prod_k \oint\frac{dt_k}{2\pi i} \frac{1}{t_k^{p_k+1}} e^{\sum_r\log D(X_r) } ,
}
where 
\eq{ \label{DdepApp}
D(X)=1+\sum_{j\geq1}t_j e^{jX}.
}
The semi-annealed approximation is defined by the key assumption that the sum over the $N^3$ candidate rules self-averages over the restricted distribution of $X$; since this is a sum of logarithms, it is better behaved than the bare REM sums discussed earlier. We therefore make the replacement $\sum_r \log D(X_r) \to N^3 \left\langle \log D(X)\right\rangle_{X<X_*}$. Then  Eq.\ref{Gd0} is amenable to a saddle-point treatment.  We obtain the saddle equation 
\eq{
p_k = N^3 t_k\left\langle\frac{e^{kX}}{D(X)}\right\rangle_{X<X_*} ,
}
which relates $t_k$ and $p_k$. In the scaling limit $\mathcal G_d(\{p_k\})$ becomes
\eq{ \label{Gd}
\mathcal G_d(\{p_k\}) \approx \frac{1}{ \prod_k t_k^{p_k}} e^{N^3 \langle \log D(X) \rangle_{X<X_*}  } ,
}
to leading exponential order. 

It is useful to note that $p_k$ contains a factor $g_k^R=\langle e^{k X}/D(X) \rangle_{X<X_*}$ which is renormalized from $g_k= \langle e^{k X}  \rangle_{X<X_*}$ by the denominator $D(X)$. Since the number of rules that are not used at all is $N^3 - \sum_{k \geq 1} p_k = N^3 \langle 1/D(X)  \rangle_{X<X_*}$, the probability that a given rule is not used is $\langle 1/D(X)  \rangle_{X<X_*}$. In the finite-tree regime this is close to unity, hence $D(X) \approx 1$ and $g_k^R \approx g_k$.


In general, however, we retain the full expression. For the surface weights we likewise introduce the corresponding quantity $\mathcal G_s(\{q_k\})$, with activities $\{ t_k' \}$, 
\eq{ \label{Gs}
\mathcal G_s(\{ q_k\}) \approx  \frac{1}{ \prod_k t'_k{}^{q_k}} e^{NT \langle \log \tilde D(Y) \rangle_{Y<Y_*}  } ,
}
\eq{ \label{Dp}
\tilde D(Y)=1+\sum_{j\geq1} t'_j e^{j Y} ,
}
and the saddle equation 
\eq{
q_k = NT t'_k\left\langle\frac{e^{k Y}}{\tilde D(Y)}\right\rangle_{Y<Y_*} ,
}
which relates $t_k'$ and $q_k$. 

Assembling results, in the semi-annealed approximation a pattern contributes
\eq{ \label{P}
\mathcal{P}(\{p_k, q_k\}) 
= \Omega_D \Omega'_D \mathcal G_d(\{p_k\}) \mathcal G_s(\{q_k\}) 
}
To identify $S$ and $\hat E$, note that $\mathcal G_d$ and $\mathcal G_s$ contain entropic components, since they count rule assignments. The number of such assignments is identified by letting all $X_r=0$, giving $\binom{N^3}{p}\binom{p}{p_1p_2\cdots}$, and similarly for the surface layer. Hence we identify the complete pattern entropy from
\eq{
e^S \equiv \Omega_D \Omega'_D \binom{N^3}{p}\binom{p}{p_1p_2\cdots} \binom{NT}{q}\binom{q}{q_1q_2\cdots}
}
and the effective energy from
\eq{
e^{-\hat E} \equiv \frac{\mathcal G_d(\{p_k\}) \mathcal G_s(\{q_k\}) }{\binom{N^3}{p}\binom{p}{p_1p_2\cdots} \binom{NT}{q}\binom{q}{q_1q_2\cdots} }
}


\subsection{Finite tree regime}

To gain intuition for the semi-annealed approximation, we first consider the finite-tree regime obtained when the derivation length $\ell$ is finite in the scaling limit. Then the entropic factors become simply $\Omega_D \Omega'_D \to (p/N)^n (q/N)^\ell$; we expect this to be smaller than unity, expressing how unlikely it is that a given set of rules will form a correct derivation. Using that $D(X)$ is close to unity, the hidden free-energetic part becomes
\eqs{
\mathcal G_d(\{p_k\}) & \approx  \frac{1}{ \prod_k t_k^{p_k}} e^{N^3 \langle \sum_k t_k e^{k X} \rangle_{X<X_*}  } \\
& \approx \prod_k \left( \frac{N^3 g_k }{ p_k  } \right)^{p_k}  e^{p } \\
& \approx N^{3p} \prod_k  \frac{g_k^{p_k} }{ p_k!  }   \\
& \approx \binom{N^{3}}{p} \binom{p}{ p_1 p_2 \cdots }  \prod_k g_k^{p_k} 
}
(and likewise for the surface part). This has a straightforward interpretation in terms of counting rule identities: first choose the symbols used by the $p$ branching rules; this gives $\binom{N^3}{p} \sim N^{3p}/p!$ if $N^3 \gg p$. Then distribute these according to the rarity distribution; this gives a multinomial coefficient $\binom{p}{p_1 \; p_2 \; \cdots }$. Then, each rule is averaged over the weight distribution; since $p_k$ rules are repeated $k$ times, this gives $\prod_k g_k^{p_k}$.

The factors $g_k$ have a very rich behavior; this is worth evaluating in detail, since it will survive the general large-corpus regime, albeit under renormalization to be discussed later.

We have
\eq{
g_k = \langle e^{k X} \rangle_{X<X_*} & = \frac{\int_{-\infty}^{X_*} dX P(X) e^{kX } }{\int_{-\infty}^{X_*} dX P(X)} \notag\\
& = \frac{\int_{-\infty}^{X_*} dX e^{kX } e^{-\ed X^2} }{\int_{-\infty}^{X_*} e^{-\ed X^2}  } \notag\\
& = \frac{ e^{k^2/(4\ed)} \text{erfc}((k-2\ed X_*)/\sqrt{4\ed} ) }{ 2 - \text{erfc}(\sqrt{\ed} X_*) } 
}
We can simplify this expression, and that for $X_*$, in the scaling limit. We use the continued fraction representation 
\eq{
\text{erfc}(z) = \frac{z}{\sqrt{\pi} } e^{-z^2} \frac{1}{z^2+h(z)} 
}
where $h(z) = 1/2 + \ldots$ for $z \gg 1$. (For $z \ll -1$ we can use erfc$(z)=2-\text{erfc}(-z)$)

First, at large $N$ we have that $\text{erfc}(\sqrt{\ed} X_*) \approx 2/N^2 \ll 1$ so that we can use its asymptotics to obtain $e^{-\ed X_*^2}/\sqrt{\ed X_*^2} \approx \sqrt{4\pi}/N^2$, which is solved
\eq{
2 \ed X_*^2 = W_0(N^4/(2\pi)) \equiv w \sim \log(N^4/(2\pi))
}
where $W_0$ is a branch of the Lambert-$W$ function satisfying $W(z)e^{W(z)}=z$. Note then that
\eq{
\ed X_* = \sqrt{\ed} \sqrt{\ed X_*^2} \sim \sqrt{2 \ed \log N},
}
which is finite in the scaling limit. As a result, the argument to $\text{erfc}$ in $g_k$ is large in magnitude (since $\ed \ll 1$) but may take either sign. We then have
\eqs{
g_k& \sim \half e^{k^2/(4\ed)} \begin{cases} 2 & k < 2 \ed X_* \\ \frac{  e^{ -(k-2\ed X_*)^2/(4\ed) } }{ \sqrt{\pi} (k-2\ed X_*)/\sqrt{4\ed}} & k > 2 \ed X_* \end{cases} \\
& =  \begin{cases} e^{k^2/(4\ed)} & k < 2 \ed X_* \\ \frac{  e^{ (4k\ed X_* - 4\ed^2 X_*^2)/(4\ed) } }{  (k-2\ed X_*) \sqrt{\pi/\ed} } & k > 2 \ed X_* 
\end{cases}
}
The asymptotics depend on $k$: for any given value of $x$, rules that are repeated often enough will feel the effect of finite $N$. 

The surface terms are analogous. Define $Y_*$ as the solution to $2\es Y_*^2 = W_0(T^2/(2\pi)) \equiv w'$. In the finite-tree limit, we have factors $h_k \equiv \langle e^{k Y} \rangle_{Y<Y_*}$ with
\eqs{
h_k & = \frac{1}{2 - \text{erfc}(\sqrt{\es} Y_*)}  e^{k^2/(4\es)} \text{erfc}\left( \frac{k-2\es Y_*}{\sqrt{4\es}}  \right) \\
& \sim  \begin{cases} e^{k^2/(4\es)} & k < 2 \es Y_* \\ \frac{  e^{ (4k\es Y_* - 4\es^2 Y_*^2)/(4\es) } }{  (k-2\es Y_*) \sqrt{\pi/\es} } & k > 2 \es Y_* \end{cases}
}
This structure completely parallels that of the hidden layers, but note that $w$ depends on $N^4$ while $w'$ depends on $T^2$.

It is useful to note that $e^{-\ed X_*^2} = \sqrt{2\pi w}/N^2$ with $w = W_0(N^4/2\pi)$ and similarly $e^{-\es Y_*^2} = \sqrt{2\pi w'}/T$ with $w' = W_0(T^2/2\pi)$. 

The Lambert-$W$ function appears repeatedly in what follows. Defined by $W(z) e^{W(z)} = z$, it has two real branches: the $W_0$ branch exists for $z > -1/e$, behaves as $W_0(z) \sim z$ as $z\to 0$ and has large $z$ behavior $W_0(z) \sim \log z - \log \log z$. The second branch $W_{-1}(z)$ exists for $-1/e < z < 0$ and has asymptotics $W_{-1}(z) \sim (\log -z) - (\log (-\log -z))$ for $|z| \ll 1$. For $z<-1/e$ neither $W_0(z)$ nor $W_{-1}(z)$ exist.

Below we will see frequent appearance of the quantities
\begin{subequations} \label{aap} \eq{ 
a & \equiv \sqrt{2 w \ed} \sim \sqrt{8 (\log N) \ed} = \sqrt{8x} \\
a' & \equiv \sqrt{2 w' \es} \sim \sqrt{4 (\log T) \es} = \sqrt{4y}
} \end{subequations}
Note also that $2 \ed X_* = \sqrt{2\ed} \sqrt{2 \ed X_*^2} = \sqrt{2\ed w} = a$ and $2 \es Y_* = \sqrt{2\es w'} = a'$.

\begin{figure*}[t!]
\includegraphics[width=\textwidth]{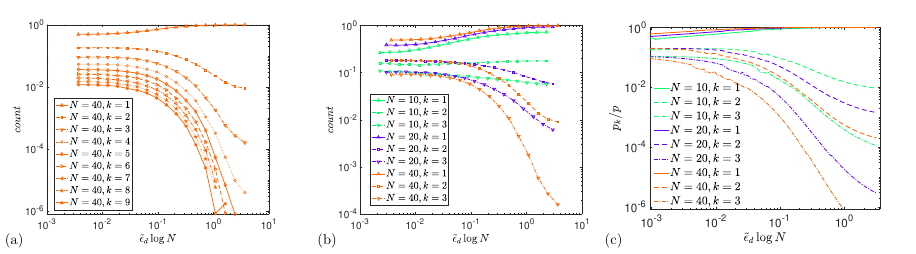}
\caption{ Multiplicity spectrum in the RLM. Numerical results in model P (a) at $N=40$ and indicated $k$; (b) for various $N$ and indicated $k$; (c) theoretical prediction in the semi-annealed approximation at fixed $\ell=20$, without fitting parameters.
}\label{fig2}
\end{figure*}

If we define integers $r,s$ such that
\eqs{
& r < a < r+1 \\
& s < a' < s+1 
} 
then we can write
\eq{ \label{gk1}
g_k \sim \begin{cases} N^{\frac{k^2}{4x}} & k \leq r \\ N^{-2 + \frac{k a}{2x} } \frac{  a }{ k-a } & k > r
\end{cases}
}
and 
\eq{ \label{hk1}
h_k \sim  \begin{cases} T^{\frac{k^2}{4y}} & k \leq s \\ T^{-1 + \frac{k a'}{2y}  } \frac{ a' }{ k-a' } & k > s \end{cases} .
}


Eqs.\ref{gk1},\ref{hk1} show explicitly that in the scaling limit, the energetic contributions have singular behavior at a whole array of temperatures, just as in the REM \cite{Derrida81}. In particular, whenever $a$ or $a'$ is a positive integer, then one of the above asymptotics changes form. 
Physically, if the reduced temperature $a$ and multiplicity $k$ are such that $k \leq r$, then the corresponding rules are abundant, and sample from the bulk of the available rule set. Instead, if $k > r$, then there are very few of the corresponding rules, and they feel the discreteness of the available rule set. The induced transitions as $a$ or $a'$ are varied through positive integers correspond to the crossover between these behaviors. A distinguished role is played by $a=1$, below which all branching rules feel discreteness of the rule set; this is the condensation transition at $x \approx 1/8$. Similarly, condensation of surface rule use occurs at $a'=1$, or $y \approx 1/4$.

%
%

\subsection{Scaling in the large$-N$ limit}


We have now all the ingredients for the semi-annealed approximation. To see how a dominant rule spectrum is attained, consider first the finite-tree regime. 
Then all $p_k$ and $q_k$ are finite, and $g_k^R \approx g_k$, $h_k^R \approx h_k$. Also, for simplicity, consider $r=s=0$, that is $a <1, a' <1$, which corresponds to $8x<1, 4y<1$ in the scaling limit. Then the leading behavior of the pattern weight is
\eqs{
\mathcal P( \{p_k, q_k \}) &\sim N^{3p+q} T^q \prod_k \left( N^{-2 + \frac{k a}{2x} } \right)^{p_k} \left( T^{-1 + \frac{k a'}{2y} }   \right)^{q_k} \\
&= N^{p+q + \frac{n a}{2x}} T^{\ell a'/2y} \\
&\sim N^{p+q+ n \sqrt{2/x}} T^{\ell \sqrt{1/y}  } .
}
As alluded to above, all contributions diverge in the scaling limit. Similar to other defect models, such as the XY model \cite{Kosterlitz73, Kosterlitz74, Zinn-Justin96, Chaikin00}, the weight of excitations depends on the temperature(s). When the multinomial factors of $e^S$ are included, this implies that rule spectra will have preferred forms in the scaling limit. Hence there is a dominant pattern at each given $\ed,\es$. 


In what follows, we extract the dominant patterns and their properties. First, we do so in finite trees, and then in large corpora;  our goal throughout is to extract the leading asymptotic forms of observables in the scaling limit. These will be compared with direct numerical simulations and will be applied to data from real languages in subsequent sections.

\subsection{Dominant rule spectrum}

In the scaling limit there is a preferred spectrum of rule multiplicities. We have $p=\sum_kp_k$, $q=\sum_kq_k$, $n=\sum_k kp_k$, and $\ell=\sum_k kq_k$. First consider the $\{ p_k \}$; the hidden and surface spectra are decoupled so long as $p/N>2/3$, which can be verified later. We maximize $\mathcal{P}$ with a Lagrange multiplier $\tilde n$ that imposes $n = \sum_k k p_k$. The result depends on whether we are in the finite-tree or large-corpus regime. 

For a finite tree, $p<N$ in the scaling limit and depletion of the available grammar is negligible. Using the Stirling approximation for $p_k!$ we obtain 
\eq{ \label{pkfinite}
p_k=N^3g_k e^{-k\tilde n}e^{n/p},
}
with $\tilde n$ fixed by $n=\sum_k kp_k$.

For a large corpus, we begin from
\eqs{
p_k = N^3 t_k g^R_k ,
}
with $g_k^R=\langle e^{k X}/D(X) \rangle_{X<X_*}$.
We maximize $\mathcal{P}$ subject to two conditions: first that $n=\sum_k kp_k$ (with associated multiplier $\tilde n$) and secondly that $N = \sum_k p_k u_k$ (with multiplier $\lambda$). This leads to 
\eqs{ 
t_k=\binom{k + m_c - 1}{m_c -1} e^{-k\tilde n+\lambda u_k} 
}
and hence
\eq{ \label{pkdep}
p_k=N^3 \binom{k + m_c - 1}{m_c -1} e^{-k\tilde n+\lambda u_k}g_k^{R} .
}

The remaining saddle equations are
\eq{ \label{saddleconstr}
n=\sum_k k p_k,
\qquad
N=\sum_k u_k p_k,
}
together with stationarity in $m_c$, which takes the form
\eq{ \label{lambdaoverlapmain}
\lambda=\frac{\sum_kp_k[\psi(k+m_c)-\psi(m_c)]}{\sum_kp_k k/(k+m_c-1)^2} ,
}
where $\psi(x)$ is the digamma function.



\begin{figure*}[t!]
\includegraphics[width=.8\textwidth]{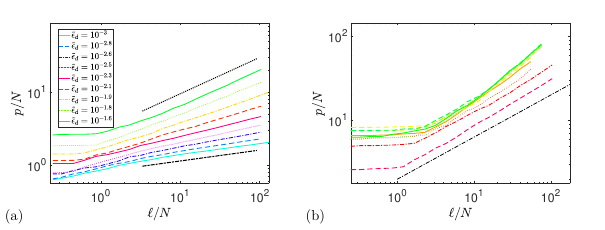}
\caption{ Rule use as a function of corpus length, in model P. (a) Number of unique rules $p$ per symbol in a corpus of length $\ell$, at $N=40$ and $\ed$ from $10^{-3}$ (blue, bottom) to $10^{-1.6}$ (green, top), equivalent to $x = 4 \times 10^{-3}$ to $x=0.093$.  Heavy dotted and dash-dotted lines indicate powers $\ell^{0.5}$ and $\ell^{0.15}$, respectively. 
(b) $p/N$ versus $\ell$ at $N=40$ and $\ed$ from $10^{-1.4}$ (red, bottom) to $10^0$ (green, top), equivalent to $x = 0.15$ to $x=3.7$. Dash-dotted line indicates power $\ell^{0.5}$. The first four curves from the bottom correspond to $1/8<x<1/2$ (the orange solid line is $x=0.49$); for larger $x$, the curves are approximately independent of $x$.
}\label{fig3}
\end{figure*}

The surface terms are analogous. For a finite tree,
\eq{
q_k =  NT h_k e^{- k \tilde m} e^{\ell/q} 
}
while for a large corpus
\eq{
q_k = NT  \binom{k + m_c' - 1}{m_c' -1} h_k^R e^{- k \tilde m} e^{\lambda' u'_k} 
}
with
\eqs{
h^R_k = \left\langle \frac{e^{k Y}}{1 + \sum_j t'_j e^{j Y} } \right\rangle_{Y<Y_*} .
}
$q=\sum_k q_k$ must be determined self-consistently, and $\tilde m$ is fixed implicitly by the constraint $\ell = \sum_k k q_k$, while $\lambda'$ is fixed implicitly by the constraint $N = \sum_k u_k' q_k$. $m_c'$ is fixed by 
\eq{ \label{lambdapoverlapmain}
\lambda'=\frac{\sum_k q_k[\psi(k+m_c')-\psi(m_c')]}{\sum_k q_k k/(k+m_c'-1)^2} .
}

\section{Rule use in finite trees}


First consider a single finite tree in the scaling limit. Then since $p\leq\ell$ while $N$ diverges, we use Eq.~\ref{pkfinite}. We need to find $\tilde n$ such that $n=\sum_k k p_k$, and $p$ such that $p=\sum_kp_k$. The result for $n=20$ is shown in Fig.~\ref{fig2}c and compared to sampled data from the RLM, where $n$ is dynamical with $\langle n\rangle\approx16$, without fitting parameters. It captures all the features of the data.

We can obtain analytical approximations for the full spectrum in several regimes, as detailed in Appendix \ref{AppC}. In the frozen regime $x<1/8$ we have 
\eq{
p_k \approx N a z^k e^{n/p} /(k-a)
}
with $z = N^{a/2x} e^{-\tilde n}$. In terms of $G = n/Na$ we find $z \approx 1 - e^{W_{-1}( -1/\log G  )}$. At small $x$, $a \ll 1$, $G \gg 1$ and $z \to 1^-$, so that $p_k/p \propto z^k/k$ is asymptotically independent of $x$. This is observed in the data (Fig.\ref{fig2}ab) for $x \lesssim 10^{-2}$. 

For $x>1/8$ we find $p \approx n$, that is, all rules are used uniquely. The $x$ dependence of $p_k$ does not have a simple form, but analytical approximations are available. 

\begin{figure*}[t!]
\includegraphics[width=\textwidth]{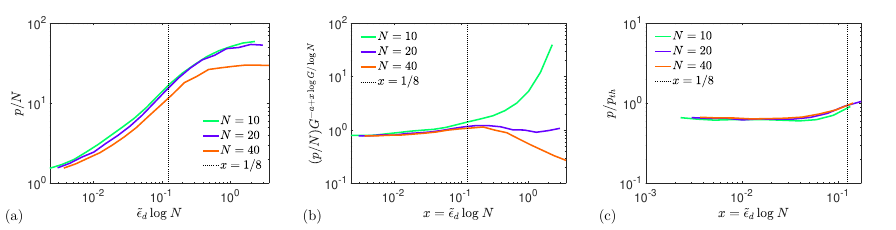}
\caption{ Rule use as a function of rescaled temperature, in model P. (a) $p/N$ versus $x=\ed\log N$ at indicated $N$, at fixed $n\approx 850$. (b) In the frozen regime, the remaining $N$ dependence is approximately collapsed by $G^{a-x \log G/\log N}$, where $G=n/Na$. The vertical dotted line indicates the freezing transition $x=1/8$. (c) A more precise collapse is obtained by retaining the full theoretical expression in the frozen phase, denoted here by $p_{th}$, and written in Eq.(\ref{pfrozen}).
}\label{fig4}
\end{figure*}

\section{Rule use in a large corpus}

Now consider the opposite regime where $p/N>1$. First, we present numerical data from the RLM at various $N$, $\ed$, and $n$, in model P (see Appendix \ref{app:WCFG} for model W). Fig.\ref{fig3}a shows the number of unique rules that appear in a corpus of total length $\ell=n+m$, versus $n/N$, at various $x=\ed \log N$ in the frozen regime. Rules are counted over complete sentences; the plateau at small $\ell$ corresponds to the typical number of unique rules in one sentence. Beyond this, $p$ is approximately power-law in $\ell$, with a `novelty' exponent that changes continuously as $x$ varies. Heavy dotted and dash-dotted lines indicate powers $\ell^{0.5}$ and $\ell^{0.15}$, respectively. 
This indicates that the language continues to exhibit novelty as corpus length grows, and the novelty rate depends upon the temperature $\ed$. At smaller temperatures, the increase of novelty with corpus length is slower. Above $x>1/8$, shown in Fig.\ref{fig3}b, the pattern is similar: the dash-dotted line indicates a power $\ell^{0.5}$.

The $N$ dependence is shown in Fig.\ref{fig4}a, at $n \approx 850$ which is the upper limit where we have data across all values of $x$ and $N$. The leading dependence is $p \propto N$ at finite $x=\ed\log N$. 

There remains a weak $N$ dependence, and a strong but nontrivial $x$ dependence. These will be discussed below.


The large-corpus saddle gives an explicit asymptotic description of these curves. It is useful to retain the finite-$N$ quantity
\eq{ \label{adep}
a=2\ed X_*=\sqrt{2w\ed},\qquad w=W_0(N^4/2\pi),
}
rather than replacing it immediately by $\sqrt{8x}$. For $N=40$, $a/\sqrt{8x}\simeq0.85$.


\subsection{Frozen regime}

In the large-corpus and frozen regime, the rule multiplicity has multiple sectors. Extensivity requires that there is a `recurrent' sector of abundant rules with $k \sim m_c \sim \ell$, which dominate the total occurrence of rules. However these abundant rules cannot contribute to novelty, where instead $k = \OO(1)$, or at least $k = o(n)$. We begin with the novelty sector, which determines the form of $p$.

For $a<1$, write $z=e^{X_*-\tilde n}$. The rules controlling novelty have $k \ll m_c$ so that 
\eqs{ 
\binom{k + m_c - 1}{m_c -1} \simeq \frac{m_c^k}{k!},\qquad u_k\simeq\frac{k}{m_c}.
}
 The normalization factor appearing in $g_k^R$ is then $D(X) = 1 + \sum_{k \geq 1} \frac{m_c^k}{k!} e^{-k\tilde n+\lambda k/m_c} e^{k X} = e^{\Theta e^{X-X_*}}$, where
\eq{ \label{ThetaDep}
\Theta=m_c z e^{\lambda/m_c}.
}
Equivalently, the renormalized energetic factor at small $k$ is simply 
\eq{ \label{gkedge}
g_k^{R}\simeq\left\langle e^{kX}e^{-\Theta e^{X-X_*}}\right\rangle_{X<X_*}.
}
The double exponential factor expresses the exhaustion of a high-weight rule once its expected use becomes order unity. It creates a soft cutoff at 
\eq{
X_*^R = X_* - \log \Theta
}
and hence an effective renormalization of $a=2\tilde\epsilon_d X_*$ to 
\eq{ \label{aR}
a_R \equiv 2\tilde\epsilon_d X_*^R = a-2 x \frac{\log\Theta}{\log N} .
}
The same quantity appears directly in the multiplicity spectrum. We have
\eq{
\frac{p_k}{N} & = N^2 \frac{m_c^k}{k!} e^{-k\tilde n+\lambda k/m_c} \int_{-\infty}^{X_*} \frac{dX}{\sqrt{\pi/\ed}} e^{kX}e^{-\Theta e^{X-X_*}} e^{-\ed X^2} \notag \\
& = \frac{a}{k!}\Theta^a e^{-\tilde\epsilon_d (\log\Theta)^2} \int_0^\Theta dt\, t^{k-1-a_R}e^{-t} e^{-\tilde\epsilon_d(\log t)^2} , \label{pk}
}
where we changed variables to $t=\Theta e^{X-X_*}$. The $e^{-t}$ factor localizes the integral to $t=\OO(1)$.  Hence, when $\tilde\epsilon_d\ll1$ and the remaining integral is nonsingular, its Gaussian curvature gives only a subleading correction, and for
$\Theta\gg1$
\[
\frac{p_k}{N} \simeq \frac{a}{k!}\Theta^a e^{-\tilde\epsilon_d(\log\Theta)^2} \Gamma(k-a_R) \left[1+\OO(\tilde\epsilon_d)\right].
\]
Summing, this gives 
\eq{ \label{pfrozen}
\frac{p}{N} \simeq a \Theta^a e^{-\tilde\epsilon_d(\log\Theta)^2} \frac{\Gamma(1-a_R)}{a_R} \left[1+\OO(\tilde\epsilon_d)\right].
}


To complete the derivation, we need $\Theta(n,N,a)$. A simple estimate is obtained by assuming that the correlation volume equation and the $\lambda$ equation are dominated by the small-$k$ rules. Then we obtain $m_c \approx n/N$ and $\lambda \approx m_c$, so that $\Theta \propto n/N$. However, in general this relation cannot be obtained from the novelty sector alone, since the constraint $n=\sum_k k p_k$ receives a finite contribution from recurrent rules with $k=\OO(m_c)$, so that a more complete treatment is necessary. In Appendix \ref{app:recurrent_saddle}, we show that throughout the frozen phase,
\eq{ \label{Theta}
\Theta = \frac{n}{Na} c(a) ,
}
where $c(a)$ is a regular function and $c(a) \approx 1-a$ as $a \to 1^-$. Hence the corpus-length dependence of novelty depends on the scaling variable $G = n/Na$, away from the condensation transition. The integral equations determining $c(a)$ are discussed in Appendix \ref{app:recurrent_saddle}.

Deep in the frozen phase, the continuum saddle on which Eq.\ref{Theta} is based breaks down, and a treatment of the full discrete sums is necessary. The main effect is that $c(a) \propto a$ as $a \to 0$, cancelling the factor from the denominator in Eq.\ref{Theta}. This occurs for $a \ll 0.3$, or $x \ll 0.01$. In what follows we adopt $c(a) \approx 1-a$ for simplicity; more refined descriptions mainly affect the prefactors of observable quantities rather than their functional forms.


Formally it is useful to take the scaling limit with $n/N$ fixed; then $\Theta$ is finite, and the renormalization of $a$ to $a_R$ is a finite$-N$ effect. However, since $\Theta \propto G \propto n/N$, in a large corpus, the renormalization of $a$ to $a_R$ can be significant. 

Note that in a strict scaling limit (i.e. when the $1/\log N$ corrections are neglected) then the four quantities $N,\ed, p, n$ appear only in the three combinations $p/N, n/N,$ and $\ed w \sim \ed \log N$. This leading $N$-dependence of $p$, $p \propto N$, is observed in Fig.\ref{fig4}a.

Including the $1/\log N$ correction, and hence renormalization of $a$, the  results explain observations of Figs.\ref{fig3},\ref{fig4}a:

First, using $c(a) \approx 1-a$, the leading exponential dependence at $a \ll 1$ is predicted to be $p/N \propto G^{a - x \log G / \log N}$. The $N$ dependence is tested by a scaling collapse in Fig.\ref{fig4}b, collapsing well the data  especially at $N=20,40$.

Second, the same expression predicts the $x$ dependence, which is indeed collapsed by the rescaling of Fig.\ref{fig4}b.


Third, at fixed $N$ the form of $p$ implies the behavior $p \sim n^{\beta(x)}$ with continuously changing exponent $\beta(x)$, observed in Fig. \ref{fig3}a. Indeed, since $\Theta \propto n$ we can define a novelty exponent by
 \[
\beta_{\rm eff} = \frac{d\log p}{d\log\Theta} \simeq a-2 x \frac{\log\Theta}{\log N}  = a_R ,
\]
which is exactly $a_R$ defined above. In the strict scaling limit, this predicts $p \sim n^{\sqrt{8x}}$, so that the novelty exponent continuously decreases as $x$ decreases, as observed in Fig. \ref{fig3}a. The finite-$N$ correction predicts that these curves have concavity, which is visible in Fig. \ref{fig3}a at the smallest $x$, although larger corpora would be needed to test it definitively. 

The terms beyond the exponential dependence further refine these predictions. Write $p_{th}$ for the full prediction of Eq.\ref{pfrozen}, with $\Theta \approx (1-a)G$. We test this by the collapse in Fig.\ref{fig4}c. Although this breaks down as $a \to 1^-$ since $\Theta$ becomes singular, away from this point the curves collapse up to the noise in the data. Note the smaller vertical range compared to Fig.\ref{fig4}b.

\begin{figure*}[t!]
\includegraphics[width=\textwidth]{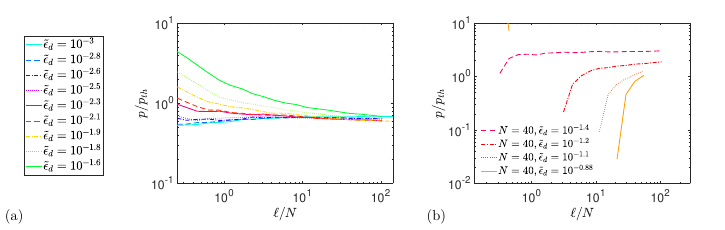}
\caption{ Rule use as a function of rescaled corpus length $\ell/N$ in both the frozen regime (a) and the transition regime (b), for $N=40$, in model P. In both cases $p$ is rescaled by the theoretical prediction, denoted by $p_{th}$, which is Eq.\ref{pfrozen} for (a) and Eq.\ref{eq:transition_three_regimes}, with $1-a_R \gg \sqrt{\ed}$ for (b). In (a),  symbols are as in Fig. \ref{fig3}a and as indicated. In (b), symbols are as in Fig. \ref{fig3}b and as indicated, but only $1/8<x<1/2$ are shown here. In (b), theory should only match at large enough $\ell/N$.
}\label{fig5}
\end{figure*}

The prediction for $p_{th}$ as a function of $n$ is tested in Fig.\ref{fig5}a, where we note that $n \approx 0.93 \ell$ in the numerics. For small $x \ll 1/8$, theory works very well for all $\ell/N$, up to an overall multiplicative factor $\approx 0.6$. As $x \to 1/8^-$ and small $\ell/N$, deviations occur due to corrections to $c(a) \approx 1-a$, are worked out in Appendix \ref{app:ccrit}; including these corrections improves the collapse by a factor $\sim 2$ (data not shown).

Overall, the theory captures the scaling dependence of $p$ on $n$, $N$, and $x$ in surprising detail.

\newcommand{\erfc}{{\rm erfc}}

\subsection{Transition regime}

We now consider $1/8<x<1/2$, which corresponds to $1<a<2$ in the scaling limit.  The important difference from the
frozen regime is that $n$ is determined by abundant rules; this allows the occurrence constraint to be evaluated explicitly.


In the transition regime, Eq.\ref{pk} continues to apply. However, the Gaussian curvature of the disorder distribution becomes relevant. This changes $\Theta$ to 
\begin{equation}
 \Theta\simeq
 \frac{n/N}{\sqrt{8\pi\log N}}\,
 N^{-\chi(x)},\qquad
 \chi(x)=2+\frac{1}{4x}-\sqrt{\frac{2}{x}} .
 \label{transition_theta}
\end{equation}
In the strict scaling limit, $\Theta \to 0$, but at finite $N$ the corrections may be important. We find that $a$ is again renormalized to $a_R$ satisfying Eq.\ref{aR}, but now with $\Theta$ following Eq.\ref{transition_theta}. The result is:
\begin{equation}
    \frac{p}{n} \simeq
     \begin{cases}
    \displaystyle
    1- \sqrt{\frac{\tilde\epsilon_d}{\pi}}\, \frac{\Gamma(2-a_R)}{a_R(a_R-1)}
    E,
    &
    a_R-1\gg\sqrt{\tilde\epsilon_d},
    \\[4mm]
    \displaystyle
    \frac12 \erfc\left[ \frac{1-a_R}  {2\sqrt{\tilde\epsilon_d}} \right],
    &
    |a_R-1|=\OO(\sqrt{\tilde\epsilon_d})
    \\[4mm]
    \displaystyle
    \sqrt{\frac{\tilde\epsilon_d}{\pi}}\,
    \frac{\Gamma(1-a_R)}{a_R}
    E,
    &
    1-a_R\gg\sqrt{\tilde\epsilon_d}.
    \end{cases}
    \label{eq:transition_three_regimes} ,
\end{equation}
where $E = e^{-(a_R-1)^2/(4\tilde\epsilon_d)}$. 
The first regime has essentially unique rule use, the second is the crossover to an $\OO(1)$ repeated fraction,
and the third is the moving-front regime, like the frozen regime.

The crossover occurs at $a_R \approx 1$, or $\log\Theta=(a-1)/(2\tilde\epsilon_d)$, corresponding to
\eq{
 \frac{n}{N} & \approx \frac{a\sqrt{\pi}}{2\sqrt{\tilde\epsilon_d}}\, e^{\frac{a^2-1}{4\ed}} \erfc\!\left( \frac{1-a}{2\sqrt{\tilde\epsilon_d}}\right) \label{transition_n} \\
 & \sim \sqrt{8\pi\log N}\, N^{\frac{8x-1}{4x}} \notag .
}
In the scaling limit $N\to\infty$ at fixed $n/N$ and fixed $x>1/8$, then $n/N$ is always below this threshold, and $p\approx n$: all rules are used uniquely at leading order. However, at any fixed $N$, a large enough corpus will enter the second regime, where rule use is nontrivial. In the latter regime, since 
 Eq.~\eqref{eq:transition_n} implies $\Theta\propto n$ at
fixed $N$ and $x$, the local novelty exponent is directly
\[
 \beta_{\rm eff}
 =\frac{d\log p}{d\log n}
 =\frac{d\log p}{d\log\Theta}.
\]
This gives $\beta_{\rm eff}\simeq a_R$, with corrections of order
$1/\log N$.  Since $a_R$ decreases logarithmically with corpus
size, the apparent novelty exponent likewise decreases slowly with
corpus size.  

We test the predictions in this regime by plotting $p/p_{th}$ in the large $\ell$ regime, that is, $1-a_R \gg \sqrt{\ed}$, in Fig.\ref{fig5}b. At large $\ell$, the curves are flat, particularly for data closest to $x=1/8^+$, showing that the theory captures the behavior shown in Fig.\ref{fig3}b. The dropoff of the curves as $\ell$ decreases corresponds to $a_R=1$, where the theoretical curves change form. 

%

\section{Surface layer}

The surface layer is solved analogously. If $p/N$ is large enough (as assumed) then it is decoupled due to the hidden feedforward nature of the sequences along the hierarchical dimension, that is from root to leaves. Comparing Eq.\ref{Gd} to Eq.\ref{Gs} and Eq.\ref{OmD1} to Eq.\ref{OmpD1}, the surface and hidden layers are mathematically isomorphic, and results for the surface layer can be obtained by a mapping. From any result for the hidden layer, first apply, at fixed $N$, the mapping
\begin{subequations} \label{hidden2surface} \eq{
a & \to a' \\
p_k/N &\to q_k/N \\
p/N &\to q/N \\
m_c &\to m_c' \\
\lambda &\to \lambda' \\
n/N &\to \ell/N \\
x &\to \half y .
} \end{subequations}
Then, any remaining unmapped appearances of $N$ should be mapped $N \to T^{1/2}$. For example,
\eq{
G = \frac{n}{Na} \to \frac{\ell}{Na'} \equiv G'
}
while novelty depends on the variable
\eq{
\Theta'=m_c'z'e^{\lambda'/m_c'}.
}
with closure
\eq{
\Theta' = G' c(a')
}
in the frozen regime. Surface rule use is characterized by an effective novelty exponent
\eq{
\beta'_{\rm eff} = a'- 2 y \frac{\log\Theta'}{\log T} .
}

The existence and qualitative similarity of the RLM transition obtained as a function of $\es$, rather than $\ed$, was shown in \cite{Lalegani24}. 


\section{Thermodynamics}

Now let us evaluate the partition function $Z$ in the semi-annealed approximation. We work in the limit of a large corpus, and initially we consider only the hidden variables. We aim to extract the terms that give an extensive free energy in the large corpus limit. 

For a given $n$, the most likely pattern contributes, from Eq.\ref{P},
\eqs{
& \prod_{k \geq 1} \frac{1}{ t_k^{p_k}}  \binom{k + m_c - 1}{m_c -1}^{p_k} e^{N^3 \langle \log D(X) \rangle_{X<X_*}  } \\
& = \prod_{k \geq 1} e^{k p_k \tilde n -\lambda u_k p_k}  e^{N^3 \langle \log D(X) \rangle_{X<X_*}  } \\
& =  e^{ n\tilde n -\lambda N }  e^{N^3 \langle \log D(X) \rangle_{X<X_*}  } ,
}
where we used the saddle-point equations. 

Thus, in the semi-annealed approximation, the hidden contribution to the partition function is
\[
 Z_d \simeq  \overline M^{\ell-m} e^{ n\tilde n -\lambda N }  e^{N^3 \langle \log D(X) \rangle_{X<X_*}  } . 
\]
We restore inverse temperature using the scaling symmetry of the model \cite{DeGiuli19}; equivalently,
$\overline M\rightarrow\overline M^\beta$, $\overline O\rightarrow\overline O^\beta$, $\tilde\epsilon_d\rightarrow\tilde\epsilon_d/\beta^2$, and
$\tilde\epsilon_s\rightarrow\tilde\epsilon_s/\beta^2$. It is convenient to rescale the rule weight $X$ in $D(X)$ so that the Gaussian measure remains at its $\beta=1$ form. Then $D(X) \to D_\beta(X) = 1+\sum_{k\ge1} \binom{k + m_c - 1}{m_c -1} e^{k(\beta X-\tilde n)+\lambda u_k}$ and similarly for $D'(Y)$. The variables $\tilde n,m_c$, and $\lambda$ depend implicitly on $\beta$, but their derivatives do not contribute when the partition function is differentiated on the saddle point. We have the hidden contribution to the partition function 
\eq{ \label{Zd}
\log Z_d \simeq \beta n \log \overline M +  n\tilde n -\lambda N + N^3 \langle \log D_\beta(X) \rangle_{X<X_*}  . 
}
The surface contribution is completely analogous, and leads to 
\[
\log Z_s \simeq \beta \ell \log \overline O +  \ell \tilde m -\lambda' N + NT \langle \log \tilde D_\beta(Y) \rangle_{Y<Y_*}  . 
\]
In total, $\log Z = \log Z_d + \log Z_s$.

We extract the energy and Boltzmann entropy from
$\langle E\rangle=-\partial_\beta\langle\log Z\rangle|_{\beta=1}$
and $\langle S_{\rm Bol}\rangle=\langle\log Z\rangle+\langle E\rangle$,
and decompose the latter as 
\begin{equation}
 \langle S_{\rm Bol}\rangle=(2\ell-2m)H_d+\ell H_{s|d}.
 \label{eq:Sdecomp}
\end{equation}
Here $H_d$ is the entropy rate of the (non-root) interior symbols, while $H_{s|d}$ is the entropy rate of the surface conditional on the
interior. At small $\tilde\epsilon_s$ we have $H_{s|d}\simeq H_s-H_d$ \cite{DeGiuli19}. In the limit of a
deterministic surface layer, $H_{s|d}=0$ and $H_s=H_d$.

We likewise decompose the energy into a hidden part and surface part, $E = n E_d + \ell E_s$. 

In what follows, we focus on the contributions from the hidden variables. This is motivated by the data-processing inequality \cite{Cover99}: the surface layer can only lose information, and alone it cannot generate long-range correlations, so it is natural to consider a deterministic surface layer. Linguistically, this amounts to the statement that the hidden symbols are so refined as to encode all possible features of each word; then for a given language, each hidden symbol has a unique observable symbol (terminal).

\subsection{Entropy}

From Eqs.\ref{Zd},\ref{eq:Sdecomp} it follows
\begin{equation}
 2nH_d= n\tilde n-\lambda N+  N^3\left\langle (1-X \p_{X}) \log D(X) \right\rangle_{X<X_*}.
 \label{eq:Hd_general}
\end{equation}
This expression is analyzed in Appendix \ref{app:recurrent_entropy}. 

\begin{figure*}[t!]
\includegraphics[width=.8\textwidth]{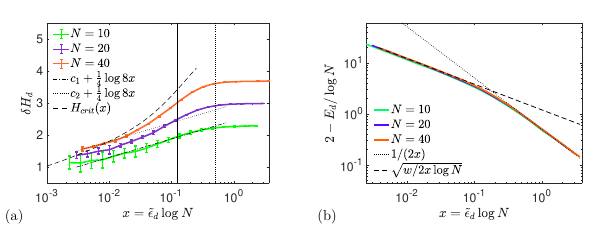}
\caption{ Hidden entropy rate and hidden energy, in model P. (a) Numerical results and frozen-regime theory. Dotted and dash-dotted lines are logarithmic with predicted prefactor $1/4$. $H_{crit}(x)$ (dashed) is the leading theoretical prediction in the critical region $x \to 1/8^-$; see Eq.\ref{Hcrit}. The vertical lines show the regime boundaries $x=1/8$ (solid) and $x=1/2$ (dotted)  (b) Energy and asymptotic theoretical results for $x>x_c$ (dotted) and $x<x_c$ (dashed) where the critical value is $x_c=1/8$ in the $N \to \infty$ limit. Theoretical results are computed for $N=40$. }\label{fig6}
\end{figure*}

\subsubsection{Entropy in the frozen region}

In the frozen regime $a<1$, we find 
\begin{equation}
 H_d=-\frac12\left(1+a\frac{\partial}{\partial a}\right)\log z(a) .
 \label{eq:H_from_z}
\end{equation}
The function $z(a)$ is in general nontrivial; its leading behavior is $z \approx (1-a)/(ae)$, which leads to
\eq{ \label{Hdlog}
 H_d 
 & \approx \text{const} + \frac14 \log 8x \qquad x < 1/8  
}
This result must break down at small enough $x$, where the continuum saddle itself disappears, but we do not observe saturation of $H_d$ in the numerics.

A precise comparison of Eq.\ref{Hdlog} with simulations is nontrivial, since naive estimates of entropy are biased \cite{Schurmann96,Grassberger03}, and a complete computation of entropy requires that all correlations must be included; for example, if correlations extend to length $\ell$, then the state space is of size $N^\ell$, which can be enormous. Define $\mathcal{H}(\ell)$ for the entropy computed over subsequences of length $\ell$; the asymptotic entropy rate is in principle determined by $H_d = \lim_{\ell \to \infty} \mathcal{H}(\ell)/\ell$. Since $H_d$ approaches its limit from above \cite{Schurmann96,Cover99}, it is convenient to estimate it by a differential entropy $\delta H_d(\ell) = \mathcal{H}(\ell) - \mathcal{H}(\ell-1)$; this only differs from the true entropy by the contribution of correlations at lengths $\ell$ and larger. In practice, we use the estimator proposed in Eq.27 of \cite{Grassberger03} to estimate $\mathcal{H}(\ell)$ for $\ell=1,2$, and then estimate $H_d$ from $\delta H_d(2) = \mathcal{H}(2)-\mathcal{H}(1)$.

The results are shown in Fig.\ref{fig6}a, in model P (see Appendix \ref{app:WCFG} for model W). As shown by the dotted and dash-dotted lines, the theoretical curve Eq.\ref{Hdlog} correctly predicts the logarithmic behavior in the frozen regime, with the correct prefactor $1/4$. The additive constant is underpredicted by the theory and was used as a fitting parameter; besides the approximations, this may also reflect measurement error. 

The full expression obtained from Eq.\ref{eq:H_from_z} and $z \approx (1-a)/(ae)$ creates a spurious divergence as $a \to 1^-$; this is rounded by the Gaussian curvature retained in the full saddle, ignored in Eq.\ref{eq:H_from_z}. The corrections lead to 
%
\eq{ \label{Hcrit}
H_{\rm crit} & = \frac12 + \frac12 \log\left[ \frac{a\sqrt{\pi}}{2\sqrt{\ed}}\, e^{u^2}\erfc(u) \right] \notag\\
& \quad + \frac12\frac{e^{-u^2}} {\sqrt{\pi\ed}\,\erfc(u)} -\frac{1-a}{4\ed} ,
}
where $u = (1-a)/\sqrt{4\ed}$. As shown by the dashed line in Fig.\ref{fig6}a, the predicted $x-$dependence tracks the upturn of $H_d$ as $x \to 1/8^-$; the additive constant was again used as a fitting parameter.

\subsubsection{Entropy in the transition regime}

In the regime $1/8<x<1/2$, then instead 
\begin{equation}
 H_d\simeq
 \left(1-\frac{1}{8x}\right)\log N+\OO(1),
 \qquad \frac18<x<\frac12 .
 \label{eq:Htransition}
\end{equation}
This result is very interesting as it implies a change of scaling $H_d \sim \log N$ for $x>1/8$ to $H_d \sim 1$ for $x<1/8$. This is visible in Fig.\ref{fig6}a, and in plots of normalized entropy, e.g. Fig. 2 of \cite{DeGiuli19}. 

The two different scalings originate directly from whether the
occurrence-weighted disorder is pinned to the extreme-value edge
or to the unconstrained Gaussian saddle.


\subsection{Energy}

The hidden energy follows directly: 
\[
\langle E_d\rangle = 2\log N -\frac{N^3}{n} \left\langle X\p_X\log D(X) \right\rangle_{X<X_*}.
\]
In the Appendix, we show that this reduces to simple expressions away from the critical region:
\eq{
\langle E_d\rangle & = \log N^2 - \begin{cases} \frac{1}{2\ed} & x \gtrsim 1/8  \\ 
\sqrt{\frac{w}{2\ed}} + \ldots& x < 1/8 \end{cases} ,
}
where we neglect subleading terms. In the frozen regime the leading term is $2 - \langle E_d \rangle/\log N = \sqrt{w/ 2\ed \log^2 N} \approx \sqrt{2/x}$. Note that the two regimes match at $2 w \ed = 1$ which reduces to $x = 1/8 + \ed/4 \log (\pi/\ed)$; in the scaling limit this becomes $x=1/8$ as expected. Hence $\langle E_d \rangle$ is continuous in the semi-annealed approximation, and clearly shows the phase transition at $x_c=1/8$. 

The theoretical predictions are compared with numerics in Fig.\ref{fig6}b, where we show unnormalized energies in model P to compare easily with predictions. They are quantitatively good, only entailing a small error near the transition point.

\begin{figure*}[t!]
\includegraphics[width=.9\textwidth]{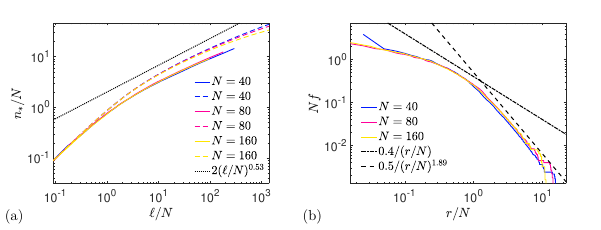}
\caption{ Heaps' law and Zipf's law in the RLM, in model P. (a) Number of unique words, $n_*$, in a corpus of total length $\ell$, from numerical simulations of the RLM (solid) at indicated $N$, and compared with theory (dashed). Here, $\ed = 10^{-2.5}$, $\es = 10^{-2}$, and $T=10^4$, corresponding to $a' = 0.528$. The large $\ell$ prediction $n_* \sim \ell^{a'}$ is shown for comparison (see text); (b) frequency of a word versus its rank $r$, at indicated $N$. Here, $\ed = 10^{-2.5}$, $\es = 10^{-2}$, and $T=10^4$, corresponding to $a' = 0.528$. The dashed line shows the large-$r$ prediction $\sim r^{-1/a'}$; for comparison, the dash-dotted line shows $\sim r^{-1}$. For $N=40$ a single outlier has the largest frequency.
}\label{fig_nstar}
\end{figure*}

%
%
\section{Application to natural language}

\subsection{Heaps' law}

The relations $p \sim (\ell-m)^\beta$ and $q \sim \ell^{\beta'}$ for the number of unique hidden and lexical rules in a corpus of total length $\ell$ are related to a known linguistic law called Heaps' law~\cite{Heaps78}. The latter states that the number of unique words in a corpus of length $\ell$ satisfies $n_* \sim \ell^b$ where $b \approx 1/2$. The two laws are not identical, because we can have multiple rules $a \to B$ for different hidden symbols $a$ that yield the same word $B$. However since the rules are iid in the RLM, we can derive the needed law. 

Indeed, a word $B$ appears in a corpus if and only if at least one lexical rule yields $B$. Since the words are uniform, the probability that $B$ doesn't appear is $(1-1/T)^q$. Then the expected number of unique words is
\eq{ \label{Heaps}
n_* = T \left( 1 - \left(1-\frac{1}{T} \right)^q \right) \approx T (1 - e^{-q/T} )
}
where the latter expression holds if $q \lesssim T$. Thus if $q \sim \ell^{\beta'}$ then we predict $n_* \sim \ell^{\beta'}$ for $q \ll T$, which is exactly Heaps law. Eq.\ref{Heaps} also predicts a final saturation of $n_*$ as $q\to T$, i.e. when the corpus exhausts the vocabulary. 

By the hidden-to-surface mapping Eq.\ref{hidden2surface}, the semi-annealed theory predicts that the effective $\beta'$, like $\beta$, will depend on $\ell$ itself. A prediction for $q$ follows by the mapping from Eq.\ref{pk}, and summing over $k$, as
\eq{
q/N & = a' \Theta'{}^{a'} e^{-\es (\log\Theta')^2} \notag \\
& \qquad \times\int_0^{\Theta'} dt\, t^{-1-a_R'} e^{-\es(\log t)^2} [ 1 - e^{-t} ] . \label{q}
}
At small $\ell/N \lesssim 1$, this predicts $q \propto \ell$, while at larger $\ell$, we get approximately $q \sim \ell^{\beta'(\ell)}$. This leads to an effective exponent
\eq{ \label{betapeff}
\beta'_{\rm eff} \equiv \frac{d \log q}{d \log \ell} \simeq a'-2\es\log\Theta' 
}
for $\ell/N \gtrsim 1$, and an effective Heaps exponent
\eq{
b_{\rm eff} \equiv \frac{d \log n_*}{d \log \ell} = \frac{q/T}{e^{q/T}-1} \beta'_{\rm eff}
}
Thus for $q \ll T$ the effective Heaps exponent is $\beta'_{\rm eff}$, while as $q \to T$ it systematically decreases; at $q=T$ it is $\approx 0.58 \beta'_{\rm eff}$. To find the crossover value $\ell_*$ where saturation is noticeable, we can use Eq.\ref{q} in the scaling limit and set $q=T$ to obtain
\eq{ \label{ellstar}
\ell_* \sim N \left( \frac{T}{N} \right)^{1/a'}
}
for $a' =\OO(1)$.

 The corresponding predictions for $n_*$ are compared with numerical data from the RLM in Fig.\ref{fig_nstar}a, computed for $T=10^4$, $\es = 10^{-2}$, $\ed = 10^{-2.5}$, and indicated $N$, in model P (see Appendix \ref{app:WCFG} for model W). We first note that both data and theory show the same nontrivial dependence on $N$, namely that $n_*/N$ is a function of $\ell/N$ before eventual saturation; this follows since $\Theta'$ depends on $\ell/N$. The numerical curves (solid) show a crossover from the predicted $n_* \propto \ell$ at small $\ell$ to a nontrivial, $\ell-$dependent curve at larger $\ell$. The theoretical curves (dashed) computed from Eq.\ref{q} match these behaviors. For these values of $\es$ and $T$, we have $a'=0.528$; the power law $n_* \sim \ell^{a'}$ is shown for comparison. Note that final saturation of $n_*$ is predicted to occur when $q$ reaches $\sim T$; even for the largest $N$ shown here, this would occur when $n_*/N \sim 62.5$, which is beyond the range of the numerical data. We confirmed in other datasets that the saturation is correctly captured by the theory. Quantitatively, the predicted Heaps curve is larger than the measured one by a factor $\sim 2$ in the large $\ell$ regime. We expect this is because not all hidden symbols are available for use by the lexical rules, reducing novelty.


A detailed analysis of Heaps' law in large texts, typically $\ell \sim 10^5$ to $\ell \sim 10^9$, Ref.\cite{Petersen12} found the following: for all languages, $n_*$ grows sublinearly with $\ell$, and the effective exponent $d \log n_* / d \log \ell$ never increases with $\ell$.

For some languages, the data are well-fit to a single power-law over two decades in $\ell$, with values $b = 0.54, 0.52, 0.60,$ and $0.51$ for English, French, German, and Spanish, respectively. For other languages, a better fit is obtained if two power-law regions are allowed, for example Chinese, Hebrew, and Russian, with crossovers from $b= 0.77$ to $b=0.26$, $b= 0.65$ to $b=0.47$, and $b= 0.65$ to $b=0.39$, for the three languages. In all languages, even those fit with a single power-law, saturation of the Heaps' exponent is visible at the largest $\ell$. 

The data are thus consistent with an expression of the form $n_* \sim \ell^{\beta'(\ell)}$ as predicted by theory, where the effective slope decreases as $\ell$ increases. The initial regime $n_* \sim \ell$ is not visible in plots of Ref.\cite{Petersen12}, but this only implies $10^5 \gtrsim N$, a weak bound. In other works where the full range of $\ell$ is shown \cite{Serrano09}, a decrease from $n_* \propto \ell$ is already visible for $\ell \sim 10^2$, giving a crude lower bound of $N$ in these works. We will discuss later an estimate $N \sim 10^4$ that comes from measurements of Zipf's law \cite{Ferrer-i-Cancho01,Gerlach13}, which we use in what follows. 

For languages fit to a single power-law, the empirical value $b \approx 1/2$ and the theoretical result $n_* \sim \ell^{a'}$ in the scaling limit identify the asymptotic value of $b$ as $a'$, if renormalization from $a'$ to $a'_R$ is negligible. To test this, we need a value of $T$. The same reference suggests $T \sim 10^6$, which gives a renormalization prefactor $\sim \es \sim 0.01$. Upon multiplication by $2\log \Theta \sim \log (\ell/N)$, this gives 0.18 at $\ell \sim 10^8$ and $N \sim 10^4$; this is small but not negligible. Therefore we expect that the quoted values of $b$ correspond to $a_R'$ averaged over the interval on which a power-law was fitted. In other words, the true value of $a'$ for these languages is closer to the quoted value of $b$, plus $\approx 0.18$, although it will depend on the fitting range of $\ell$ and the estimated value of $a'$ itself. Taking the midpoint of the corpus size range for $\ell$, and using language-specific estimates of $T$ and $a'$,  this gives refined estimates of $a' \approx 0.78, 0.77, 0.82, 0.67$ for English, French, German, and Spanish, respectively. 

For languages fit to two power-law regimes, we repeat this procedure, taking $a_R'$ from the first regime. This identifies $a' \approx 0.79, 0.71, 0.50$ for Chinese, Russian, and Hebrew, respectively, where we assume $N \sim 10^4$ in $\Theta$, except for Russian where~$N \sim 6 \times 10^4$~\cite{Gerlach13}.

Heaps' law can also be rationalized by a sample-space-reducing model in which the number of accessible states decreases as a text is written \cite{Mazzolini18,Mazzolini18a}. Such models, however, fail to match the observed correlation structure between words. The same long-range structure is also reflected in the slow dependence of predictive entropy on context length, considered below.

\begin{table}[h!]
\begin{center}
\begin{tabular}{|| c | c | c | c | c | c | c||} 
 \hline
 Language & $a'_{\rm est}$ & $T_{\rm est}$ & $N_{\rm est}$ & $\alpha_{\rm est}$ \cite{Petersen12} & $\alpha_{\rm est}$ \cite{Gerlach13}  & $\alpha_{\rm pred}$\\ [0.5ex] 
 \hline\hline
English & 0.78 & $2 \times 10^6$ & $8 \times 10^3$ & $1.68 \pm 0.10$ & 1.565 & 1.63 \\ \hline
French & 0.77 & $2 \times 10^6$ & $8 \times 10^3$ &  $1.69 \pm 0.06$  & 1.562 & 1.62 \\ \hline
German & 0.82 & $2 \times 10^6$ & $2 \times 10^4$ & $1.63 \pm 0.16$ & 1.617 & 1.69 \\ \hline
Spanish & 0.67 & $2 \times 10^6$ & $9 \times 10^3$ & $1.61 \pm 0.15$ & 1.562 & 1.55 \\ \hline
Chinese & 0.79 & $2 \times 10^5$ & -- & $1.49 \pm 0.15$ & -- & 1.61 \\ \hline
Russian & 0.71 & $3 \times 10^6$ & $6 \times 10^4$ & $1.55 \pm 0.17$ & 1.515 & 1.62 \\ \hline
Hebrew & 0.50 & $6 \times 10^5$ & -- & $1.34 \pm 0.09$  & -- & 1.42 \\ \hline
\end{tabular}
\end{center}
\caption{Language properties from analysis of Heaps' law and Zipf's law, using data from  Refs.\cite{Petersen12,Gerlach13} (for English, we used the fiction dataset in Ref.\cite{Petersen12} since it shows saturation of Heaps' law). $T_{\rm est}$ is estimated from the saturation of $n_*$ in Heaps' law, Fig.2 in Ref.\cite{Petersen12}. $N_{\rm est}$ is taken from fits of Ref.\cite{Gerlach13}, when available. $a'_{\rm est}$ is obtained from measured values of the Heaps' exponent in Ref.\cite{Petersen12} as described in the text. The Ref.\cite{Petersen12} estimate of $\alpha$ is given as $\alpha_-$ in that work. The Ref.\cite{Gerlach13} estimate of $\alpha$ is obtained from their Zipf's law exponents, denoted $\gamma^*$ in that work, using $\alpha = 1+1/\gamma^*$. The values of $\alpha_{\rm pred}$ are obtained from Eq.\ref{alpha}, evaluated at $f=10^{-7}$ which is approximately the geometric midpoint of the range used to obtain $\alpha_-$ in Ref.\cite{Petersen12}. For Chinese and Hebrew where no value of $N$ was available, $N = 10^4$ was taken. 
} \label{table1}
\end{table}

\subsection{Zipf's law}

Heaps law is related to Zipf's law \cite{Zipf13,Cancho03}, the power-law relationship between the frequency of a word $f$ and its rank, $r$, with $f \propto r^{-\zeta}$ and $\zeta \approx 1$ often reported. Zipf's law is also represented in the spectrum of word frequencies $\PP(f)$, where $\PP(f) \propto f^{-\alpha}$ in the pure Zipfian case, with $\alpha = 1+1/\zeta$.

The surface multiplicity spectrum $\{ q_k \}$ gives directly the distribution of lexical rule frequencies $f=k/\ell$. Applying the hidden-to-surface mapping Eq.~\ref{hidden2surface} to Eq.~\ref{pk}, for $1\ll k\ll\Theta'$ the integral is localized near $t\simeq k$, giving
\eq{ \label{qkZipf}
q_k/N \propto
\frac{\Gamma(k-a'_R)}{\Gamma(k+1)}
e^{-\es(\log k)^2}
\sim k^{-1-a'_R}e^{-\es(\log k)^2},
}
with $a'_R\equiv a'-2\es\log\Theta'$.
Thus the spectrum of frequencies $f=k/\ell$ is lognormal. We can write
\eq{
\PP(f) 
& \propto f^{-1-a' + 2\es \log (\Theta'/\ell) }e^{-\es(\log f)^2} , \quad f \ll \Theta'/\ell
}
which is a lognormal. For $f \gtrsim e^{-1/\sqrt{\es}}$ it is close to a power-law $\PP(f) \sim f^{-\alpha}$. We define an effective exponent that runs with $f$
\eq{
\alpha_{\rm eff}(f) & = - \frac{d \log \PP(f)}{d \log f} \\
& = 1 + a' - 2\es \log (\Theta'/f\ell) \notag \\
& = 1 + a' + 2\es \log (fN) + 2\es \log (a'/c(a')) \label{alpha}
}
The $f-$dependence is mild but can change the exponent by $\sim 0.1$ over several decades in $f$.
For $f \gtrsim \Theta'/\ell$ the power-law steepens into an exponential decay (derivation not shown). 

We can also derive the rank-frequency Zipf law. Indeed, if a lexical rule is used $k$ times, then its rank is $r(k)=\sum_{j\geq k}q_j$, up to an $\OO(1)$ ambiguity.  
Summing $q_k$ and converting to an integral, 
\eqs{
r(k)/N & \propto\int_{\log k}^\infty dv\,e^{-a'_R v-\es v^2} \propto \erfc(u) 
}
with $u = \frac{a_R'}{\sqrt{4\es}} + \sqrt{\es} \log k$. The corresponding local Zipf exponent is, using $f=k/\ell$, therefore
\eq{ \label{Zipfexp}
\zeta_{\rm eff}(k)
\equiv -\frac{d\log f}{d\log r}
& = \sqrt{\frac{\pi}{4\es} } e^{u^2} \erfc(u) \\ 
& = \frac{1}{\sqrt{4\es}} \frac{u}{u^2 + 1/2 + \ldots} , \notag 
}
where 
the last line is an expansion in large $u$. 
In the strict scaling limit this reduces to $\zeta=1/a'$. 
At finite $T$, however, $\zeta_{\rm eff}$ runs slowly with frequency, so that an exact power law is
not expected over the entire frequency range; departures from a single Zipf exponent are indeed seen empirically \cite{Ferrer-i-Cancho01,Corral15}.

The quantities $q_k$ count lexical rules rather than words, since several hidden symbols may emit the same observable
symbol. We find empirically that the Zipfian power-law regime survives. 

Numerical rank versus frequency plots from the RLM are shown in Fig.\ref{fig_nstar}b for $T=10^4$, $\es = 10^{-2}$, $\ed = 10^{-2.5}$, and indicated $N$, in model P (see Appendix \ref{app:WCFG} for model W). As predicted by theory, these are functions of $r/N$ only. The numerical curves (solid) show a smooth crossover from a slowly decreasing curve to the $f \propto r^{-1/a'}$ curve predicted by theory (dashed). The beginning of the power-law regime occurs at $r \approx N$. In the intermediate range just before, a curve $f \propto r^{-1}$ is viable over a factor of $\sim 5$ in rank. 

The empirical situation is subtle. Although a pure power-law $f \propto r^{-\zeta}$ with $\zeta \approx 1$ was reported by Zipf, later analyses on large corpora \cite{Ferrer-i-Cancho01,Gerlach13} showed that there are two regimes, with $f \propto r^{-1}$ at $r < r_*$ and $f \propto r^{-\zeta}$ at $r > r_*$, with $\zeta$ non-universal but closer to 2. Refs.\cite{Ferrer-i-Cancho01,Gerlach13} estimate $r_* \sim 10^4$ in English, French, Spanish, and German, and $r_* \sim 10^5$ in Russian. 

The double power-law behavior also extends to $\PP(f)$. Ref.\cite{Petersen12} found that many languages have a small-frequency power-law regime with $\alpha \approx 1.7$, which at $f \sim f_* \sim 10^{-4}$ crosses over to a steeper power-law with $\alpha \approx 2$. Since theory predicts crossovers at $r_* \sim N$ and $f_* \sim 1/N$, the empirical estimates are consistent with $N \sim 10^4$ except perhaps Russian where $N \sim 10^5$. In Ref.\cite{Petersen12}, the small-frequency power-law regime is fit over the range $10^{-8} \lesssim f \lesssim 10^{-6}$. Using estimated values of $a'$, $N$, and $T$, from Eq.\ref{alpha} we obtain estimates for $\alpha_{\rm pred} \equiv \alpha(f=10^{-7})$ evaluated at the midpoint of the fitted range; these are shown in the last column of Table \ref{table1}. These are within the error bars of direct measurements in Ref.\cite{Petersen12} for all languages, except French, for which the prediction is slightly lower.




Overall, both Heaps' law and Zipf's law are predicted by the RLM. In the strict scaling limit, both laws are predicted to be pure power laws. At finite $T$, the effective exponents receive corrections proportional to $\es$; the Heaps exponent also varies with corpus size. Comparison with data from natural languages indicates that the rescaled surface temperature is typically in the range $a' \in (0.67,0.82)$, with Hebrew a possible exception. Empirical analyses of Zipf's law, combined with theory, suggest that $N \sim 10^4$ in the considered languages.





\section{Application to Large Language Models}

\begin{figure*}[t!]
\includegraphics[width=\textwidth]{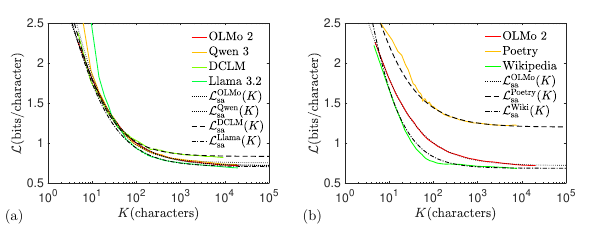}
\caption{ Code length in bits/character for indicated Large Language Models (LLMs), digitized from Ref.\cite{Scheibner25} (colors), and fits to the RLM semi-annealed prediction within a grammar-mixture model, Eq.\ref{LK} (black). In (a), the models are tested on the C4 corpus. In (b), OLMo is tested on three datasets: the C4 dataset (same as in (a)); the Project Gutenberg Poetry corpus; and English Wikipedia. Parameters for the fits are displayed in Table \ref{table2}.}\label{fig_L}
\end{figure*}

LLMs capture long-range correlations of natural language, evident by their ability to write coherent text and code over very large context lengths. To apply the RLM to Large Language Models (LLMs), where more than $10^9$ parameters are typically used, the $N \to \infty$ limit is natural. Within the RLM, however, distinct sentences are independent for any given grammar. There are two modelling possibilities: either we consider entire documents as single trees and work in the corresponding regime of the RLM, or we consider the LLM as modelling an ensemble of grammars, where an individual element of the ensemble corresponds approximately to a single document. The latter is a coarse-grained model of the former, and simpler to model; we adopt it in what follows as a simple extension of the RLM.

To see the connection between the two approaches, suppose the core elements of a language are modelled by a grammar $\GG_0$, and that the various document-level properties are specified by discrete latent variables $\{ \eta \}$  that modify $\GG_0$ to $\GG(\eta)$. Then one could construct a super-grammar, still a stochastic context-free grammar, that for each derivation first selects the latent variables, and then generates sentences as subtrees. Then each derivation corresponds to an entire document, and the sentences are correlated through their shared latent variables. 

The ensemble picture is useful because it explicitly separates these global correlations from the regular tree structure of a CFG. Indeed, let $\eta$ denote the unknown latent variables and $D_\ell=(o_1,\ldots,o_\ell)$ a corpus of length $\ell$.  Marginalizing
over latent variables gives
\[
\PP(D_\ell)=\int d\eta \,\PP(\eta)\PP(D_\ell| \eta).
\]
The entropy of this mixture satisfies
\eq{
H(D_\ell)=H(D_\ell|\eta)+I(\eta;D_\ell),
\label{grammarMI}
}
where $H(D_\ell|\eta)$ denotes the entropy averaged over fixed
grammars and $I(\eta;D_\ell)$ is the mutual information between
the corpus and its particular grammar.  Thus the excess block entropy associated
with not knowing the grammar is precisely the information about
$\mathcal G(\eta)$ which can be learned from a corpus of length $\ell$.


Since $H(D_\ell|\eta)$ is asymptotically extensive and additive over distinct trees, it can only contain subextensive contributions from individual large trees, which are rare. Instead, $I(\eta;D_\ell)$ can contain large global subextensive contributions from any information that is not extensive, such as the rule use spectrum. For simplicity we assume that the $\ell$ dependence is dominated by $I(\eta;D_\ell)$. 

As a concrete model of the latent variables, suppose that each rule is modulated by a binary variable acting as a switch for the rule, and that this variable is observable when the rule is used. Since each rule reveals $\log 2$ nats of information, then we have simply
\eq{ \label{Ibinary}
I(\eta;D_\ell) \approx (\log 2) p(\ell) ,
}
where $p(\ell)$ is the number of unique rules used in a corpus of length $\ell$. For simplicity, we adopt Eq.\ref{Ibinary} in what follows, although richer modelling possibilities yield similar results. For example, one could model the $\eta$ as a second set of document-level RLM weights that modulate the bare grammar $\GG_0$. In that model, one finds (see Appendix \ref{app:document_grammar_mixture}) $I(\eta;D_\ell) \approx \Gamma(a) (1/2\tilde\epsilon_{\rm doc})^a p(\ell)/2$, where $\tilde\epsilon_{\rm doc}$ is the RLM temperature of the document-level weights, assumed large here for simplicity. Both models yield the same $\ell$ dependence of $I$ via $p(\ell)$, which is what we test in what follows.

\subsection{Dependence of predictive entropy on context length}

Recently, Ref.\cite{Scheibner25} measured the code length $\mathcal{L} = -\langle \log P(o_0 | o_{-K},o_{-K+1},\ldots o_{-1}) \rangle$ of natural language, as a function of context length $K$. $\mathcal{L}$ measures the average negative log-probability of the next token, given its context, according to the LLM, and averaged over the dataset. For an ideal predictor it is equal to the conditional entropy of the next token, given the context, and approaches the entropy rate as $K \to \infty$. It was checked explicitly in \cite{Scheibner25} that it has the same behavior as conditional entropy over different datasets and models. In all cases it was observed that $\mathcal{L}$ decays very slowly with $K$, but it does not take any known elementary functional form. 

We digitized data from Ref.\cite{Scheibner25} manually using PlotDigitizer software \footnote{In the software, the user selects two points along each axis to calibrate; then the user manually clicks on data points.}; in noisy regions, we estimated the center point of the data in the corresponding interval. In the large $K$ regime, the noise from the data is approximately of magnitude $\pm 0.02$ bits/character. 

These data are reproduced as the colored curves in Fig.\ref{fig_L}, showing Figs. 1 and 2 from Ref.\cite{Scheibner25}. Fig.\ref{fig_L}a shows data from models OLMo 2 \cite{OLMo24}, Llama 3.2 \cite{Grattafiori24}, Qwen3 \cite{Yang25}, and DCLM \cite{Li24b}, all evaluated on the C4 corpus; these models have $10^9$, $10^9$, $8 \times 10^9$, and $1.7 \times 10^9$ parameters, respectively. Fig.\ref{fig_L}b shows data from OLMo 2 and different datasets. All plots are shown in units of bits/character.

To model this result, we regard the preceding $K$ tokens as data from which the latent variables $\eta$ governing the text can be inferred.
Writing $D_K=(o_{-K},\ldots,o_{-1})$, the conditional entropy of the next token obeys
\[
H(o_0|D_K) = H(o_0|D_K,\eta) + I(\eta;o_0|D_K).
\]
For sufficiently large $K$, we assume that the residual
$K$-dependence of the known-grammar contribution is small, so that the
slow decay of the conditional entropy is dominated by uncertainty in
the effective grammar.  By the chain rule,
\[
I(\eta;o_0|D_K) = I(\eta;D_{K+1})-I(\eta;D_K).
\]
Using Eq.\ref{Ibinary} then gives
\eq{
I(\eta;o_0|D_K) \simeq (\log 2) [ p(K+1)-p(K) ].
\label{eq:predictive_entropy}
}
Treating the measured LLM code length as a proxy for this conditional
entropy, we obtain a prediction for the $K$-dependence of the predictive entropy, at any fixed $N$ and $\ed$. To convert to bits/character we multiply by $1/(n_{c/t}\log 2)$ where $n_{c/t}$ is the number of characters per token in this dataset, of order $\sim 5$ but depending on the model and dataset \cite{Scheibner25}. We also need to rescale the argument appropriately. Then, if $K_c = K n_{c/t}$ is now in characters, 
\eq{
\mathcal L_{\rm sa}(K_c) = \mathcal L_\infty + \frac{p(K_c/n_{c/t}+1)-p(K_c/n_{c/t})}{n_{c/t}},
\label{LK}
}
in bits/character. 
We expect that $\mathcal{L}_\infty$ includes an effect of surface ambiguity (e.g. synonyms) and is therefore {\it a priori} not obtainable from the hidden-layer parameters $\ed$ and $N$. Using Eq.\ref{pfrozen} this is an explicit function of $N$, $\ed$, and $K_c$. 

Since $dp(K)/dK \approx a_R(K) p(K)/K$ this predicts
\eqs{
\mathcal L_{\rm sa}(K_c) \sim \mathcal L_\infty + a_R(K_c/n_{c/t}) \frac{p(K_c/n_{c/t})}{K_c},
}
at large $K_c$, at leading order. In the scaling limit we have $p(K)/K \sim K^{a-1}$ but the various corrections may be important.




The black curves in Fig.\ref{fig_L} show fits to the data, obtained by least-squares fit and then manually adjusted (if necessary) to obtain a good fit in the large $K$ regime. They are generally extremely good over many decades in $K$. The parameters obtained from the fits are shown in Table \ref{table2}. In all cases, the inferred $a<1$, as required to be in the frozen regime. 
 Note that Eq.~\ref{pfrozen} assumes $\Theta\gg1$, whereas for the fitted parameters $\Theta = \OO(1)$ at $K\sim10^2$. Thus, the agreement at large $K$ provides a more meaningful test of the theory, and we place more emphasis on this regime. 

We note that if $p$ is replaced by its leading form $p \approx N (K/N)^a$, the fitted $a$ values are very similar. Thus, within the model, $a$ is the exponent of the decay in $\mathcal{L}_{\rm sa}(K)-\mathcal{L}_{\rm sa}(\infty) \sim K^{a-1}$. 

The only exceptions to excellent fits are as follows: for the Qwen and Llama models, the initial decay appears to be faster than power-law; for the Poetry dataset, the data are irregular at small $K$; for Wikipedia, the `knee' is sharper than predicted by theory; and for the Qwen and Llama models, the data continue to decay for the last factor of 5 in $K$, while the theory is closer to asymptoting.

The particular coefficient $\log 2$ in Eq.\ref{eq:predictive_entropy} corresponds to the binary choice model; other models will vary the coefficient. Empirically we find that equally good fits are obtained over a wide range of prefactors, with changes in $N$ absorbing most of the coefficient. Indeed, the leading behavior $p \propto N^{1-a}$ means that the main effect of $N$ is to control the prefactor of the decay towards $\mathcal{L}_\infty$. The values of $N$ shown in Table \ref{table2} should then be considered as representative when Eq.\ref{eq:predictive_entropy} has an $\OO(1)$ coefficient, but are not quantitative.

For example, if we change the right-hand side of Eq.\ref{eq:predictive_entropy} to $0.05 (\log 2) [ p(K+1)-p(K) ]$, then fits are equally good and have nearly the same values of $a$, but have $N \sim 10^4$ as found from analysis of Heaps' law and Zipf's law. This has several different interpretations: (i) either the measured decay in entropy with context reflects a subset of the complexity of real language, so that the inferred $N$ is relevant only within that context, and Eq.\ref{eq:predictive_entropy} should have an $\OO(1)$ coefficient; or (ii) LLMs do capture the true $N$ of natural language, and these can all reduce entropy with context, but in an effective binary-choice grammar ensemble the observed text conveys a smaller amount of information regarding the latent variables of a rule than one bit. To discriminate between these scenarios likely requires a systematic analysis of different models and datasets within the grammar ensemble picture.

\begin{table}[h]
\begin{center}
\begin{tabular}{||c c | c c c c||} 
 \hline
 Model & Dataset & $N$ & $a$ & $\mathcal{L}_\infty$ & $\mathcal{L}_\infty$ \\ 
  &  & & &  (nats/token) & (bits/char) \\ 
 \hline\hline
 OLMo & C4 & 68 & 0.363 & 2.40 & 0.72 \\ 
 \hline
 LLama & C4 & 65 & 0.29 & 2.35 & 0.76 \\  
 \hline
Qwen & C4 & 74 & 0.246 & 2.48 & 0.84\\ 
 \hline
DCLM & C4 & 67 & 0.213 & 2.70 & 0.71 \\ 
 \hline
 OLMo & Poetry & 53 & 0.331 & 3.28 & 1.20\\  
 \hline
 OLMo & Wiki & 139 & 0.09 & 2.20 & 0.69\\ 
 \hline
\end{tabular}
\end{center}
\caption{Parameters obtained from code length fits to Eq.\ref{LK} shown in Fig.\ref{fig_L}. See text for discussion.} \label{table2}
\end{table}

\section{Conclusion}

In this work we have developed a theory of the Random Language Model in a (double) scaling limit where the grammar temperature $\ed \to 0$ and the number of hidden symbols $N\to \infty$ at fixed $x = \ed \log N$. In this limit, the model exhibits a condensation transition at $x_c=1/8$, inherited from the Random Energy Model, and governed by the breakdown of the law of large numbers in sums over grammar weights. This resolves the apparent paradox between finite-$N$ numerical results, which show smooth crossover behavior, and the existence of a sharp transition in the thermodynamic limit. We focus on the regime of a large corpus, $G \gg 1$ where $G \sim n/(N\sqrt{8x})$ is a rescaled corpus length that emerges from the theory.

The key technical step is the reorganization of the partition function as a sum over patterns, characterized by their rule multiplicity spectrum $\{ p_k, q_k \}$. This leads to a semi-annealed description in which the dominant contribution arises from a saddle-point pattern. Within this framework, the number of distinct rules $p$ emerges as a natural order parameter. In the high-temperature (``transition'' and ``babbling") regimes, and in the strict scaling limit, then $p\approx n$ and all rules are used uniquely. In contrast, below $x_c$ the model enters a frozen phase in which weight condenses onto a small subset of rules, $p/N \sim G^{\beta(x,G)}$ where $\beta(x,G) < 1$ is nontrivial, leading to low entropy and strongly constrained structure. The transition between these regimes is continuous in finite systems but sharp in the scaling limit.

The semi-annealed theory provides explicit predictions for the rule spectrum, entropy, and energy, and explains the scaling of the number of distinct rules with corpus size and temperature. In particular, it rationalizes the emergence of Heaps' law and its continuously varying exponent as a function of control parameters, Zipf's law, and connects these behaviors to the proximity of the system to the condensation transition. More generally, the theory shows how coarse-grained statistical regularities of language can arise from an ensemble of generative grammars through a well-defined statistical-mechanical mechanism.

The framework also provides a natural interpretation of recent observations in large language models: document-level grammar uncertainty predicted by the semi-annealed theory yields a slowly decaying finite-context contribution to predictive entropy, capturing empirical code lengths over many orders of magnitude in context length.


Several directions remain open. A more complete treatment of the quenched problem, beyond the semi-annealed approximation, would further clarify the nature of fluctuations around the saddle-point pattern. It would also be of interest to extend the analysis to richer grammatical structures beyond context-free rules \cite{Nakaishi24,Toji26}, and to connect the present static theory to dynamical or learning-based formulations, in which the parameters $x$ and $N$ emerge from training processes. Similarly, although the present results suggest that many features of natural language are captured by the purely random ensemble, it would be useful to incorporate grammatical parameters, and their known phylogenetic relations, into the theory \cite{Longobardi01,Longobardi23,Yarahmadi26}. 

More broadly, our results suggest that universal statistical properties of language may be understood as consequences of an underlying condensation transition in the space of generative models.

Finally, the semi-annealed construction is not specific to language, but applies more broadly to systems in which macroscopic behavior is controlled by rare configurations selected from a disordered ensemble. Examples include glasses \cite{Berthier11b}, optimization landscapes in machine learning, random Markov systems \cite{Mosam21,Mosam26}, random growth processes \cite{Bernard26}, and evolutionary fitness landscapes, where dominant states arise from extreme-value statistics but must be described in terms of coarse-grained, permutation-invariant observables. \\


{\bf Acknowledgments: } EDG acknowledges conversations with Jean-Philippe Bouchaud, Alessio Giorlandino, Sebastian Goldt, and Alessandro Treves, and NSERC Discovery Grant RGPIN-2020-04762. ChatGPT 5.6 Sol was used in revision of the manuscript. \\

\vfill 

\bibliography{../../language,../../Glasses}

\vfill

\begin{widetext} 

\appendix 

\section{Field-theory of the Random Language Model} \label{FieldTheory}

Here we recall the construction of \cite{DeGiuli19a}, which builds a field theoretic representation of the RLM. We will use this field theory to show the breakdown of simple `paramagnetic' saddle-point solutions, governed by the Random Energy Model. This discussion parallels that of \cite{Giorlandino26}, with the difference that here we use the formalism of \cite{DeGiuli19a}, which will also be employed in Appendices to later sections, while in \cite{Giorlandino26} the conventional belief propagation equations \cite{Mezard09} on a fixed tree were used \footnote{The idea to relate belief propagation equations to the Random Energy Model is due to Alessio Giorlandino and Sebastian Goldt.}. In a first reading, this section can be skipped. 

The main idea of the construction in \cite{DeGiuli19a} is to find a field theory whose Feynman diagrams generate exactly the syntax trees of the RLM, with the correct weights. Such a field theory will operate on a fixed grammar $\GG = \{M_{abc}, O_{aB} \}$ and thus be a disordered field theory. Note that the later sections of \cite{DeGiuli19a}, which solve the model in the replica-symmetric approximation, contain a subtle error corrected in \cite{De-Giuli22}. This does not affect the following discussion.

Fix a stochastic context-free grammar in Chomsky normal form $\GG = \{M_{abc}, O_{aB} \}$. We consider the partition function of $m$ sentences with total length $\ell$, which can be written as
\eq{
\ZZ(\GG; m, \ell) = \sum_{\{\ell_i\}, \sum_{i=1}^m \ell_i = \ell} \prod_i \ZZ(\GG; \ell_i) ,
}
where the tree-averaged partition function for a single sentence of length $\ell_i$ is
\eq{
\ZZ(\GG; \ell_i) = \sum_{\TT, \ell(\TT)=\ell_i} \PP(\TT| \GG) Z(\TT,\GG) .
}
We have $\sum_{i} |\p \Omega_{\TT_i}| = \sum \ell_i = \ell$, and $\sum_i |\Omega_{\TT_i}| = \sum (\ell_i-1) = \ell - m$, so that $\ZZ_{tree} \equiv \prod_i \PP(\TT_i|\GG)=p_E^{\ell} (1-p_E)^{\ell-m} Z_{tree}^{-m}$ just gives a trivial factor.

The first result of \cite{DeGiuli19a} is that $\ZZ(\GG; m,\ell)$ has a field-theoretic representation in terms of two complex scalar `fields' $L_a, R_a, a = 1,\ldots,N$ living on the space of hidden symbols. This takes the form 
\eq{ \label{W2}
\ZZ(\GG; m, \ell) = m! \oint' \frac{d\zeta}{\zeta^{1+m}} \oint' \frac{d\xi}{\xi^{1+\ell}} \oint' \frac{d\eta}{\eta^{1+\ell-m}} \;\FF(\GG),
} 
where $\oint' = \oint/(2\pi i)$, where 
\eq{
\FF(\GG) = \int DL \int DR \; e^{-\frac{1}{g} \sum_a \left[ L_a L_a^\dagger + R_a R_a^\dagger \right] } e^{I} ,
} 
where ${}^\dagger$ denotes complex conjugate and
\eq{ \label{I}
I & = \zeta h (L_1+R_1) + \xi \sum_a O_a (L_a^\dagger + R_a^\dagger) \notag\\
& \qquad + \eta \sum_{a,b,c} M_{abc} (L_a^\dagger + R_a^\dagger) L_b R_c.
}
with $O_a = \sum_B O_{aB}$. The measure $DL=\prod_a d\text{Re}[L_a] d\text{Im}[L_a]/(\pi g)$ is normalized such that $\int DL \; e^{-\frac{1}{g} \sum_a L_a L_a^\dagger} = 1$, and similarly for $R$. The constants $g$ and $h$ are chosen to satisfy $(2h)^m g^{2\ell-m} = p_E^\ell (1-p_E)^{\ell-m} Z_{tree}^{-m}$, i.e. $g=\sqrt{p_E(1-p_E)}, h=\sqrt{p_E/(1-p_E)}/(2Z_{tree})$.

Eq.\ref{W2} defines a complex-valued $3+2N-$component disordered field theory in 0 dimensions. The fact that the $L_a,R_a$ fields are complex-valued corresponds to the trees being directed: roots are distinguished from leaves, and for example $M_{abc}$ corresponds specifically to a branch from $a \to bc$ that is aligned from root towards the leaves. This entails that the propagator of the field theory distinguishes `past' from `future', if we identify time as flowing from the root to the leaves.

It follows by a straightforward rescaling that in this representation, $\ZZ(\GG; m,\ell)$ is governed by a saddle-point when $\ell \to \infty$ and $m \to \infty$ at finite average sentence length $\ell/m$. This is the limit of a large text. It is convenient to write $H_a = L_a^\dagger+R_a^\dagger$ for a head. The saddle-point equations are
\begin{subequations} \label{SP} \eq{
L_a & = g \xi O_a + g \eta \sum_{b,c} M_{abc} L_b R_c \label{SP1}\\
L_a^\dagger & = g h \zeta \delta_{a1} + g \eta \sum_{a',b,c} M_{a'bc} H_{a'} \delta_{ab} R_c \label{SP2} \\
R_a & = g \xi O_a + g \eta \sum_{b,c} M_{abc} L_b R_c \label{SP3} \\
R_a^\dagger & = g h \zeta \delta_{a1} + g \eta \sum_{a',b,c} M_{a'bc} H_{a'} L_b \delta_{ac}  \label{SP4} \\
\ell-m & = \eta \sum_{a,b,c} M_{abc} H_a L_b R_c \label{SP5} \\
\ell & = \xi \sum_{a} O_a H_a \label{SP6} \\
m & = \zeta h (L_1+R_1) \label{SP7}
} \end{subequations}
for all $a$ \footnote{Here we have fixed a typo in the second and fourth equations, with $H_{a'}$ rather than $H_a$ as published in \cite{DeGiuli19a}.}. 

Eqs.\ref{SP} bear a formal resemblance to belief propagation equations \cite{Mezard09}, but they are not identical. The former are defined for one fixed tree, and in terms of quantities with fixed site indices. In contrast, here the fields are defined on the symbols, and the partition function that is generated automatically samples all tree topologies.
Nevertheless, this identification allows us to interpret $L_a$ and $R_a$ as downward messages, while $\Ld_a$ and $\Rd_a$ are upward messages.

Subtracting the third equation from the first we see that $R_a=L_a$ although the left-right symmetry does not hold for the conjugates, in general. We can however sum the $\Ld$ and $\Rd$ equations (after swapping $b$ and $c$ in the latter) to get
\eq{
H_a & = 2 g h \zeta \delta_{a1} + g \eta \sum_{a',c}  [M_{a'ac} + M_{a'ca}  ] H_{a'} L_c \label{SP8} 
}
which is closed in $H_a$ rather than $\Ld_a$ and $\Rd_a$ separately. This implies that unless we probe observables that break the left-right symmetry, we can reduce the theory by eliminating the $R_a$ field.

Evaluating $\ZZ(\GG; m,\ell)$ on the saddle-point, we obtain to leading exponential order
\eq{
\left.\ZZ(\GG; m,\ell)\right|_{SP} & = \frac{m!}{\zeta^m \xi^\ell \eta^{\ell-m}} e^{-\frac{1}{g} \sum L_a [\Ld_a+\Rd_a] } e^{m + \ell + \ell - m} \notag  \\
& = \frac{m!}{\zeta^m \xi^\ell \eta^{\ell-m}} e^{-\frac{1}{g} g (2\ell-m) } e^{2\ell} \notag \\
& = \left( \frac{m \eta}{\zeta} \right)^m \left( \xi \eta \right)^{-\ell} 
}
using
$\sum L_a [\Ld_a+\Rd_a] = \sum_a g \xi O_a H_a+ g \eta \sum_{b,c} M_{abc} H_a L_b R_c = g \ell + g (\ell-m) = g (2\ell-m)$. This expression for $\ZZ$ includes the terms that give a finite limit in $\lim_{m \to \infty} \log \ZZ/m$.
Further simplification requires solution of the saddle-point equations.

Ref. \cite{DeGiuli19a} (with corrections discussed in \cite{De-Giuli22}) considered the disorder average of Eq.\ref{W2} in the replica-symmetric approximation. This quantitatively predicted the growth of an order parameter as $\ed$ is reduced, depending on $\ed \log N$. It did not however describe fully the low-temperature phase.

\subsection{ REM approach }

To establish the connection to the REM, consider the `paramagnetic' ansatz 
\eq{
L_a & = g \xi O_a + \bar L\\
H_a & = 2 g h \zeta \delta_{a1} + \bar H 
}
so that the main equations above become
\eq{
\bar L & =  g \eta \sum_{b,c} M_{abc} (g \xi O_b + \bar L) (g \xi O_c + \bar L) \\
\bar H  & = g \eta \sum_{a',c} (M_{a'ca}+M_{a'ac}) (2g h \zeta \delta_{a'1} + \bar H )  (g \xi O_c + \bar L)  
}
and we also have normalization equations for $\eta,\xi,$ and $\zeta$:
\eq{
\ell-m & = \eta \sum_{a,b,c} M_{abc} (2 g h \zeta \delta_{a1} + \bar H ) (g \xi O_b + \bar L)(g \xi O_c + \bar L)  \\
\ell & = \xi \sum_{a} O_a (2 g h \zeta \delta_{a1} + \bar H )  \\
m & = 2\zeta h (g \xi O_1 + \bar L) 
}
Let us make the simplifying assumption that $O_1=0$, i.e. the start symbol never directly becomes an observable. In the general case this is a $1/N$ correction that is irrelevant in the large $N$ limit to be considered. Then $\zeta = m/(2h\bar L)$ and $\xi = \ell/(\bar H \sum_a O_a)$, and the $\eta$ equation is trivially solved.

Now return to the main equations. The paramagnetic ansatz will close if the law of large numbers holds for all terms carrying a free index. We have the sums
\begin{subequations} \eq{
S_a^{(1)} &= \sum_{b,c} M_{abc} \\
S_a^{(2)} &= \sum_{b,c} M_{abc} O_b \\
S_a^{(3)} &= \sum_{b,c} M_{abc} O_c \\
S_a^{(4)} &= \sum_{b,c} M_{abc} O_b O_c \\
S_a^{(5)} &= \sum_{c} (M_{1ac}+M_{1ca}) \\
S_a^{(6)} &= \sum_{c} (M_{1ac}+M_{1ca}) O_c ,
} \end{subequations} 
which all resemble REM partition functions, but with different system sizes and variances. Let us write
\eq{
Z_{REM}(\NN,\lambda) = \sum_{j=1}^{\NN} e^{-\beta E_j} 
}
where $\langle E_j \rangle = 0, \langle (\beta E_j)^2 \rangle = \lambda^2/2$. Derrida's result is that the LLN breaks down when 
\eq{
\frac{\beta_c}{\beta} = \frac{2\sqrt{\log 2}}{\beta J} = \frac{2\sqrt{\log \NN}}{\lambda} 
}
is smaller than unity. The corresponding parameters for the above sums are
\begin{subequations} \eq{
\NN^{(1)} & = N^2, \lambda^{(1)} = \sqrt{1/\ed} \\
\NN^{(2)} & = N^2 T, \lambda^{(2)} = \sqrt{1/\ed + 1/\es} \\
\NN^{(3)} & = N^2 T, \lambda^{(3)} = \sqrt{1/\ed + 1/\es} \\
\NN^{(4)} & = N^2 T^2, \lambda^{(4)} = \sqrt{1/\ed + 2/\es} \\
\NN^{(5)} & = N, \lambda^{(5)} = \sqrt{1/\ed} \\
\NN^{(6)} & = N T, \lambda^{(6)} = \sqrt{1/\ed + 1/\es} 
} \end{subequations} 
In this identification, we are assuming that for each sum the elements are independent. This is strictly true only for $S_a^{(1)}$; for the others, there are correlations because for example $M_{abc}O_b$ is correlated with $M_{a'bc'}O_b$ through the shared value $O_b$. In the large $N$ limit, we expect these correlations to be irrelevant, because they occur for a vanishing fraction of terms in the sums. 

Note that, in the scaling limit, we have that all $\lambda^{(i)} \sim \sqrt{\log N}$ while $\sqrt{\log \NN^i} \sim \sqrt{\log N}$ as well. Hence $\beta_c/\beta$ is finite: the scaling limit of the RLM corresponds to the thermodynamic limit of the REM.

We see that the first breakdown will occur due to the root terms (i.e. those involving fixed symbol index 1), when $\ed \log N < 1/4$ or $\frac{\ed \es}{\ed+\es} \log NT < 1/4$. However it is not obvious whether the associated terms are actually relevant in the equations. A short calculation shows that, starting from the large $\ed$ limit where all terms are self-averaging, the root terms have a relative weight
\eq{
\frac{2g h \zeta \delta_{a'1}}{\bar H} \sim \frac{m}{\ell-2m},
}
which is finite as $N \to \infty$ but suppressed for large trees, since $\ell/m$ is the average length of a sentence.


After this first breakdown, which has a weak effect because it is suppressed by the average sentence length, the LLN will break down when $S_a^{(1)}$ is not self-averaging; this occurs at $\ed \log N = 1/8$. This sum controls the bulk propagation of messages and the breakdown of the LLN for it will destabilize the paramagnetic solution.

The result of the analysis is then the following: starting at large $\ed$, for the bulk of the tree (i.e. for observables that do not depend on the terminals, and are therefore independent of $\es$ and $T$), a first weak breakdown will occur at $\ed \log N = 1/4$, due to the influence of the start symbol. A complete breakdown will occur at $\ed \log N = 1/8$. 

For observables that depend on the terminals, the result depends on $T/N$ and $\ed/\es$ but if $\es$ is fixed to be small (which is usually taken in the numerics, e.g. $\es = 0.01$) then a weak breakdown will occur at $\frac{\ed \es}{\ed+\es} \log NT = 1/4$, and a complete breakdown at $\frac{\ed \es}{\ed+\es} \log (N^2T) = 1/4$. 

These breakdowns are all of the LLN. As in the REM there will be corresponding breakdowns in the CLT at larger temperatures \cite{Ben-Arous05}. For a sum over $N^K$ variables, the LLN breaks down at $x_{\rm LLN}=1/4K$ while the CLT breaks down at $x_{\rm CLT}=1/K$. Thus each CLT transition occurs at a value of $x$ that is larger by a factor 4.




\section{Counting derivations} \label{AppA}

Suppose we have $p$ rules $a_r \to b_r c_r$, $r = 1, \ldots, p$ and $q$ rules $t_r \to B_r$. Write
\eq{
M_{abc} = \sum_r \delta_{(abc),(a_r,b_r,c_r)}, \quad  O_{aB} = \sum_r \delta_{(aB),(t_r,B_r)} .
}
The partition function $\ZZ(\GG; m,\ell)$ then counts how many configurations can be made with this grammar. We will evaluate this, averaged over the identities of the rules, in the annealed approximation. Using the diagrammatic representation the grammar contributes through a term 
\eq{
e^{\eta \sum_{abc} M_{abc} H_a L_b R_c} ,
}
which can be grammar-averaged, assuming independence of rules, to obtain
\eqs{
\overline{ e^{\eta \sum_{abc} M_{abc} H_a L_b R_c} } & = \prod_{r=1}^{p} \left( \frac{1}{N^3} \sum_{a_r=1}^N \sum_{b_r=1}^N \sum_{c_r=1}^N e^{\eta \sum_{abc} \delta_{(abc),(a_r,b_r,c_r)}  H_a L_b R_c} \right) \\
& = \prod_{r=1}^{p} \left( \frac{1}{N^3} \sum_{a_r=1}^N \sum_{b_r=1}^N \sum_{c_r=1}^N e^{\eta H_{a_r} L_{b_r} R_{c_r}} \right) \\
& = ( S/N^3)^p
}
with $S= \sum_{a,b,c} e^{\eta H_{a} L_{b} R_{c}}$.  Likewise the average over terminal rules gives 
\eqs{   
\overline{ e^{\xi \sum_{a,B} O_{aB} H_a} } 
& = \prod_{r=1}^{q} \left( \frac{1}{NT}  \sum_{t_r=1}^N \sum_{B_r=1}^T  e^{\xi H_{t_r} } \right) \\
& = ( S'/N)^q
}
with $S' = \sum_a e^{\xi H_a}$.

The same arguments as in \cite{DeGiuli19a} imply that $Z$ is dominated by a saddle-point. The saddle-point equations are
\begin{subequations}\eq{
L_a & = g q S'^{-1} \xi e^{\xi H_a} + g p S^{-1} \eta \sum_{b,c} L_b R_c e^{\eta H_{a} L_{b} R_{c}} \label{appSP1}\\
R_a & = g q S'^{-1} \xi e^{\xi H_a} + g p S^{-1} \eta \sum_{b,c} L_b R_c e^{\eta H_{a} L_{b} R_{c}} \label{appSP2}\\
\Ld_a & = g h \zeta \delta_{a1} + g p S^{-1} \eta \sum_{a',c} H_{a'} R_c e^{\eta H_{a'} L_{a} R_{c}} \label{appSP3} \\
\Rd_a & = g h \zeta \delta_{a1} + g p S^{-1} \eta \sum_{a',b} H_{a'} L_b e^{\eta H_{a'} L_{b} R_{a}} \label{appSP4} \\
\ell-m & = p S^{-1} \eta \sum_{a,b,c}  H_a L_b R_c e^{\eta H_{a} L_{b} R_{c}} \label{appSP5} \\
\ell & = q S'^{-1} \xi \sum_{a} H_a e^{\xi H_a}  \label{appSP6} \\
m & = \zeta h (L_1+R_1) \label{appSP7}
} \end{subequations}
Subtracting the first two equations implies $L_a=R_a$. Adding the third and fourth closes the equations in terms of $H_a= \Ld_a+\Rd_a$:
\eq{
H_a & = 2 g h \zeta \delta_{a1} + 2 g p S^{-1} \eta \sum_{a',b} H_{a'} L_b e^{\eta H_{a'} L_{b} L_{a}} \label{appSP8} 
}
Note that 
\eqs{
\sum_a L_a H_a & = 2 g h \zeta L_1 + 2 g p S^{-1} \eta \sum_{a,a',b} H_{a'} L_a L_b e^{\eta H_{a'} L_{b} L_{a}} \\
& = gm + 2 g (\ell-m)
}
so that the partition function at the saddle-point is
\eqs{
\overline{ \ZZ(\GG; m,\ell) } & = \frac{m!}{\zeta^m \xi^\ell \eta^{\ell-m} } e^{-\frac{1}{g} \sum_a L_a H_a} e^{\zeta h (L_1+R_1) + q \log (S'/N) + p \log (S/N^3) } \\
& = \frac{m!}{\zeta^m \xi^\ell \eta^{\ell-m} } e^{-(2\ell-m)} e^{m}  (S'/N)^q (S/N^3)^p .
}
The saddle-point equations have no disorder, so they will have a simple solution $H_a=H_*$ for all $a>1$, and likewise for $L_a$. Then we have the system
\eqs{
L_1 & = g q S'^{-1} \xi e^{\xi H_a}  + g p S^{-1} \eta \sum_{b,c} L_b L_c e^{\eta H_{1} L_{b} L_{c}} \\
L_* & = g q S'^{-1} \xi e^{\xi H_a}  + g p S^{-1} \eta \sum_{b,c} L_b L_c e^{\eta H_{*} L_{b} L_{c}} \\
H_1 & = 2g h \zeta + 2g p S^{-1} \eta \sum_{a',c} H_{a'} L_c e^{\eta H_{a'} L_{1} L_{c}}  \\
H_* & = 2g p S^{-1} \eta \sum_{a',c} H_{a'} L_c e^{\eta H_{a'} L_{*} L_{c}}  \\
\ell-m & = p S^{-1} \eta \sum_{a,b,c}  H_a L_b L_c e^{\eta H_{a} L_{b} L_{c}} \\
\ell & = q S'^{-1}\xi \sum_{a} H_a e^{\xi H_a}  \\
m & = 2\zeta h L_1 \\
S & = \sum_{a,b,c} e^{\eta H_{a} L_{b} L_{c}} \\
S' & = \sum_a e^{\xi H_a}
}
This system has a one-parameter scaling symmetry
\eq{
\xi \to c \xi, \quad H \to (1/c) H, \quad L \to c L, \quad \eta \to (1/c) \eta, \quad \zeta \to (1/c) \zeta
}
for any $c$, which allows us to fix $\zeta=1$ and hence $L_1 = m/(2h)$. 

There are various possible regimes of interest depending upon the scaling of $\ell$ and $m$ with $N$. We assume $p \sim N$.


The most important regime is when the average tree length $\ell/m$ is large, because then correlation lengths can be large. This should also be the simplest, because the root plays a less important role. 
In this case we can seek a solution with $H_1 \sim H_*$ and $L_1 \sim L_*$. In the large $N$ limit the contributions of the root would then be negligible compared to the bulk in all sums, giving
\eqs{
L_1 & = g q S'^{-1}  \xi e^{\xi H_1}  + g p S^{-1} \eta (N-1)^2 L_*^2 e^{\eta H_{1} L_*^2 } \\
L_* & = g q S'^{-1} \xi e^{\xi H_*} + g p S^{-1} \eta (N-1)^2 L_*^2 e^{\eta H_{*} L_*^2} \\
H_1 & = 2g h  + 2g p S^{-1} \eta (N-1)^2 H_{*} L_* e^{\eta H_{*} L_{1} L_*}  \\
H_* & = 2g p S^{-1} \eta (N-1)^2 H_{*} L_* e^{\eta H_{*} L_{*}^2 }  \\
\ell-m & = p S^{-1} \eta (N-1)^3 H_* L_*^2 e^{\eta H_{*} L_{*}^2 } \\
\ell & = q S'^{-1} \xi (N-1) H_* e^{\xi H_*} \\
S & = (N-1)^3 e^{\eta H_{*} L_*^2 } , \\
S' &= (N-1) e^{\xi H_*}
}
which reduces to 
\eqs{
L_* & = g (q/N) \xi  + g (p/N) \eta L_*^2 \\
\ell-m & = p \eta H_* L_*^2 \\
\ell & = q \xi H_*  \\
H_* & = 2g (p/N) \eta H_{*} L_*   
}
and then to
\eqs{
L_* & = g \ell /(N H_*)  + (g/N) (\ell-m) /H_*\\
H_* & = 2(g/N) (\ell-m)/L_*.
}
For $m/\ell \to 0$ in the large $N$ limit these are degenerate but consistent and give $H_*L_* = 2g\ell/N$. It follows $\eta = \ell/(p H_*L_*^2) = N/(2g p L_*)$ and then we have, using $S = N^3 e^{\ell/p}$,
\eqs{
L_1 & = g \ell/(N H_*)  + g p N^{-1} e^{-\ell/p} \eta L_*^2 e^{\eta H_{1} L_*^2 } \\
H_1 & = 2g h  + 2g p N^{-1} e^{-\ell/p} \eta H_{*} L_* e^{\eta H_{*} L_{1} L_*}  
}
which become
\eqs{
L_1 & = g \ell/(N H_*)  + \half e^{-\ell/p}  L_* e^{N H_{1} L_* /(2gp) } \\
H_1 & = 2g h  + e^{-\ell/p} H_{*} e^{N H_{*} L_{1} /(2gp) }  
}
and then
\eqs{
H_* L_1 & = g \ell/N  + \half e^{-\ell/p} (2g\ell/N) e^{N H_{1} L_* /(2gp) } \\
H_1 L_* & = 2g h L_* + e^{-\ell/p} (2g\ell/N) e^{N H_{*} L_{1} /(2gp) }  
}
These are nontrivial but consistent if $m/\ell \to 0$ as $N \to \infty$. Since $L_1 \sim m$ we get $H_* \sim 1/m$ and we see that $H_1 \sim 1, L_* \sim 1$. This allows a weak divergence of $m$ with $N$. For example, in $S$ we have
\eqs{
S = N^3 e^{\eta H_* L_*^2} + 2N^2 e^{\eta H_* L_* L_1} + \ldots
}
which can be dominated by the second term if, assuming $\eta H_* L_*^2 \approx \ell/p$ as above, $N \ll 2 e^{(L_1/L_*-1)(\ell/p)}$, i.e. $m > 1 + (p/\ell) \log N$. Therefore to avoid this, $m$ cannot grow faster than $(p/\ell) \log N$.


The partition function becomes
\eqs{
\overline{ \ZZ(\GG; m,\ell) } & = m! e^{-2\ell+2m} (N/(2g p L_*))^m (\ell N/(2g q p H_* L_*))^{-\ell} (e^{\ell/q})^q (e^{\ell/p})^p  \\
& = m! e^{2m} (N/(2g p L_*))^m (N^2 /(4g^2 pq))^{-\ell} 
}
so that the leading behavior of $\log\overline{ \ZZ(\GG; m,\ell) }$ is 
\eq{
\log\overline{ \ZZ(\GG; m,\ell) } = \ell \log (4g^2) + \ell \log (q/N) + (\ell-m)\log (p/N) + \ldots
}
with corrections proportional to $m$. This result matches expectations: the term $\ell \log (4g^2)$ is the contribution of trees (with entropy $\ell \log 4$ and energy $-\ell \log g^2$), while $(\ell-m)\log (p/N) + \ell \log (q/N)$ is the entropy of derivations on the trees. The latter is exactly what is predicted by counting derivations.

\subsection{Corrections to leading order}

To examine the corrections, we have
\eqs{
H_* L_1 = \frac{2gp}{N} \left[ \ell/p + \log (N/2g\ell) + \log(H_1L_*-2ghL_*) \right]
}
and then
\eqs{
\ell e^{-\ell/p} e^{N H_{1} L_* /(2gp) } = \ell + 2p \log (N/2g\ell) + 2p \log(H_1L_*-2ghL_*) 
}
giving $H_1 L_* \approx 2g\ell/N$ up to logarithmic corrections. Similarly $H_* L_1 \approx 2g\ell/N$. To obtain the corrections we write $H_1 L_* = 2g\ell/N + a$, $H_* L_1 = 2g\ell/N + b$, giving
\eqs{
b & = (g\ell/N) \left[ e^{N a /(2gp) } -1 \right] \\
a & = 2g h L_* +  (2g\ell/N) \left[ e^{N b /(2gp) } -1 \right] .
}
Assuming that $a$ and $b$ are small (since $2g\ell/N \sim 1$) we get $b \approx a \ell/(2p)$ and then $2gh L_* = a - \ell b/p = a ( 1 - \ell^2/2p^2)$. This yields
\eqs{
& N h L_*^2 ( 1 - \ell^2/(2p^2))^{-1} + 2 p L_* - mp/h = 0 \\
}
and finally 
\eq{
L_* & = \frac{2p^2 - \ell^2}{4pNh} \left[ - 1 + \sqrt{ 1 + \frac{4pmN}{2p^2 - \ell^2} } \right] 
}
Self-consistency requires $Na/(2gp) \lesssim 1$ which amounts\footnote{We have $g^2 \approx 1/4$ to get large trees}  to $4pmN/(2p^2-\ell^2) \lesssim 1$. This requires $m \lesssim p/2N$, meaning that the number of trees cannot be too large. Equivalently, we can safely ignore the root terms only when trees are very large. 



\subsection{Quenched grammar average}

The above results can be extended to computation of $\overline{ \log \ZZ }$, the quenched average over grammars. Using the replica method 
\eqs{
\overline{ \log \ZZ } = \overline{ \lim_{q \to 0} d(\ZZ^q)/dq } = \lim_{q \to 0} d\overline{\ZZ^q}/dq
}
we need to replicate all $\{ \eta, \zeta, \xi, L_a, R_a \}$, adding a replica index $\alpha = 1,\ldots,q$. The key point is that the exponential coupling gets replaced by 
\eqs{
e^{\eta H_a L_a R_a} \to e^{\sum_\alpha \eta_\alpha H_a^\alpha L_a^\alpha R_a^\alpha} ,
}
and likewise for $e^{\xi H_a}$, which does not affect the structure of the saddle-point equations. If we make the same assumption $L_a^\alpha = L_*^\alpha$ for $a>1$, then it is easily shown that all replicas are identical, so that $\overline{\log \ZZ} = \log \overline{\ZZ}$. 

Thus the only way to find more general solutions is to break the symmetry among \emph{symbols}. Preliminary attempts to find RLM solutions breaking the symbol symmetry were reported in \cite{Lalegani24a}.

\section{Contingency tables} \label{AppB}


Suppose we have $p$ rules $r= 1, \ldots, p$ with $\pi_r$ copies of each rule. We want to count the number of ways to place these into $m_c$ piles $j = 1,\ldots,m_c$ such that each pile has $\ell_c$ rules, counted with multiplicity. Consistency requires that $\sum_r \pi_r = \ell_c m_c = n$. 

Let $A_{rj}$ be the number of rule $r$ placed into pile $j$. Then, considered as a matrix, $A$ has both imposed row sums and imposed column sums. Understanding such ``contingency tables" is a classical problem of statistics \cite{Good63}, and has been studied in combinatorics \cite{Barvinok10}. We use statistical physics methods that allow us to go beyond the leading order. We want to compute
\eq{
Z_c & = \sum_{ \{ A_{rj} \} } \prod_{r=1}^{p-1} \delta_{\pi_r - \sum_j A_{rj}} \prod_{j=1}^{m_c} \delta_{\ell_c - \sum_r A_{rj} } \\
& = \sum_{ \{ A_{rj} \} } \prod_{r=1}^{p-1} \int_0^{2\pi} \frac{d\lambda_r}{2\pi} e^{ i\lambda_r (\pi_r - \sum_j A_{rj}) } \prod_{j=1}^{m_c} \int_0^{2\pi} \frac{d\gamma_j}{2\pi} e^{i\gamma_j (\ell_c - \sum_r A_{rj}) }  \notag\\
& = \prod_{r=1}^{p-1} \int_0^{2\pi} \frac{d\lambda_r}{2\pi} e^{ i\lambda_r \pi_r  } \prod_{j=1}^{m_c} \int_0^{2\pi} \frac{d\gamma_j}{2\pi} e^{i\gamma_j \ell_c  } \prod_{r} \prod_{j} (1 - e^{-i\lambda_r \Theta_{p-r} - i\gamma_j  } )^{-1} \notag
}
where the Kronecker $\delta$'s enforce the constraints. Note that $r$ only runs from $1$ to $p-1$ in the constraints because one of the constraints is redundant, due to the consistency relation. In the third line, $\Theta_{p-r}$ is 1 except when $r=p$. 
The regime of interest is when $\ell_c$ is large, so we can seek a saddle-point.  The action is 
\eq{
S = -\sum_{r < p} i\lambda_r \pi_r - \sum_{j \leq m_c} i\gamma_j \ell_c + \sum_r \sum_j \log(1 - e^{-i\lambda_r \Theta_{p-r}  - i\gamma_j  })
}
and the saddle-point equations are
\eq{
\pi_r & = \sum_{j} (1 - e^{-i\lambda_r - i\gamma_j  } )^{-1} e^{-i\lambda_r - i\gamma_j  } \\
\ell_c & = \sum_{r \leq p} (1 - e^{-i\lambda_r \Theta_{p-r} - i\gamma_j  } )^{-1} e^{-i\lambda_r \Theta_{p-r} - i\gamma_j } 
} 
The $\gamma_j$ equations have a solution $\gamma_j=\gamma$ for all $j$. This leads to
\eqs{
\pi_r & = m_c  (e^{i\lambda_r + i\gamma  } -1 )^{-1} \\
\ell_c & = \sum_{r<p} (e^{i\lambda_r + i\gamma  } -1 )^{-1} +  (e^{i\gamma  } -1 )^{-1}
}
and then
\eqs{
i\lambda_r & = -i\gamma + \log (1 + m_c/\pi_r) \\
\ell_c & = \sum_{r<p} (\pi_r/m_c) +  (e^{i\gamma  } -1 )^{-1} = (n-\pi_p)/m_c +  (e^{i\gamma  } -1 )^{-1} 
}
The second equation yields $i\gamma = \log (1 + m_c/\pi_p)$. Evaluating the action we have
\eq{
S|_{SP} & = -\sum_{r < p} [-i\gamma + \log (1 + m_c/\pi_r)] \pi_r - m_c [\log (1 + m_c/\pi_p)] \ell_c \notag\\
&\qquad + \sum_{r<p} m_c \log(1 - 1/(1+m_c/\pi_r)) + m_c \log(1 - 1/(1+m_c/\pi_p)) \notag \\
& = i\gamma (n-\pi_p) - \sum_{r < p} \pi_r \log (1 + m_c/\pi_r) - n \log (1 + m_c/\pi_p) \notag\\
&\qquad + \sum_{r \leq p} m_c \log(m_c/(\pi_r+m_c)) \notag \\
& =  - \sum_{r \leq p} \pi_r \log (1 + m_c/\pi_r) - \sum_{r \leq p} m_c \log(1+\pi_r/m_c) 
}
All dependence on the special index $r=p$ has disappeared, as it must. 

Note that, using Stirling,
\eqs{
\prod_{k \geq 1} \binom{k + m_c - 1}{m_c-1}^{p_k} &= \prod_r \binom{\pi_r + m_c - 1}{m_c-1} \\
& \sim \frac{(\pi_r+m_c-1)^{\pi_r+m_c-1}}{(m_c-1)^{m_c-1}\pi_r^{\pi_r}} \\
& = \prod_r \left(1+\frac{m_c-1}{\pi_r} \right)^{\pi_r} \left(1+\frac{\pi_r}{m_c-1} \right)^{m_c-1} \\
& \sim e^{-S|_{SP}}
}
for $m_c \gg 1$. This leading estimate agrees with a classical approximation of Good \cite{Good63}. It implies that when $m_c \gg 1$ the argument given in the main text, which does not enforce that each pile have strictly $\ell_c$ rules, gives the same result as the leading saddle-point contribution of the full computation. 

The corrections are obtained by evaluating the Hessian at the saddle point. Starting from the nonlinear part of the action
\eqs{
S_2 = \sum_{r<p} \sum_j \log(1 - e^{-i\lambda_r - i\gamma_j  }) + \sum_j \log(1 - e^{- i\gamma_j  })
}
we have
\eqs{
\p_r S_2 & = i\sum_j (e^{i\lambda_r + i\gamma_j  }-1)^{-1} \\
\p_j S_2 & = i\sum_{r<p} (e^{i\lambda_r + i\gamma_j  }-1)^{-1} +  i(e^{i\gamma_j  }-1)^{-1}
}
and then
\eqs{
\p_{r'} \p_r S_2 & = \delta_{rr'} \sum_j (e^{i\lambda_r + i\gamma_j  }-1)^{-2} e^{i\lambda_r + i\gamma_j  } \\
\p_j \p_r S_2 & = (e^{i\lambda_r + i\gamma_j  }-1)^{-2} e^{i\lambda_r + i\gamma_j  } \\
\p_{j'} \p_j S_2 & = \delta_{jj'} [ \sum_{r<p} (e^{i\lambda_r + i\gamma_j  }-1)^{-2} e^{i\lambda_r + i\gamma_j  }  + (e^{i\gamma_j  }-1)^{-2} e^{i\gamma_j  }  ]
}
The Hessian matrix, evaluated on the saddle-point, is
\eq{
H = \begin{bmatrix} \nabla_{r} \nabla_{r'} S_2 &  \nabla_{r} \nabla_{j} S_2 \\ \nabla_{j'} \nabla_{r'} S_2 &  \nabla_{j} \nabla_{j'} S_2 \end{bmatrix} =  \begin{bmatrix} \delta_{rr'} m_c v_r & v_r \\ v_r & \delta_{jj'} a \end{bmatrix}
}
where $r$ and $r'$ run from $1$ to $p-1$. Here $v_r = \pi_r (\pi_r+m_c)/m_c^2$ and $a= \sum_{r=1}^p v_r$. Its determinant is
\eq{
\det H & = a^{m_c-1} m_c^{p-1} \prod_{r=1}^p v_r \notag\\
& = \left[\sum_{k \geq 1} p_k k (k+m_c) \right]^{m_c-1} m_c^{-2m_c+2+p-1 -2p} \prod_{k \geq 1} ( k (k+m_c))^{p_k} 
}
so that
\eq{
Z_c \approx e^{-S|_{SP}} (2\pi)^{-(m_c+p-1)/2} |\det H|^{-1/2}
}
Since $S_{SP} \sim n$ while $\log \det H \sim \max\{p,m_c\}$ the latter is subdominant as expected.

\section{Coverage of hidden symbols at the surface} \label{app:surface_coverage}

Let $\mathcal A$ be the set of hidden symbols that occur at
leaves of the hidden tree, and write $R=|\mathcal A|$.
We estimate $R$ within a homogeneous occupancy approximation.
Critical branching determines the tree topology, but does not
by itself imply homogeneous sampling of hidden labels; the
latter is an additional assumption of this estimate.

A branching rule used $k$ times has two child positions.
In an unconditioned critical binary Galton--Watson process,
each child terminates with probability $1/2$.
Neglecting correlations between termination decisions and
the grouping of occurrences into rule types, the probability
that a given child position is observed as a leaf at least
once is    $\tau_k=1-2^{-k}$. Thus the expected number of exposed rule-child positions,
before identifying positions carrying the same hidden label,
is
\begin{equation}
    s_{\partial}
    =2\sum_{k\geq1}p_k(1-2^{-k})
    \equiv \kappa_{\partial}p,
    \qquad
    1\leq\kappa_{\partial}\leq2,
    \label{eq:boundary_exposures}
\end{equation}
where $p=\sum_kp_k$.
The coefficient depends on the multiplicity spectrum:
singleton rules give $\kappa_{\partial}=1$,
twice-used rules give $\kappa_{\partial}=3/2$,
and frequently reused rules approach
$\kappa_{\partial}=2$.
Finite-tree conditioning and correlations between repeated
rule occurrences are neglected in this closure.

Assuming independent uniform hidden labels at the candidate
child positions, the probability that a specified symbol is
never exposed is approximately
\begin{equation}
    P_{\rm absent} \simeq \prod_{k\geq1} \left(1-\tau_k + \tau_k (1 -1/N) \right)^{2p_k} \approx e^{-s_\partial /N}
\end{equation}
for $p\ll N^2$. 
Consequently, the typical coverage is estimated by
\begin{equation}
    R\simeq
    N\left[1-e^{-\frac{s_{\partial}}{N}} \right].
    \label{eq:boundary_coverage}
\end{equation}
In particular,
\begin{equation}
    R\simeq\kappa_{\partial}p
    \quad (p\ll N),
    \qquad
    R\simeq N
    \quad (N\ll p\ll N^2).
\end{equation}

The expected number of missing symbols is approximately $N e^{-s_{\partial}/N}$, so complete coverage with high probability requires $s_{\partial}\gtrsim N\log N$, or $p/N \gtrsim (1/\kappa_\partial) \log N$.

\section{Weighted versus Probabilistic Grammars} \label{app:WCFG}

\begin{figure*}[t!]
\includegraphics[width=\textwidth]{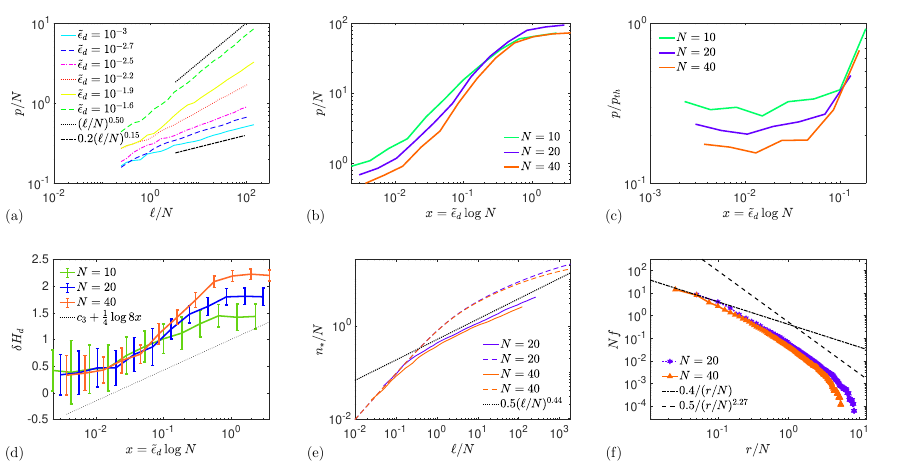}
\caption{ Representative results for a WCFG with $T=10^3$ and $\es = 10^{-2}$. (a) $p$ vs $\ell$ at $N=40$; (b) $p$ vs $x$ at $\ell \sim 4000$; (c) $p$ vs $p_{th}$ from Eq.\ref{pfrozen} at $\ell \sim 4000$, the theoretical prediction of Eq.\ref{pfrozen}; (d) hidden entropy rate $H_d$; (e) Heaps' law at $\ed = 10^{-2.5}$. Here, $a'=0.44$ and the asymptotic result is $n_* \sim \ell^{a'}$; and (f) Zipf's law. The asymptotic result for $r>N$ is $f \sim r^{-1/a'}$.}\label{fig_WCFG}
\end{figure*}

The main text shows numerical results for model P, a probabilistic grammar. Here we show robustness of the numerical results in model W, a weighted grammar. In Model W, the partition function depends on each tree:
\eq{
Z(\TT,\GG) = \sum_{\{ \sigma_i, o_t \} } e^{-E(\sigma,o)} , 
}
and the probability of a derivation, given the tree and the grammar, in this weighted CFG, is 
\eqs{
\PP(\sigma,o | \TT,\GG) = \frac{e^{-E(\sigma,o)}}{Z(\TT,\GG)} .
}
To sample a weighted grammar, one can use Markov chain Monte Carlo, iterating over derivations on a single tree, but it is simpler to numerically find $Z(\TT,\GG)$ exactly. This is described below.

Fig.\ref{fig_WCFG} shows representative results for a WCFG, sampled using the method below. Compared to a PCFG, it has larger sample-to-sample fluctuations, but otherwise behaves similarly. Fig.\ref{fig_WCFG} shows (a) $p$ vs $\ell$; (b) $p$ vs $x$; (c) $p$ vs $p_{th}$, the theoretical prediction of Eq.\ref{pfrozen}; (d) hidden entropy rate $H_d$; (e) Heaps' law; and (f) Zipf's law. The main differences from model P are (i) novelty is smaller by a factor of $\sim 3$; (ii) Heaps' law is closer to the asymptotic power-law form without renormalization; (iii) Zipf's law has a smoother crossover between regimes.

The smaller values of $p$, and in particular the relatively large region where $p/N \lesssim 1$, means that the surface layer is not entirely decoupled from the hidden layer. In practice, this means that the number of hidden symbols available to become observable symbols is smaller than $N$, and hence surface novelty is smaller. This qualitatively explains why the naive Heaps' law prediction is $\sim 5$ times larger than the observed value, see  Fig.\ref{fig_WCFG}e. A refinement of the theory could treat this quantitatively.

\subsection{Exact sampling of weighted grammars}
\label{app:weighted_sampling}

For the weighted-grammar simulations performed with the direct
sampler, we prescribe the distribution of sentence lengths and
then sample tree topologies and rule assignments jointly,
conditional on length. Let $d=(\sigma,o)$ denote a complete ordered
derivation, $\ell(d)$ its number of terminals, and
$e^{-E(d)}$ the product of its branching and lexical
rule weights. The root is fixed to the start symbol $1$.
Writing $L$ for the exclusive length cutoff, the simulated law is
\begin{equation}
 P_{\rm samp}(d\mid\mathcal G)
 =\pi_L(\ell(d))\,
   \frac{e^{-E(d)}}{\mathcal Z_1(\ell(d))},
 \qquad 1\leq\ell(d)<L,
 \label{eq:wcfg_sampling_law}
\end{equation}
where
\begin{equation}
 \pi_L(\ell)=
 \frac{C_{\ell-1}p_E^\ell(1-p_E)^{\ell-1}}
 {\displaystyle\sum_{r=1}^{L-1}
  C_{r-1}p_E^r(1-p_E)^{r-1}},
 \qquad
 C_j=\frac{1}{j+1}\binom{2j}{j}.
 \label{eq:wcfg_gw_lengths}
\end{equation}
This is the binary Galton--Watson leaf-count law conditioned
on the cutoff. Thus $p_E$ controls the length distribution,
while grammar weights determine the conditional distribution
of shapes and labels.

\paragraph{Length-dependent partition functions.}
Let $\mathcal D_a(\ell)$ contain all complete derivations with
root symbol $a$ and $\ell$ terminals. Define
\begin{equation}
 \mathcal Z_a(\ell)
 =\sum_{d\in\mathcal D_a(\ell)} e^{-E(d)}.
 \label{eq:wcfg_length_partition}
\end{equation}
Unlike a partition function on a fixed topology, this quantity
sums over all ordered binary shapes of the specified length.
In particular,
$\mathcal Z_1(\ell)=\sum_{\mathcal T:\,\ell(\mathcal T)=\ell}
Z(\mathcal T,\mathcal G)$.
For one terminal the root emits directly, whereas for
$\ell\geq2$ every derivation has a unique root rule
$a\to bc$ and a unique left-subtree length $k$. Therefore
\begin{align}
 \mathcal Z_a(1)&=\sum_B O_{aB},
 \label{eq:wcfg_length_base}\\
 \mathcal Z_a(\ell)&=
 \sum_{k=1}^{\ell-1}\sum_{b,c}
 M_{abc}\mathcal Z_b(k)\mathcal Z_c(\ell-k).
 \label{eq:wcfg_length_recursion}
\end{align}
The subtree weights multiply and their assignments are summed
independently, conditional on $b,c$ and $k$.
This decomposition is exact for a fixed grammar, including
when rule identities recur in different subtrees.
Equivalently, the formal generating functions
$F_a(z)=\sum_{\ell\geq1}\mathcal Z_a(\ell)z^\ell$ obey
\begin{equation}
 F_a(z)=z\sum_B O_{aB}
       +\sum_{b,c}M_{abc}F_b(z)F_c(z),
\end{equation}
whose coefficient equations are precisely
Eq.~\eqref{eq:wcfg_length_recursion}.

\paragraph{Numerical evaluation and sampling.}
The partition functions are evaluated in increasing order of
$\ell$, storing their logarithms. All sums use the stable
operation
\begin{equation}
 \operatorname{LSE}_j x_j
 =x_{\max}+\log\sum_j e^{x_j-x_{\max}}.
\end{equation}
For efficiency the code caches
$G_{ab}(m)=\sum_c M_{abc}\mathcal Z_c(m)$, so that
$\mathcal Z_a(\ell)=\sum_{k,b}\mathcal Z_b(k)G_{ab}(\ell-k)$.
The same contractions are performed in logarithmic form,
avoiding products of large or small weights. Only lengths
up to the largest length drawn in the current batch need
to be tabulated.

After drawing $\ell$ from Eq.~\eqref{eq:wcfg_gw_lengths},
an internal subtree with root $a$ and length $\ell\geq2$
is sampled by choosing
\begin{equation}
 P(k,b,c\mid a,\ell,\mathcal G)
 =\frac{M_{abc}\mathcal Z_b(k)\mathcal Z_c(\ell-k)}
 {\mathcal Z_a(\ell)}.
 \label{eq:wcfg_split_sampling}
\end{equation}
The two subtrees are then sampled recursively.
At length one, the terminal is drawn with probability
$O_{aB}/\mathcal Z_a(1)$.
Multiplication of these conditional probabilities cancels
all non-root subtree partition functions, giving exactly
$W(d\mid\mathcal G)/\mathcal Z_1(\ell)$.

For model P, the row normalizations imply
$\mathcal Z_a(\ell)=C_{\ell-1}$ for every $a$, by induction
using the Catalan recursion. In that case the conditional
shape distribution is uniform and the procedure recovers
the usual cutoff-conditioned Galton--Watson PCFG law.
For model W, the shape distribution is instead weighted by
$Z(\mathcal T,\mathcal G)$ at each prescribed length.

\section{Analytical results for finite trees} \label{AppC}

\subsection{Frozen regime}

To orient ourselves consider the frozen regime $r=0$, which means that $x < 1/8$. Then $\tilde n$ is fixed by
\eq{
n & = N e^{n/p} a \sum_{k\geq 1} \frac{k z^k}{k-a} \notag \\
& \approx N e^{n/p} a \sum_k z^k \notag \\
& = N e^{n/p} a z/(1-z)
}
where $z = N^{a/2x} e^{-\tilde n}$ and we have assumed $a \ll 1$, which is $8x \ll 1$. This gives
\eqs{
z^{-1} = 1 + e^{n/p}/G .
}
where $G = n/Na$. We still have to fix $p$ through
\eq{
p & = \sum_k p_k = N e^{n/p} a \sum_{k\geq 1} \frac{z^k}{k-a} \notag \\
& \approx N e^{n/p} a \sum_k z^k/k \notag \\
& = N e^{n/p} a \log(1/(1-z))
}
whence
\eq{
n/p = W\left(G / \log(1/(1-z))\right) 
}
and then
\eqs{
z^{-1} & = 1 + \frac{1}{G} e^{W\left(G/\log(1/(1-z)) \right) } \\
& = 1 + \frac{1}{\log(1/(1-z)) W\left(G/\log(1/(1-z)) \right)} 
}
which becomes
\eqs{
\frac{z}{1-z} & = \log\left(\frac{1}{1-z}\right) W\left(\frac{G}{\log(1/(1-z))}\right) 
}
This can be rewritten as
\eqs{
\log \left(\frac{1}{1-z}  \right) \frac{1-z}{z} \log \left(\frac{1-z}{z} G \right) = 1
}
This simplifies if $z \to 1^-$ in which it becomes 
\eqs{
\log(1-z) & \approx W_{-1}\left(-\frac{1}{\log G}\right) \\
& \approx -\log\log G -\log\log\log G 
}
so that $1-z \approx 1/(\log G \log \log G )$ which finally gives $n/p = W(G / (\log\log G+\log\log\log G))$. The $z \to 1^-$ regime is reached for small $x$. 

The result is that the rule use spectrum should follow $p_k/p \propto z^k/k$ at small $x$, asymptotically independent of $x$ (since $z \to 1^-$). This is observed in the data (Fig.\ref{fig2}ab).

\subsubsection{Critical region:} 
This result breaks down as $x \to 1/8^-$ because then the denominator $k-a\sim k-\sqrt{8x}$ in the sums determining $n$ and $p$ has a singularity for $k=1$; this singularity can be extracted separately, extending the theory to the critical region. This changes the $n$ and $p$ equations to
\eq{
n & = N e^{n/p} a \sum_{k\geq 1} \frac{k z^k}{k-a} \notag \\
& \approx N e^{n/p} a \left[ \frac{z}{1-a} + \sum_{k \geq 2} z^k \right] \notag \\
& = N e^{n/p} a \left[ \frac{z}{1 -a } + \frac{z^2}{1-z} \right] 
}
and
\eq{
p & = \sum_k p_k = N e^{n/p} a \sum_{k\geq 1} \frac{z^k}{k-a} \notag \\
& \approx N e^{n/p} a \left[ \frac{z}{1-a} + \sum_{k \geq 2} z^k/k \right] \notag \\
& = N e^{n/p} a \left[ \frac{z}{1 -a } + \log(1/(1-z)) - z \right] 
}
In the critical region $\delta x = 1/8-x \ll 1$, we have
\eqs{
G \approx e^{n/p} z [1/ (4\delta x) + 1], \qquad p/n = 1 + \OO(\delta x)
}  
Therefore $x=1/8$ is where $p/n \to 1$ from below, in finite trees.

\subsection{High temperature regime: } For small trees it is convenient to make an approximation considering only the $k=1$ and $k=2$ terms in the sums. At high temperature and in small trees, most rules are used uniquely; if we keep only the $k=1$ and $k=2$ terms we have
\eq{
n & \approx N^3 e^{n/p} e^{-\tilde n} [ g_1 + 2 g_2 e^{-\tilde n}] \\
p & \approx N^3 e^{n/p} e^{-\tilde n} [ g_1 + g_2 e^{-\tilde n}] 
}
so that 
\eq{
e^{-\tilde n} & = \frac{g_1}{4g_2} \left[ -1 + \sqrt{1 + \frac{8n g_2 e^{-n/p}}{N^3 g_1^2} } \right] \\
& \approx \frac{n}{N^3 g_1} e^{-n/p} \notag
}
and
\eqs{
n/p = 1 + g_2 e^{-\tilde n}/(g_1 + g_2 e^{-\tilde n})
}
This result can be used in the high-temperature regime $x>1/8$.


\section{Analytical results for large corpora} \label{AppD}

Here we derive analytical results in the large corpus regime. First, we show how multiplicity sums are converted into disorder integrals.

\subsection{Conversion of multiplicity sums to disorder integrals}
\label{Conversion}

We first record a general relation that will be used repeatedly below.
For the large-corpus saddle define
\begin{equation}
    C_k \equiv
    {k+m_c-1\choose m_c-1}
    e^{-k\tilde n+\lambda u_k},
    \qquad
    D(X)=1+\sum_{k\geq 1} C_k e^{kX}.
\end{equation}
Eq.\eqref{pkdep} can then be written
\begin{equation}
    p_k
    =
    N^3
    \left\langle
    \frac{C_k e^{kX}}{D(X)}
    \right\rangle_{X<X_*}.
    \label{eq:pk-local}
\end{equation}
Consequently, for any function $F(k)$,
\begin{equation}
    \sum_{k\geq1} p_k F(k)
    =
    N^3
    \left\langle
    \frac{\sum_{k\geq1}F(k)C_k e^{kX}}
         {D(X)}
    \right\rangle_{X<X_*}.
    \label{eq:k-sum-general}
\end{equation}
Thus the multiplicity sums may be performed locally at fixed
disorder $X$, before carrying out the disorder average.

In the recurrent sector $k=O(m_c)$, with $m_c\gg1$, write
\begin{equation}
    k=m_c \kappa,
    \qquad
    r=\frac{\lambda}{m_c},
    \qquad
    s=e^{X-X_*},
    \qquad
    z=e^{X_*-\tilde n}.
\end{equation}
Using Stirling's approximation, the terms in $D(X)$ have the
large-$m_c$ form
\begin{equation} \label{saddleform}
    C_k e^{kX} \sim e^{ m_c\left[ h(\kappa)+\kappa\log(zs)+r\frac{\kappa}{1+\kappa} \right] },
\end{equation}
where    $h(\kappa)=(1+\kappa)\log(1+\kappa)-\kappa\log \kappa$. For fixed $X$, or equivalently fixed $s$, the sum over $k$ is
therefore sharply concentrated at $k=m_c \kappa(s)$, where
\begin{equation}
    h'(\kappa)+\log(zs)+\frac{r}{(1+\kappa)^2}=0.
    \label{eq:local-recurrent-saddle}
\end{equation}
Hence, for observables controlled by the recurrent sector,
Eq.~\eqref{eq:k-sum-general} reduces at leading order to
\begin{equation}
    \frac{1}{N}\sum_k p_k F(k) \simeq N^2 \int_{-\infty}^{X_*}dX\,P(X)\,
    F\!\left(m_c \kappa(e^{X-X_*})\right) .
    \label{eq:k-to-X}
\end{equation}
The exponential factor has cancelled from the numerator and denominator.

The remaining integral is simplified by introducing
$s=e^{X-X_*}$.  Since
\begin{equation}
    P(X_*+\log s)
    =
    P(X_*)\,
    s^{-a}
    e^{-\tilde\epsilon_d(\log s)^2},
    \qquad
    a=2\tilde\epsilon_d X_*,
\end{equation}
and extreme-value normalization gives $N^2P(X_*)\simeq a,$ we obtain
\begin{equation}
    N^2P(X)\,dX \simeq  a\,ds\, s^{-1-a} e^{-\ed(\log s)^2}.
    \label{eq:edge-measure-s}
\end{equation}
Thus, neglecting the finite-$N$ Gaussian curvature of the edge,
\begin{equation}
    \frac{1}{N}\sum_k p_k F(k) \simeq a\int_0^1 ds\,s^{-1-a}
    F\!\left(m_c \kappa(s)\right) ,
    \label{eq:k-to-s-general}
\end{equation}
which will be used repeatedly below.



\subsection{Frozen regime}
\label{app:recurrent_saddle}

Here we determine the dependence of the novelty scale $\Theta$ on
the corpus length.  The calculation leading to Eq.\eqref{pfrozen} involves
rules with $k=\OO(1)$, whereas the constraints in Eqs.\eqref{saddleconstr},\eqref{lambdaoverlapmain}
also receive a finite contribution from recurrent rules with
$k=\OO(m_c)$. For the latter write
\begin{equation}
    k=m_c \kappa(s),\qquad
    s=e^{X-X_\ast},\qquad
    r=\frac{\lambda}{m_c}.
\end{equation}
The saddle equation for $\kappa$ is 
\begin{equation}
    \log\frac{1+\kappa}{\kappa}
    +\log(zs)
    +\frac{r}{(1+\kappa)^2}=0.
    \label{eq:rec_kappa}
\end{equation}
Applying Eq.\ref{eq:k-to-s-general}, the correlation-volume constraint and the $m_c$ stationarity
condition become
\begin{equation}
    1= a\int_0^1ds\,s^{-1-a}
    \frac{\kappa(s)}{1+\kappa(s)},
    \label{eq:rec_corr}
\end{equation}
and, using
\eqs{
    \psi(k+m_c)-\psi(m_c)
    \simeq \log(1+\kappa),
    \qquad
    \frac{k}{(k+m_c-1)^2}
    \simeq
    \frac{1}{m_c}\frac{\kappa}{(1+\kappa)^2},
}
\begin{equation}
    r=
    \frac{\displaystyle
    \int_0^1ds\,s^{-1-a}\log[1+\kappa(s)]}
    {\displaystyle
    \int_0^1ds\,s^{-1-a}
    \kappa(s)/[1+\kappa(s)]^2}.
    \label{eq:rec_r}
\end{equation}
Equations~\eqref{eq:rec_kappa}--\eqref{eq:rec_r} are intensive:
they contain $z$, $r$, and $a$, but not the corpus length $n$.
They therefore determine $z$ and $\lambda/m_c$ at fixed $a$.

The remaining occurrence constraint gives
\begin{equation}
    \frac{n}{N}
    =
    m_c a\int_0^1ds\,s^{-1-a}\kappa(s),
\end{equation}
and hence
\begin{equation}
    \Theta=m_cze^{\lambda/m_c}
    =\frac{c(a)}{a} \frac{n}{N},
\end{equation}
where
\begin{equation}
    c(a)\equiv
    \frac{z e^{\lambda/m_c}}
    {\displaystyle\int_0^1ds\,s^{-1-a}\kappa(s)}.
    \label{eq:C_general}
\end{equation}

The important point is that $c(a)$ is nonsingular throughout the
regular frozen scaling regime.  Indeed, as $s\to0$, Eq.~\eqref{eq:rec_kappa} gives $\kappa(s)\simeq z e^{\lambda/m_c}s$, so that the integrand appearing in Eq.~\eqref{eq:C_general} behaves as
\eqs{
    s^{-1-a}\kappa(s) \simeq z e^{\lambda/m_c}s^{-a}.
}
For every fixed $a<1$ this is integrable at $s=0$.  Thus, away
from the boundaries of the frozen scaling regime, $c(a)$ contains
only order-one dependence on $a$, which establishes the scaling variable
\begin{equation}
    G=\frac{n}{Na},
\end{equation}
since $\Theta=\OO(G)$ throughout this regime.

Close to the freezing transition, the leading behaviour can be extracted from the the small-$s$ part of the integral: $\int_0^1ds\,s^{-a}=1/(1-a).$ In this limit $\kappa\ll1$ over the dominant range, and
Eqs.~\eqref{eq:rec_corr}--\eqref{eq:rec_r} give
\begin{equation}
    z e^{\lambda/m_c}\simeq\frac{1-a}{a},
    \qquad
    \frac{\lambda}{m_c}\simeq1,
    \qquad
    m_c\simeq\frac nN.
\end{equation}
Consequently $c(a) \simeq 1-a$ and $z \approx (1-a)/(ae)$ as $a \to 1^-$.

If $n/N$ is fixed and $a$ is sufficiently small, then in the full discrete equations, $m_c$ decreases to $\OO(1)$, where the $m_c \gg 1$ saddle no longer applies. A description of this regime in terms of Poisson-Dirichlet statistics \cite{Pitman97} can be found and has a similar final result for $p$, namely $p/N \approx \Gamma(m_c+a)/(\Gamma(m_c)\Gamma(1+a))$ (derivation not shown). When $m_c \approx n/N$ then $p/N \approx (n/N)^a/\Gamma(1+a)$, which has the same corpus-length behavior as above, in the scaling limit, without the renormalization of $a$. This expression does however lead to $p/N \to 1$ as $a \to 0$, as required.

\subsubsection{Critical correction to $c(a)$.}
\label{app:ccrit}

The result $c(a)\simeq1-a$ becomes nonuniform as $a\to1^{-}$. In this region, occurrence is dominated by rules that appear uniquely $(k=1)$ and the above saddle approach is inappropriate. However it is straightforward to just keep the $k=1$ terms in their exact form. 

%
%

This leads to
\[
\frac{n}{N}\simeq a\Theta\int_0^1ds\,s^{-a}e^{-\tilde\epsilon_d(\log s)^2}.
\]
Using $G=n/(Na)$ and $\Theta=c_{\rm crit}G$ gives
\eq{
c_{\rm crit}^{-1}=\int_0^1ds\,
s^{-a}e^{-\tilde\epsilon_d(\log s)^2} =
\frac{\sqrt{\pi}}{2\sqrt{\tilde\epsilon_d}}\,
e^{\frac{(1-a)^2}{4\tilde\epsilon_d} } \erfc\left[ \frac{1-a}{2\sqrt{\tilde\epsilon_d}} \right].
}
Hence
\begin{equation}
c_{\rm crit}(a,\tilde\epsilon_d)
= 2\sqrt{\frac{\tilde\epsilon_d}{\pi}}\, \frac{ e^{-(1-a)^2/(4\tilde\epsilon_d)} }{ \erfc\left[(1-a)/(2\sqrt{\tilde\epsilon_d})\right] }.
\end{equation}

This expression matches the frozen result away from the
critical region.  For
$1-a\gg\sqrt{\tilde\epsilon_d}$,
\[
c_{\rm crit} = (1-a) +\frac{2\tilde\epsilon_d}{1-a} +\cdots,
\]
and therefore $c_{\rm crit}\simeq1-a$ to leading order.
Within the finite-$N$ critical layer,
$1-a=O(\sqrt{\tilde\epsilon_d})$, the Gaussian curvature
instead cuts off the apparent singularity.  In particular,
\[
c_{\rm crit}(1,\tilde\epsilon_d)=2\sqrt{\frac{\tilde\epsilon_d}{\pi}}.
\]
Thus $\Theta$ remains finite at the rounded transition,
rather than vanishing as predicted by the sharp-edge
approximation.

The same expression follows from the transition-side
occurrence constraint, Eq.\eqref{eq:transition_n}, continued to $a<1$.
Thus the finite-$N$ correction provides a common
description of the first moment on both sides of the
freezing transition.

\subsection{Transition regime}


First consider the occurrence constraint $n=\sum_k kp_k$.
Using
$\sum_{k\geq1}kt^ke^{-t}/k!=t$, we obtain
\eq{
 \frac{n}{N} &\simeq
 a\Theta^a e^{-\tilde\epsilon_d(\log\Theta)^2}
 \int_0^\Theta dt\,
 t^{-a_R}e^{-\tilde\epsilon_d(\log t)^2} \notag\\
 & =
 \frac{a\Theta\sqrt{\pi}}{2\sqrt{\tilde\epsilon_d}}\, e^{(a-1)^2/4\tilde\epsilon_d} \erfc\!\left( \frac{1-a}{2\sqrt{\tilde\epsilon_d}}\right).
 \label{eq:transition_n}
}
Thus $\Theta$ is proportional to $n$ at fixed $N$ and $x$.
This is the main simplification in the transition regime.

The same approximation applied to the second constraint in
Eq.\eqref{saddleconstr} gives
$N\simeq m_c^{-1}\sum_k kp_k=n/m_c$, and hence
$m_c\simeq n/N$.  Expanding Eq.~(\ref{lambdaoverlapmain}) for the same
$k/m_c\ll1$ sector gives $\lambda/m_c\simeq1$.
Consequently $z\simeq\Theta/[e n/N]$.


Away from the finite-$N$ crossover
$|a-1|\lesssim\sqrt{\tilde\epsilon_d}$, the error function in
Eq.~\eqref{eq:transition_n} may be expanded.  For $a>1$ this gives
\begin{equation}
 \Theta\simeq
 \frac{n/N}{\sqrt{8\pi\log N}}\,
 N^{-\chi(x)},\qquad
 \chi(x)=2+\frac{1}{4x}-\sqrt{\frac{2}{x}},
 \label{app:transition_theta}
\end{equation}
where in the final form we used $a\simeq\sqrt{8x}$.
Thus $\chi(1/8)=0$ and $\chi(1/2)=1/2$.  Close to
$x=1/8$ the full error-function expression in
Eq.~\eqref{eq:transition_n} should instead be retained.

We next calculate the number of distinct rules.  Summing $p_k$
rather than $kp_k$ gives
\begin{equation}
 \frac{p}{N}\simeq
 a\Theta^a e^{-\tilde\epsilon_d(\log\Theta)^2}
 \int_0^\Theta dt\,
 t^{-1-a_R}(1-e^{-t})
 e^{-\tilde\epsilon_d(\log t)^2}.
 \label{eq:transition_p}
\end{equation}
Equations~\eqref{eq:transition_n} and
\eqref{eq:transition_p} give a closed finite-$N$ description of
rule discovery in the transition regime. It is useful to extract their limiting behavior.  Recall that
\begin{equation}
    a_R=a-2\tilde\epsilon_d\log\Theta .
\end{equation}
The character of the rule spectrum changes according to
whether the moving depletion front lies above or below the
Gaussian saddle controlling the occurrence sum.

First consider $a_R>1$.  In this regime almost all
occurrences belong to rules which are used only once.  To
calculate the leading correction, subtract \eqref{eq:transition_p} from \eqref{eq:transition_n}:
\begin{align}
    \frac{n-p}{N} &\simeq a\Theta^a e^{-\tilde\epsilon_d(\log\Theta)^2}
    \int_0^\Theta dt\, t^{-1-a_R} \left(t-1+e^{-t}\right) e^{-\tilde\epsilon_d(\log t)^2}.
    \label{eq:transition_np}
\end{align}
For
\begin{equation}
    a_R-1\gg\sqrt{\tilde\epsilon_d},
    \qquad 1<a_R<2,
\end{equation}
the integral in Eq.~\eqref{eq:transition_np} is localized
at $t=O(1)$.  The Gaussian curvature may therefore be
neglected within the integral and the upper limit extended
to infinity, giving
\begin{equation}
    \int_0^\infty dt\, t^{-1-a_R} \left(t-1+e^{-t}\right) = \frac{\Gamma(2-a_R)}
    {a_R(a_R-1)} .
\end{equation}
On the other hand the occurrence integral in Eq.\eqref{eq:transition_n} is
controlled by its interior Gaussian saddle.  Hence
\begin{equation}
    \frac{p}{n}
    \simeq 1- \sqrt{\frac{\tilde\epsilon_d}{\pi}}\,
    \frac{\Gamma(2-a_R)}
    {a_R(a_R-1)}
    e^{-\frac{(a_R-1)^2}{4\tilde\epsilon_d} } .
    \label{eq:transition_unique}
\end{equation}
Thus, for fixed $a_R>1$, repeated rule use is
exponentially rare in the scaling limit and $p\simeq n$.

The crossover can be resolved explicitly when
\begin{equation}
    |a_R-1|=O(\sqrt{\tilde\epsilon_d}).
\end{equation}
Writing $v=-\log t$, the occurrence measure is
proportional to $e^{(a-1)v-\tilde\epsilon_d v^2}$ and is therefore concentrated around
\begin{equation}
    v_*=\frac{a-1}{2\tilde\epsilon_d}.
\end{equation}
The depletion threshold $\Theta u\sim1$ corresponds to
$v=\log\Theta$.  Since
\begin{equation}
    \sqrt{\tilde\epsilon_d}
    \left( \log\Theta-v_* \right) = \frac{1-a_R}
    {2\sqrt{\tilde\epsilon_d}},
\end{equation}
the fraction of occurrences lying on the effectively
unique-rule side of the front is, to leading order,
\begin{equation}
    \frac{p}{n} \simeq \frac{1}{2}
    \erfc\left[
        \frac{1-a_R}
        {2\sqrt{\tilde\epsilon_d}}
    \right].
    \label{eq:transition_crossover}
\end{equation}
In particular $p/n\simeq1/2$ at $a_R=1$.

The corresponding corpus scale is obtained from
\begin{equation}
    \log\Theta_c
    =
    \frac{a-1}{2\tilde\epsilon_d}.
\end{equation}
Using Eq.\eqref{eq:transition_n}, and taking $a>1$ outside the finite-$N$
rounding region, this gives
\begin{equation}
    \left(\frac{n}{N}\right)_c
    \simeq
    \sqrt{8\pi\log N}\,
    N^{\,2-1/(4x)} .
    \label{eq:transition_repetition_crossover}
\end{equation}
Thus an $\OO(1)$ fraction of repeated occurrences appears
only when the moving front reaches the Gaussian saddle,
$a_R=1+\OO(\sqrt{\tilde\epsilon_d})$.

At still larger corpus size the front lies inside the frozen
tail,
\begin{equation}
    1-a_R\gg\sqrt{\tilde\epsilon_d},
    \qquad 0<a_R<1.
\end{equation}
Eq.\eqref{eq:transition_p} is then itself localized at $t=\OO(1)$, so
that
\begin{align}
    \frac{p}{N}
    &\simeq
    a\Theta^a
    e^{-\tilde\epsilon_d(\log\Theta)^2}
    \int_0^\infty dt\,
    t^{-1-a_R}(1-e^{-t})
    \nonumber\\
    &=
    a\Theta^a
    e^{-\tilde\epsilon_d(\log\Theta)^2}
    \frac{\Gamma(1-a_R)}{a_R}.
    \label{eq:transition_moving_front}
\end{align}
Dividing by Eq.\eqref{eq:transition_n} gives
\begin{equation}
    \frac{p}{n}
    \simeq
    \sqrt{\frac{\tilde\epsilon_d}{\pi}}\,
    \frac{\Gamma(1-a_R)}{a_R}
    e^{-\frac{(1-a_R)^2}{4\tilde\epsilon_d}} .
    \label{eq:transition_moving_front_ratio}
\end{equation}
In this regime the local novelty exponent is again set by the moving edge,
\begin{equation}
    \beta_{\rm eff} = \frac{d\log p}{d\log n}
    \simeq a_R,
\end{equation}
up to corrections of order $1/\log N$.

We therefore find three regimes within the transition phase:
\begin{equation}
    \frac{p}{n}
    \simeq
    \begin{cases}
    \displaystyle 1- \sqrt{\frac{\tilde\epsilon_d}{\pi}}\,
    \frac{\Gamma(2-a_R)}
    {a_R(a_R-1)}
    e^{-(a_R-1)^2/(4\tilde\epsilon_d)},
    &
    a_R-1\gg\sqrt{\tilde\epsilon_d},
    \\[4mm]
    \displaystyle
    \frac12
    \erfc\left[
        \frac{1-a_R}
        {2\sqrt{\tilde\epsilon_d}}
    \right],
    &
    |a_R-1|=\OO(\sqrt{\tilde\epsilon_d}),
    \\[4mm]
    \displaystyle
    \sqrt{\frac{\tilde\epsilon_d}{\pi}}\,
    \frac{\Gamma(1-a_R)}{a_R}
    e^{-(1-a_R)^2/(4\tilde\epsilon_d)},
    &
    1-a_R\gg\sqrt{\tilde\epsilon_d}.
    \end{cases}
    \label{app:transition_three_regimes}
\end{equation}
The first regime has essentially unique rule use, the
second is the crossover to an $\OO(1)$ repeated fraction,
and the third is the moving-front regime, like the frozen regime.

\subsection{Entropy}
\label{app:recurrent_entropy}

Here we derive the leading hidden entropy in the frozen and
transition regimes.  We begin from
\begin{equation}
 2nH_d = n\tilde n-\lambda N
 +N^3\left\langle
 (1-X\p_X)\log D(X)
 \right\rangle_{X<X_*}.
 \label{eq:Hd_app}
\end{equation}

\subsubsection{Frozen region}

In the frozen regime the leading entropy is controlled by the
recurrent sector rather than by the $k=\OO(1)$ rules responsible
for novelty.  Using the recurrent saddle introduced above,
the local large-$m_c$ form of $D(X)$ is
\[
 \log D(X) \simeq m_c\left[ h(\kappa)+\kappa\log(zs)+r\frac{\kappa}{1+\kappa} \right].
\]
Since $\kappa(s)$ extremizes the expression in brackets,
differentiation with respect to $X$ gives $\p_X\log D(X)\simeq m_c \kappa(s)$. Using the same coordinates as above, Eq.~\eqref{eq:Hd_app} becomes
\[
 \begin{split}
 2nH_d \simeq {}& n\tilde n-\lambda N +Nm_ca\int_0^1ds\,s^{-1-a} \left[ h(\kappa)+\kappa\log(zs) +r\frac{\kappa}{1+\kappa}-X\kappa \right].
 \end{split}
\]
Now $\log(zs)=X-\tilde n$.  The term proportional to
$\tilde n$ cancels the first term by the occurrence constraint,
while the term proportional to $r$ cancels $-\lambda N$ by
Eq.~\eqref{eq:rec_corr}.  Hence
\[
 2nH_d = Nm_ca\int_0^1ds\,s^{-1-a}h(\kappa).
\]
Using the occurrence constraint once more gives
\begin{equation}
 2H_d=
 \frac{\displaystyle
 \int_0^1ds\,s^{-1-a}h(\kappa)}
 {\displaystyle
 \int_0^1ds\,s^{-1-a}\kappa}.
 \label{eq:H_recurrent}
\end{equation}
Thus the leading frozen entropy is entirely determined by the
recurrent multiplicity entropy $h(\kappa)$.

This expression can be reduced further without solving explicitly
for $\kappa(s)$.  Write
\[
 J=\int_0^1ds\,s^{-1-a}\kappa(s).
\]
Using Eq.~\eqref{eq:rec_kappa} together with
$h(\kappa)-\kappa h'(\kappa)=\log(1+\kappa)$ gives
\[
 h(\kappa) = - \kappa \log(zs) -r\frac{\kappa}{(1+\kappa)^2} +\log(1+\kappa).
\]
The last two terms cancel after integration by
Eq.~\eqref{eq:rec_r}, so that
\[
 \int_0^1ds\,s^{-1-a}h(\kappa)
 =  -J\log z-\int_0^1ds\,s^{-1-a}\kappa\log s .
\]
To evaluate the remaining term, let $\Phi(s)$ denote the
saddle exponent in Eq.~(\ref{saddleform}), namely $\Phi(s) = h(\kappa)+\kappa\log(zs)+r\kappa/(1+\kappa)$.  Stationarity gives
$\p_{\log s}\Phi=\kappa$, while Eq.~\eqref{eq:rec_kappa} gives
\[
 \Phi= \log(1+\kappa)+r\left(\frac{\kappa}{1+\kappa}\right)^2 .
\]
Equations~\eqref{eq:rec_corr} and \eqref{eq:rec_r} then imply
\[
 a\int_0^1ds\,s^{-1-a}\Phi(s)=r.
\]
Differentiating this relation with respect to $a$ along the
recurrent saddle, the derivatives through $r$ cancel because
\[
 a\int_0^1ds\,s^{-1-a}\p_r\Phi = a\int_0^1ds\,s^{-1-a}\frac{\kappa}{1+\kappa} =1.
\]
The derivative through $z$ gives
$aJ\,\p_a\log z$, while at fixed $z$ and $r$ an integration
by parts using $\p_{\log s}\Phi=\kappa$ gives
\[
 \left.
 \p_a\left[ a\int_0^1ds\,s^{-1-a}\Phi \right]\right|_{z,r}
 = -\int_0^1ds\,s^{-1-a}\kappa\log s ,
\]
 where the boundary terms vanish. It follows that
\[
 \int_0^1ds\,s^{-1-a}\kappa\log s
 = a J\,\frac{\p\log z}{\p a}.
\]
Substituting above finally gives
\begin{equation}
 H_d = -\frac12 \left(1+a\frac{\p}{\p a}\right)\log z(a).
 \label{eq:H_from_z_app}
\end{equation}
The derivative is taken along the intensive recurrent solution
determined in the preceding subsection.

Using the leading solution in the frozen phase, $z \approx (1-a)/(ae)$, this leads to
\eq{ \label{app:Hd}
H_{d} = \frac12 + \frac12 \log\frac{a}{1-a} +\frac{1}{2(1-a)} + \ldots
}

\subsubsection{Entropy in the critical region.}

The sharp-edge result becomes nonuniform as $a\to1^{-}$.
To obtain the finite-$N$ rounding, we retain the Gaussian
curvature in the edge measure,
\[
N^2P(X)\,dX
\simeq
a\,ds\,s^{-1-a}
e^{-\ed(\log s)^2}.
\]
In the critical region $1-a=\OO(\sqrt{\ed})$, the recurrent saddle has $\kappa\ll1$.  To leading order,
\[
r=1+o(1),\qquad
\kappa(s)=Bs+o(B),\qquad
B=ze^r .
\]
The correlation-volume constraint therefore gives $1\simeq a B I_0$ with
\[
I_0 = \int_0^1 ds\, s^{-a}e^{-\ed(\log s)^2} = \frac{\sqrt{\pi}}{2\sqrt{\ed}}\,
e^{u^2}\erfc(u),
\]
with $u=\frac{(1-a)}{2\sqrt{\ed}}$.

The entropy expression may be treated in the same way.
Since $h(\kappa)=\kappa(1-\log \kappa)+\OO(\kappa^2)$, the leading critical contribution is
\[
2H_{\rm crit} = 1+\log(aI_0) + \p_a \log I_0 .
\]
Thus
\[
2H_{\rm crit} =1+ \log\left[ \frac{a\sqrt{\pi}}{2\sqrt{\ed}}\, e^{u^2}\erfc(u) \right] + \frac{e^{-u^2}} {\sqrt{\pi\ed}\,\erfc(u)} -\frac{1-a}{2\ed}.
\]
For $1-a\gg\sqrt{\ed}$ this reduces exactly to Eq.\ref{app:Hd}, hence it furnishes an improvement in the frozen regime.


\subsubsection{Entropy in the transition regime}

We now consider $1/8<x<1/2$.  Here the assumptions are
those of the transition calculation above: $m_c\gg1$, the
multiplicities carrying the occurrence constraint satisfy
$k\ll m_c$, and hence
\[
D(X)\simeq e^t,\qquad
m_c\simeq\frac{n}{N},\qquad
\frac{\lambda}{m_c}\simeq1.
\]
In particular $\p_X\log D=t$ and
$n=N^3\langle t\rangle$.  Substitution in Eq.~(\ref{eq:Hd_app}) gives
\[
2nH_d = n\tilde n-\lambda N+n -N^3\langle Xt\rangle .
\]
Since $\lambda N\simeq m_cN\simeq n$, the middle two terms
cancel.  Using the same change of variables as in the
transition calculation therefore gives
\begin{equation}
2H_d= \tilde n-
\frac{\displaystyle
\int_0^\Theta dt\,
t^{-a_R}e^{-\tilde\epsilon_d(\log t)^2}
\left[X_*+\log(t/\Theta)\right]}
{\displaystyle
\int_0^\Theta dt\,
t^{-a_R}e^{-\tilde\epsilon_d(\log t)^2}} .
\end{equation}
Since $\tilde n=X_*-\log z$ and
$z\simeq\Theta/(e m_c)$, this simplifies to
\begin{equation}
2H_d = 1+\log m_c-
\frac{\displaystyle
\int_0^\Theta dt\,
t^{-a_R}e^{-\tilde\epsilon_d(\log t)^2}\log t}
{\displaystyle
\int_0^\Theta dt\,
t^{-a_R}e^{-\tilde\epsilon_d(\log t)^2}} .
\label{D52}
\end{equation}

For fixed $x>1/8$, away from the finite-$N$ rounding of
the freezing transition, the denominator is controlled by
an interior Gaussian saddle.  Writing $u=\log t$, its
exponent is $(1-a_R)u-\tilde\epsilon_d u^2 $ so that
\[
\log t \simeq \log\Theta-\frac{a-1}{2\tilde\epsilon_d}.
\]
Eq.~(\ref{D52}) then becomes
\[
2H_d \simeq 1+\log\frac{m_c}{\Theta} +\frac{a-1}{2\tilde\epsilon_d}.
\]
The occurrence constraint derived above gives, using
$m_c\simeq n/N$,
\[
\frac{m_c}{\Theta} \simeq a\sqrt{\frac{\pi}{\tilde\epsilon_d}}\, e^{\frac{(a-1)^2}{4\tilde\epsilon_d} },
\]
and hence
\[
2H_d \simeq 1+\log\left( a\sqrt{\frac{\pi}{\tilde\epsilon_d}} \right) +\frac{a^2-1}{4\tilde\epsilon_d}.
\]
Finally, using
$a^2=2w\tilde\epsilon_d$,
$w=W_0(N^4/2\pi)$, and
$we^w=N^4/(2\pi)$, the terms involving $w$ combine to
give
\begin{equation}
H_d \simeq \left(1-\frac{1}{8x}\right)\log N+\frac12, \qquad
\frac18<x<\frac12 .
\label{app:Htrans}
\end{equation}
Close to $x=1/8$ the Gaussian saddle approaches the
upper cutoff, and the full ratio of integrals in Eq.~(\ref{D52}) should be retained.

\subsection{Energy}

The hidden energy follows directly from the same on-shell
partition function.  By stationarity,
\[
\langle E_d\rangle= 2\log N -\frac{N^3}{n} \left\langle X\p_X\log D(X) \right\rangle_{X<X_*}.
\]

In the frozen regime,
$\p_X\log D\simeq m_c \kappa(s)$, with
$s=e^{X-X_*}$.  Using the occurrence constraint therefore
gives
\[
\langle E_d\rangle = 2\log N - \frac{ \displaystyle
\int_0^1 ds\,s^{-1-a}\kappa(s)
\left[X_*+\log s\right]
}{
\displaystyle
\int_0^1 ds\,s^{-1-a}\kappa(s)
}.
\]
For fixed $a<1$ the second term in the numerator gives only
an $\OO(1)$ correction, since the relevant disorder remains
within an $\OO(1)$ distance of the extreme-value edge.
Hence
\begin{equation}
\langle E_d\rangle
\simeq 2\log N-X_*+ \OO(1) = 2\log N-\sqrt{\frac{w}{2\tilde\epsilon_d}}+\OO(1)
\simeq
\left(2-\sqrt{\frac{2}{x}}\right)\log N+\OO(1),
\qquad x<\frac18 .
\end{equation}

Above freezing, the multiplicities carrying the occurrence
constraint satisfy $k\ll m_c$, so that
$\p_X\log D=t$, where
$t=\Theta e^{X-X_*}$.  The factor entering the energy is
therefore proportional to $e^{X-\tilde\epsilon_dX^2}$, truncated at $X=X_*$.  Thus
\[
\langle E_d\rangle = 2\log N - \frac{ \displaystyle \int_{-\infty}^{X_*}dX\,
X e^{X-\tilde\epsilon_dX^2}
}{
\displaystyle \int_{-\infty}^{X_*}dX\,e^{X-\tilde\epsilon_dX^2}
}.
\]
Away from the finite-$N$ crossover, the Gaussian saddle
$X=1/(2\tilde\epsilon_d)$ lies below the upper cutoff, and
therefore
\begin{equation}
\langle E_d\rangle \simeq 2\log N-\frac{1}{2\tilde\epsilon_d} = \left(2-\frac{1}{2x}\right)\log N,
\qquad x>\frac18 .
\end{equation}
The frozen and transition expressions coincide at
$x=1/8$ in the scaling limit.

At finite $N$ the crossover is rounded.  The truncated
Gaussian integrals can be evaluated explicitly, giving
\begin{equation}
\langle E_d\rangle \simeq 2\log N-\frac{1}{2\tilde\epsilon_d} +
\frac{ e^{-(a-1)^2/(4\tilde\epsilon_d)} }{ \sqrt{\pi\tilde\epsilon_d}\, \erfc[(1-a)/(2\sqrt{\tilde\epsilon_d})] }.
\end{equation}
This expression interpolates through the critical region;
outside it the last term is subleading on the transition
side, while on the frozen side the disorder is instead
pinned to the extreme-value edge as above.

\section{A document-level ensemble of grammars}
\label{app:document_grammar_mixture}


Here consider a grammar mixture in which document-level structure is controlled by a second RLM set of log-weights. The resulting grammar mixture gives a contribution to block entropy which is controlled by the number of distinct rules encountered in a corpus.

\subsection{Two levels of quenched disorder}

Let $\GG_0$ denote a universal grammar, drawn once and shared by all
documents.  For each document $d$, write its branching-rule log-weights as
\begin{equation}
 X_{abc}^{(d)}=X_{abc}^{(0)}+\eta_{abc}^{(d)},
 \label{eq:doc_weights}
\end{equation}
where the two contributions are independent Gaussian variables,
\begin{equation}
 \overline{(X_{abc}^{(0)})^2}=\frac{1}{2\tilde\epsilon_0},
 \qquad
 \overline{(\eta_{abc}^{(d)})^2}=\sigma^2
 =\frac{1}{2\tilde\epsilon_{\rm doc}}.
 \label{eq:doc_variances}
\end{equation}
The $\eta^{(d)}$ are drawn independently between documents but are
quenched within a document: repeated uses of a rule share the same
perturbation.  We keep the terminal weights unchanged for simplicity.
Summing independent Gaussians shows that the marginal grammar of a
random document is an ordinary RLM, with
\begin{equation}
 \frac{1}{\tilde\epsilon_{\rm eff}}
 =\frac{1}{\tilde\epsilon_0}
 +\frac{1}{\tilde\epsilon_{\rm doc}}.
 \label{eq:doc_effective_temp}
\end{equation}
The novelty spectrum of the marginal document ensemble is therefore
governed by the usual RLM theory at $\tilde\epsilon_{\rm eff}$.
The correlations between different documents, however, depend on the
variance carried by $\GG_0$ and cannot be obtained from
$\tilde\epsilon_{\rm eff}$ alone.

Let $D_K$ be a corpus containing $K$ branching-rule uses, including
its hidden derivations for the moment.  Conditional on the known
universal grammar, the distribution of this corpus is a mixture,
\begin{equation}
 P(D_K\mid \GG_0)
 =\int d\eta\,P(\eta)P(D_K\mid \GG_0,\eta).
 \label{eq:doc_mixture}
\end{equation}
The exact entropy identity is
\begin{equation}
 \begin{split}
 H(D_K\mid \GG_0)={}&
 \mathbb E_{\eta}H(D_K\mid \GG_0,\eta) +I(\eta;D_K\mid \GG_0).
 \end{split}
 \label{eq:doc_entropy_identity}
\end{equation}
Both entropies on the right and left are evaluated at the same
$N,\tilde\epsilon_0,\tilde\epsilon_{\rm doc}$.  The first term on the
right, after averaging over $G_0$, is the mean fixed-grammar RLM
entropy at $\tilde\epsilon_{\rm eff}$.  


To estimate the last term of Eq.~\eqref{eq:doc_entropy_identity},
consider first an idealized rule-sampling process.  Conditional on the
universal grammar, let $q_r$ be the probabilities of the candidate
rules and $\eta_r$ their independent document perturbations.  In the
normalized version of this process,
\begin{equation}
 q_r(\eta)=\frac{q_re^{\eta_r}}
 {\sum_s q_se^{\eta_s}}.
 \label{eq:doc_q}
\end{equation}
The likelihood of an ordered sample with occupancies $k_r$ is
proportional to
\begin{equation}
 P(\{k_r\}\mid\eta,q)
 \propto\frac{e^{\sum_r k_r\eta_r }\prod_rq_r^{k_r}} {(\sum_s q_se^{\eta_s})^K}.
 \label{eq:doc_likelihood}
\end{equation}

A transparent estimate is obtained by a Poisson approximation to the occurrences of rare rules. We replace Eq.\ref{eq:doc_likelihood} by 
\eq{
 P(\{k_r\}\mid\eta,q) \approx \prod_r P(k_r | \eta_r)
}
which is then separable over rules. To define the parameters of the model, introduce a nominal exposure $t_r=Kq_r$, and take
conditionally independent counts
\begin{equation}
 k_r\mid\eta_r\ \sim\
 \operatorname{Poisson}\left(t_re^{\eta_r-\sigma^2/2}\right).
 \label{eq:doc_poisson}
\end{equation}
The factor $e^{-\sigma^2/2}$ fixes the mean intensity of each rule, i.e. $ \mathbb E_\eta (t_re^{\eta_r-\sigma^2/2}) = t_r$, using $ \mathbb E_\eta e^{\eta_r} = e^{\sigma^2/2}$.
For this auxiliary sampling model, information about distinct
$\eta_r$ is additive.  Define the one-rule information function
\begin{equation}
 \mathcal J_\sigma(t)
 =I\!\left(\eta;k\right),\qquad
 \eta\sim\mathcal N(0,\sigma^2),\quad
 k\mid\eta\sim\operatorname{Poisson}
     \left(te^{\eta-\sigma^2/2}\right).
 \label{eq:doc_one_rule_info}
\end{equation}
Then within the Poissonized model,
\begin{equation}
 I(\eta;\{k_r\}\mid\{q_r\}) \approx \sum_r I(\eta_r; k_r) =\sum_r\mathcal J_\sigma(Kq_r).
 \label{eq:doc_poisson_info}
\end{equation}
This formula includes information from observing zero occurrences;
indeed $\mathcal J_\sigma$ is the mutual information of the entire
count variable, not the information conditional on a particular
positive count.  It may be evaluated from a one-dimensional integral:
\begin{align}
 Z_k(t)&=\int\frac{d\eta}{\sqrt{2\pi\sigma^2}}
 \exp\left[-\frac{\eta^2}{2\sigma^2}
 +k(\eta-\sigma^2/2)-te^{\eta-\sigma^2/2}\right],
 \label{eq:doc_Zk}\\
 P(k\mid t)&=\frac{t^k}{k!}Z_k(t),\nonumber\\
 \mathcal J_\sigma(t)&=\sum_{k\ge0}P(k\mid t)
 D_{\rm KL}[P(\eta\mid k,t)\Vert P(\eta)].
 \label{eq:doc_Jscalar}
\end{align}
The posterior in the last line is proportional to the
integrand defining $Z_k$.

In the frozen regime, the ranked rule probabilities are described,
in the simplest independent-rule approximation, by a
Poisson--Dirichlet distribution with index $0<a<1$.  Its mean
frequency density satisfies
\begin{equation}
 \mathbb E\sum_r f(q_r)
 =\frac{1}{\Gamma(a)\Gamma(1-a)}
 \int_0^1dq\,q^{-1-a}(1-q)^{a-1}f(q).
 \label{eq:doc_PD_intensity}
\end{equation}
For fixed $\sigma^2>0$, the information function obeys
$\mathcal J_\sigma(t)=\sigma^2t/2+O(t^2)$ as $t\to0$,
and $\mathcal J_\sigma(t)=\tfrac12\log t+O(1)$ as
$t\to\infty$.  Consequently, the integral of
$t^{-1-a}\mathcal J_\sigma(t)$ converges at both ends.
Applying Eq.~\eqref{eq:doc_PD_intensity} to
Eq.~\eqref{eq:doc_poisson_info} and changing variables
to $u=Kq$, we find
\begin{equation}
 \mathbb E I(\eta;\{k_r\}\mid\{q_r\})
 \simeq\frac{K^a}{\Gamma(a)\Gamma(1-a)}
 \int_0^\infty du\,u^{-1-a}\mathcal J_\sigma(u).
 \label{eq:doc_I_scaling}
\end{equation}
On the other hand, the mean number of distinct rules
under the unperturbed sampling probabilities is
\begin{equation}
 p_0(K)=\mathbb E\sum_r(1-e^{-Kq_r})
 \simeq\frac{K^a}{\Gamma(1+a)}.
 \label{eq:doc_p_scaling}
\end{equation}
Thus the auxiliary model gives the result
\begin{equation}
\mathbb E I(\eta;\{k_r\}\mid\{q_r\}) \simeq c_{\rm doc}(a,\sigma^2)\,p_0(K),
 \label{eq:doc_main_result}
\end{equation}
where
\begin{equation}
 c_{\rm doc}(a,\sigma^2)
 =\frac{a}{\Gamma(1-a)}
 \int_0^\infty du\,u^{-1-a}\mathcal J_\sigma(u).
 \label{eq:doc_c_exact_poisson}
\end{equation}
This integral determines a finite coefficient for any
fixed $0<a<1$ and finite $\sigma^2>0$. 


\subsection{Weak document disorder}

The coefficient can be obtained explicitly when
$\sigma^2\ll1$.  In the range of exposures contributing
to Eq.~\eqref{eq:doc_c_exact_poisson}, the log-likelihood
is locally quadratic and the information about one Gaussian
log-weight becomes
\begin{equation}
 \mathcal J_\sigma(t)
 \simeq\frac12\log(1+\sigma^2t).
 \label{eq:doc_Fisher_J}
\end{equation}
Substitution into Eq.~\eqref{eq:doc_c_exact_poisson}
yields
\begin{align}
 c_{\rm doc}(a,\sigma^2) &\simeq\frac{a}{2\Gamma(1-a)} \int_0^\infty du\,u^{-1-a} \log(1+\sigma^2u)\nonumber\\
 &=\frac{\Gamma(a)}{2}(\sigma^2)^a.
 \label{eq:doc_c_weak}
\end{align}
The nonanalytic dependence on $\sigma^2$ originates in
the heavy tail of the rule-frequency distribution.  The
limits of small document variance and large corpus do
not commute.  For finite $K$ the information begins
analytically in $\sigma^2$, whereas the asymptotic
$K^a$ regime probes progressively smaller rule
frequencies.

\end{widetext} 
\end{document}